\documentclass[aps,prd,preprintnumbers,nofootinbib,superscriptaddress,showpacs,letterpaper,twocolumn,floatfix]{revtex4-1}
\pdfoutput=1
\usepackage{amsmath,amssymb,amsbsy,longtable,tabularx}
\usepackage[utf8]{inputenc}
\usepackage{graphicx,color,ulem,verbatim}
\usepackage[colorlinks=true, linkcolor=blue, filecolor=blue, urlcolor=blue, citecolor=blue, pdftex, plainpages=false]{hyperref}

\newcommand\buaff{Department of Physics and Center for Computational Science, Boston University, Boston, Massachusetts 02215, USA}
\newcommand\coaff{Department of Physics, University of Colorado, Boulder, Colorado 80309, USA}

\definecolor{orange}{rgb}{1.0, 0.5, 0}

\newcommand{\MSbar}{\ensuremath{\overline{\mbox{MS}}} }
\newcommand{\vev}[1]{\ensuremath{\left\langle #1 \right\rangle} }

\newcommand\clearrow{\global\let\rowmac\relax}
\clearrow

\begin{document}

\title{Gradient flow step-scaling function for SU(3) with ten fundamental flavors }
\author{A.~Hasenfratz}\affiliation\coaff
\author{C.~Rebbi}\affiliation\buaff
\author{O.~Witzel}\email[Corresponding author: ]{oliver.witzel@colorado.edu}\affiliation\coaff

\date{\today}

\begin{abstract}
We calculate the step-scaling function, the lattice analog of the renormalization group $\beta$-function, for an SU(3) gauge theory with ten fundamental flavors. We present a detailed  analysis including the study of systematic effects of our extensive data set generated with ten dynamical flavors using the Symanzik gauge action and three times stout smeared M\"obius domain wall fermions. Using up to $32^4$ volumes, we calculate renormalized couplings for different gradient flow schemes and determine the step-scaling $\beta$ function for a scale change $s=2$ on up to five different lattice volume pairs.  In an accompanying paper we discuss that gradient flow can promote lattice dislocations to instanton-like objects, introducing nonperturbative lattice artifacts to the step-scaling function. Motivated by the observation that Wilson flow   sufficiently suppresses these artifacts, we choose  Wilson flow with the Symanzik operator as our preferred analysis.  We  study systematic effects by calculating the step-scaling function based on alternative flows (Zeuthen or Symanzik), alternative  operators (Wilson plaquette, clover), and also explore the effects of the perturbative tree-level improvement.  Further we investigate the effects due to the finite value of $L_s$.
\end{abstract}
\maketitle

\section{Introduction}
Strongly coupled gauge-fermion systems play a central role in  different  beyond the Standard Model scenarios. Two prominent examples are models for composite dark matter \cite{Kribs:2016cew,Brower:2019oor} and composite Higgs models \cite{DeGrand:2015zxa,Nogradi:2016qek,Witzel:2019jbe,Brower:2019oor,Cacciapaglia:2020kgq}. The latter require a large scale separation to accommodate that only a light 125 GeV Higgs boson has been experimentally observed but no other heavier resonances.  Hence composite Higgs models favor gauge-fermion systems  featuring near-conformal dynamics, i.e.~a slowly running (walking) gauge coupling. To identify promising candidate systems, it is  essential to understand the nature of near-conformal models. Gauge-fermion systems are characterized by the gauge group ${\cal G}$ and the number $N_f$ of fermion flavors in representation ${\cal R}$. Given these three characteristics, the infrared properties of gauge-fermion systems can be derived by determining the renormalization group (RG) $\beta$-function.  Calculations based on perturbation theory predict that systems with SU(3) gauge group and fermions in the fundamental representation undergo a transition: as the number of flavors is increased, a chirally broken system with fast running coupling changes to a conformal system where the gauge coupling is irrelevant \cite{Banks:1981nn}. For even larger number of flavors, gauge-fermion systems become infrared free and possibly asymptotically safe~\cite{Shrock:2013cca,Antipin:2017ebo,Leino:2019qwk}.

While perturbation theory provides guidance about the $N_f$ dependence,  the inherently nonperturbative nature of gauge-fermions systems limit the reliability of perturbative predictions. 
Particularly challenging is to identify the lowest, critical number of flavors $N_f^c$ where a gauge-fermion system of gauge group  ${\cal G}$ and representation ${\cal R}$ becomes conformal. It is known that for a system inside the conformal window, the infrared fixed point moves to stronger coupling as $N_f$ decreases toward $N_f^c$. Nonperturbative methods are required to determine $N_f^c$ and gain field theoretical insight into how a chirally broken system changes to a conformal system.

Lattice field theory provides a nonperturbative framework  to determine the RG $\beta$-function from first principles. 
We have studied the finite volume step-scaling function \cite{Fodor:2012td} for SU(3) with ten and twelve fundamental flavors \cite{Hasenfratz:2019dpr,Hasenfratz:2018wpq,Hasenfratz:2017qyr,Hasenfratz:2017mdh} using domain wall  (DW) fermions.  Our work complements earlier studies of the eight and twelve flavor systems with staggered fermions \cite{Cheng:2014jba,Hasenfratz:2014rna,Lin:2015zpa,Hasenfratz:2016dou,Fodor:2016zil,Fodor:2017gtj}.
Recently, we explored a new method, the continuous gradient flow $\beta$-function \cite{Hasenfratz:2019hpg,Hasenfratz:2019puu,Fodor:2017die} presenting results for SU(3) with two and twelve fundamental flavors.  In the infinite volume continuum limit both methods determine the renormalization group (RG) $\beta$-function, though in different renormalization schemes.
 Our recent DW results reveal that the two flavor system exhibits a fast running $\beta$ function close to the perturbative 1-loop prediction, whereas for $N_f=12$ our step-scaling calculation shows that the $\beta$ function is small in magnitude and identifies an infrared fixed point (IRFP) in the range $5.2 \le g_c^2 \le 6.4$  using the $c = 0.250$ renormalization scheme.

In this work we present a detailed analysis of our step-scaling calculation for ten fundamental flavors. Compared to our results published in Refs.~\cite{Hasenfratz:2017qyr,Hasenfratz:2017mdh}, we performed additional simulations at stronger bare couplings and added further volumes to improve  the infinite volume continuum limit extrapolation. The additional simulations allowed us to increase the explored coupling range  for $c=0.300$ from  $g_c^2\approx6.5$ in Refs.~\cite{Hasenfratz:2017qyr,Hasenfratz:2017mdh} to  $g_c^2 \gtrsim 11$.  At that strong coupling we also discovered previously unaccounted  lattice artifacts. In  an accompanying paper we discuss that gradient flow on coarse configurations can promote dislocations to instanton-like objects. This introduces a nonperturbative lattice artifact  to the step-scaling beta function which leads to incorrect continuum limit extrapolations  \cite{Hasenfratz:2020vta}. We find that the perturbatively preferable Symanzik and Zeuthen flows introduce many more of these artifacts. In order to minimize  this artifact, we choose Wilson flow as our preferred analysis.

\begin{figure}[p]
  \includegraphics[width=0.99\columnwidth]{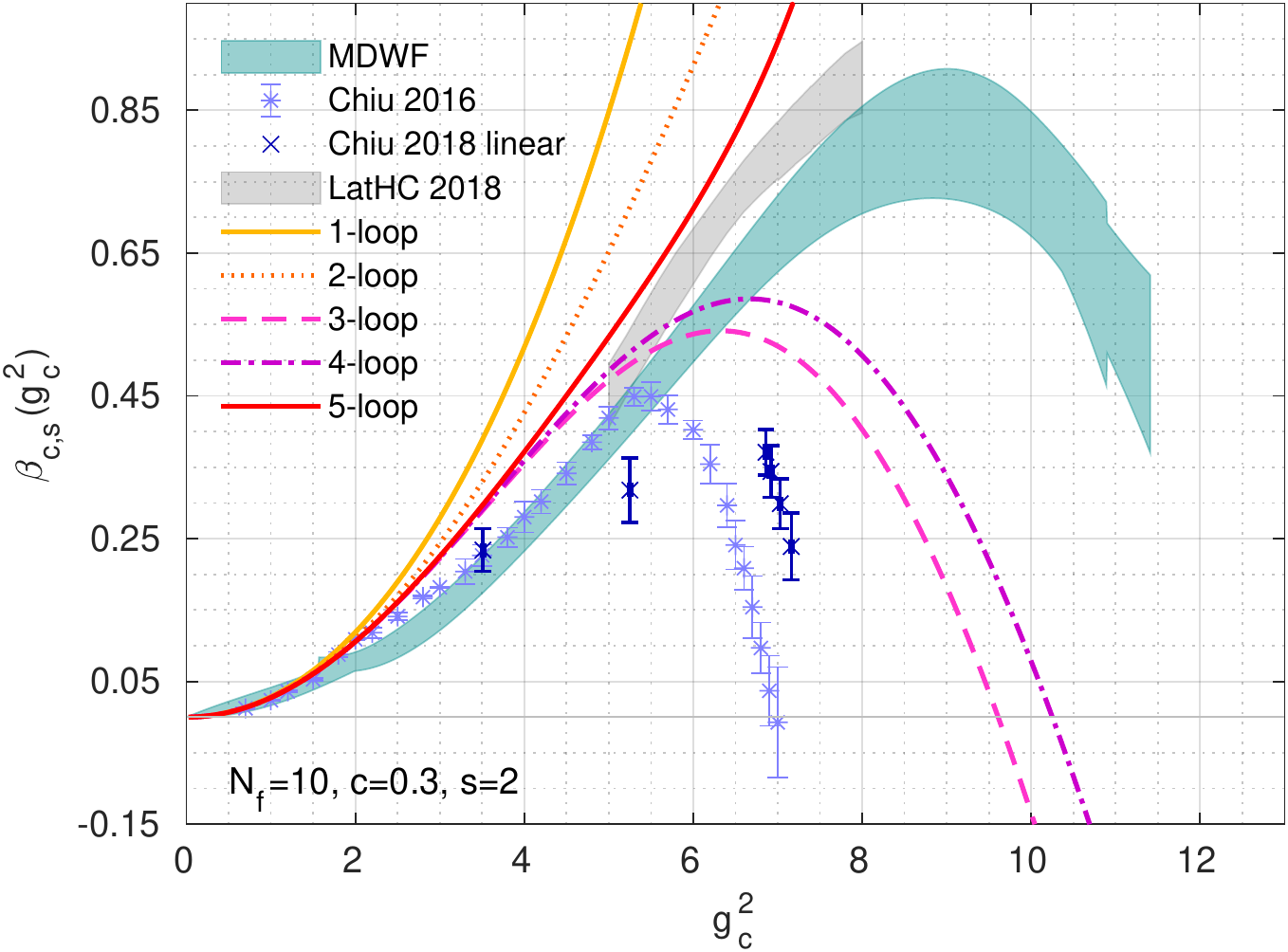} \\[2mm]
  \includegraphics[width=0.99\columnwidth]{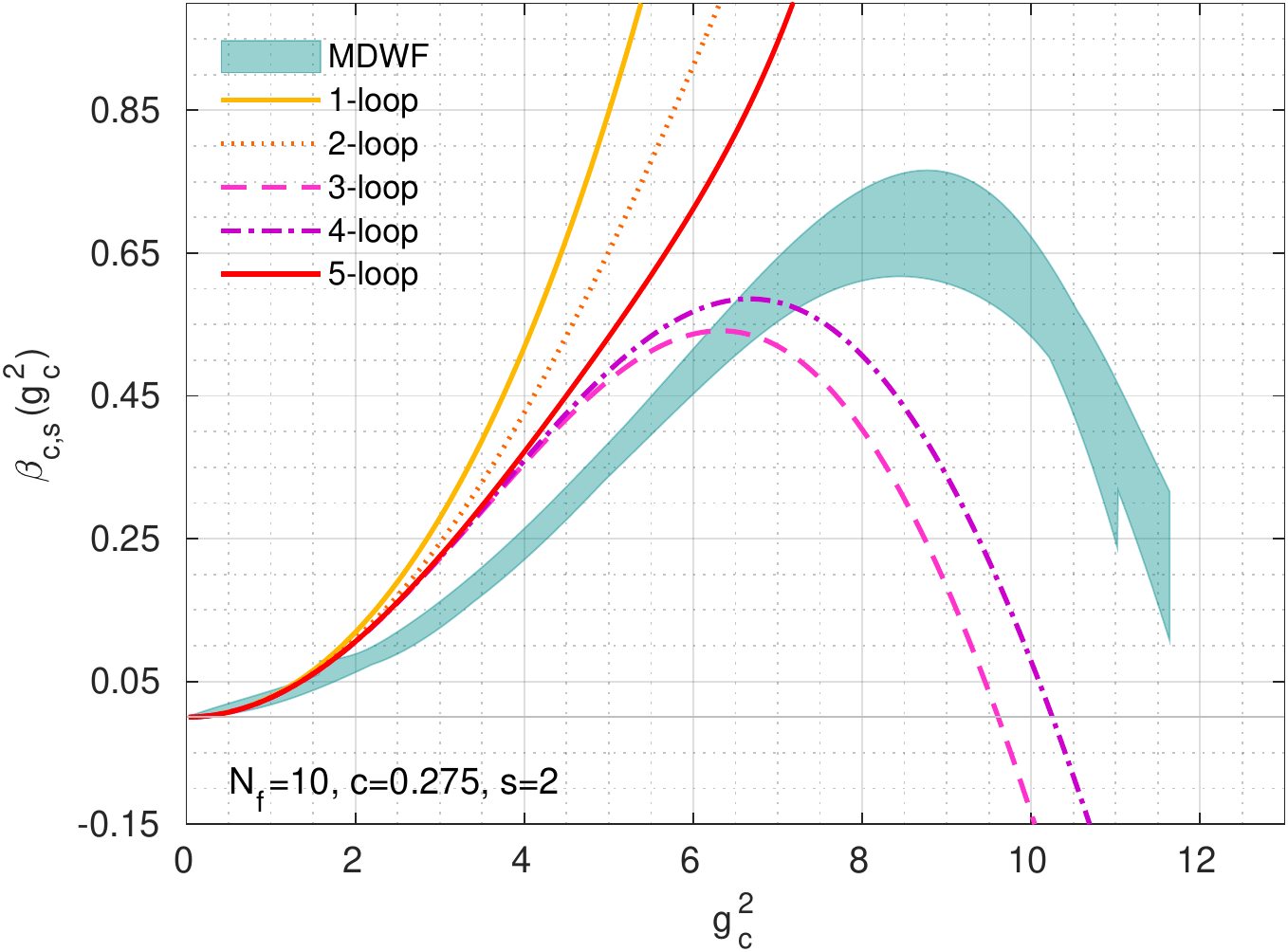} \\[2mm]
  \includegraphics[width=0.99\columnwidth]{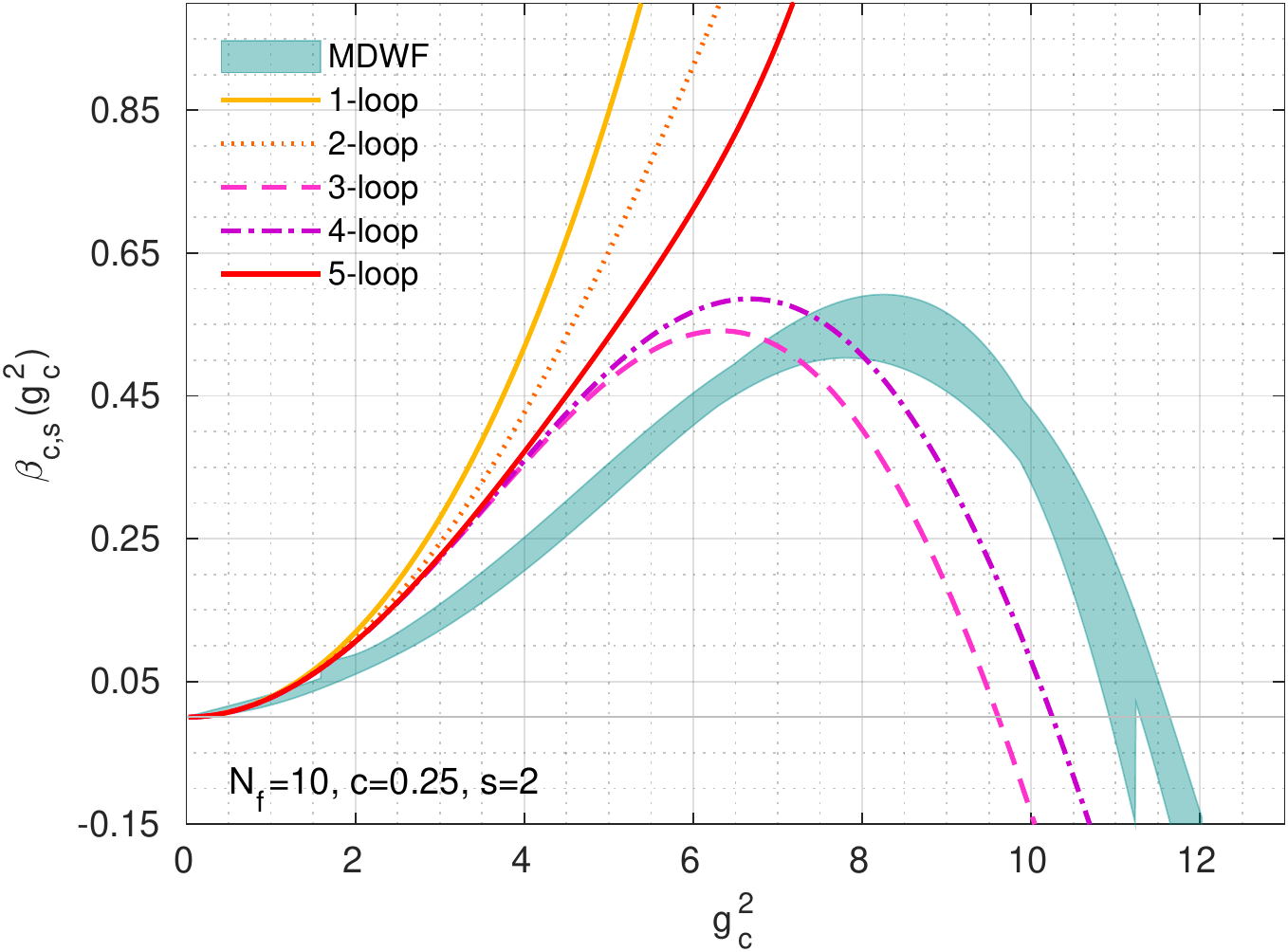}
  \caption{Our final results for GF step-scaling function for SU(3) with ten fundamental flavors using the renormalization schemes $c=0.300$, 0.275, and 0.250 (from top to bottom). The green bands show our result based on   domain wall fermions in comparison to perturbative predictions (yellow/orange/pink/purple/red) lines \cite{Baikov:2016tgj,Ryttov:2010iz,Ryttov:2016ner,Ryttov:2016hal} and other lattice determinations \cite{Chiu:2016uui,Chiu:2017kza,Chiu:2018edw,Fodor:2017gtj,Fodor:2018tdg,Fodor:2019ypi}  in the $c=0.300$ scheme.
  Lattice corrections due to small flow time in the $c=0.250$ scheme could be  significant, affecting the continuum limit shown on the last panel.}
  \label{Fig.beta_Nf10_final}
\end{figure}

In Fig.~\ref{Fig.beta_Nf10_final} we present our final result of the  continuum limit  extrapolated GF step-scaling $\beta$ function in the  renormalization schemes $c=0.300$, 0.275, and 0.250. 
Our predictions are labeled ``MDWF''  (for M\"obius domain wall fermions) and shown by  green bands. For the $c=0.300$ scheme we also show the nonperturbative lattice determinations by Chiu (blue symbols) \cite{Chiu:2016uui,Chiu:2017kza,Chiu:2018edw} and LatHC (gray band) \cite{Fodor:2017gtj,Fodor:2018tdg,Fodor:2019ypi}.\footnote{We estimate the values of the LatHC result (gray band) for $s=2$ based on Fig.~5 of Ref.~\cite{Fodor:2019ypi} as the numerical values are not yet published. The blue data points are from  private communication with T.-W.~Chiu and from Ref.~\cite{Chiu:2018edw}.} In addition we display by the  yellow/orange/pink/purple/red lines the $\MSbar$ perturbative predictions at 1 -- 5-loop order \cite{Baikov:2016tgj,Ryttov:2010iz,Ryttov:2016ner,Ryttov:2016hal}. 

Comparing the different nonperturbative lattice predictions in the $c=0.300$ scheme, we find that our result is in perfect agreement at weak coupling ($g_c^2 \lesssim 5.8$) with the findings by Chiu and sits  just below LatHC's result in the range $5.0 \lesssim g_c^2 \lesssim 8$. At present only our calculation has reached  the $8.0 \lesssim g_c^2 \lesssim 11.0$ range where we observe a down-turn of the $\beta$ function pointing to a possible IRFP around $g_c^2\sim 13$. 
Our nonperturbative results suggest that $N_f=10$ is  likely conformal. 

The  bottom two panels of Fig.~\ref{Fig.beta_Nf10_final} show our continuum limit predictions in the  $c=0.275$ and $0.250$ schemes. The results reveal that the GF step-scaling $\beta$-function exhibits a dependence on the  renormalization scheme parameter $c$. However, cutoff effects on the finite volume step-scaling function are more severe at smaller $c$. Unfortunately, our available data set does not allow to rigorously scrutinize our findings for $c=0.250$ but we nevertheless include it in this publication for future reference.

In the remainder of this paper we present the details of our calculation and describe how we arrived at our final results shown in Fig.~\ref{Fig.beta_Nf10_final}. Section \ref{Sec.StepScaling} gives a brief description of the step-scaling function and  Sec.~\ref{Sec.Numerical} summarizes the numerical details of our simulations. We continue by discussing a novel nonperturbative lattice artifact of the gradient flow in Sec.~\ref{Sec.FlowTopo} with a detailed investigation of this topology-related artifact presented in our companion work \cite{Hasenfratz:2020vta}. There we  also demonstrate that nonzero topology is correlated with an increased value of the gradient flow coupling $g_{GF}^2$ and the corresponding step-scaling $\beta$-function. 
Afterwards we explain in Sec.~\ref{Sec.Nf10} how we arrive at our final results and how we check for systematic effects including a discussion on the finite value of the fifth dimension of domain wall fermion simulations. Finally, in Sec.~\ref{Sec.Summary} we summarize our results and comment on the comparison with perturbative predictions.

\section{Step-scaling function}
\label{Sec.StepScaling}

The finite volume gradient flow coupling in volumes $V=L^4$ is
\begin{align}
 \label{Eq.gc2}
g^2_{GF} (t;L,\beta) = \frac{128\pi^2}{3(N^2 - 1)} \frac{1}{C(t,L/a)} \vev{t^2 E(t)},
\end{align}
where  $E(t)$ is the energy density, $N=3$ for SU(3), and $C(t,L/a)$ is a perturbatively computed tree-level improvement term\footnote{Numerical values for $C(t,L/a)$ are listed in Table III in the Appendix of Ref.~\cite{Hasenfratz:2019dpr}.} \cite{Fodor:2014cpa}. Without tree-level improvement $C(c,L/a)$ is replaced by the term $1/(1+\delta(t/L^2))$ that compensates for the zero modes of the gauge fields in periodic volumes~\cite{Fodor:2012td}. 
In the finite volume renormalization scheme the flow time $t$ is set by renormalization scheme parameter $c$ and the lattice size $L$
\begin{align}
  t=(c L)^2/8.  
  \label{Eq.RenCon}
\end{align}
The discrete step-scaling $\beta$ function of scale change $s$ is defined as in Ref.~\cite{Fodor:2012td}
\begin{equation}
  \beta_{c,s}(g^2_c;L,\beta) = \frac{g^2_c(sL; \beta)- g^2_c(L; \beta)}{\log\;s^2} \,,
  \label{Eq.beta_cs}
\end{equation}
where $g_c^2(L,\beta) =g^2_{GF}(t=(c L)^2/8; L,\beta)$.

The gradient flow transformation  can be performed with different flow actions. In this work we consider Wilson, Symanzik and Zeuthen flows \cite{Ramos:2014kka,Ramos:2015baa}. 
Similarly, the energy density $E(t)$  at gradient flow time $t$ can be approximated by different lattice operators. We consider the Wilson plaquette (W), Symanzik (S) and clover (C) operators. We use the shorthand notation [flow][operator] to refer to the various combinations.  When the tree-level improvement term $C(c,L/a)$ is included we add ``n'' to our shorthand notation, e.g.~nWS for Wilson flow, Symanzik operator, and tree-level improved coupling.

The renormalized coupling $g^2_c$ is defined at a  bare coupling $\beta$, therefore it is contaminated by  cutoff effects. The infinite cutoff continuum limit requires $t/a^2 \to \infty$ or equivalently $L/a \to \infty$. At fixed  value of $g^2_c$ this means tuning the bare coupling $g_0^2= 6/ \beta \to 0$, the Gaussian fixed point.

In practice we perform simulations at many values of the bare coupling $\beta$ and combine them to cover the investigated range of the renormalized coupling. 
At fixed $g_c^2$ we take the continuum limit ($L/a\to \infty$) of the discrete step-scaling function $\beta_{c,s}(g^2_c;L)$ and obtain the continuum step-scaling $\beta$-function $\beta_{c,s}(g_c^2)$ in the renormalization scheme $c$.

\section{Numerical simulation details}
\label{Sec.Numerical}

We determine the  gradient flow step-scaling function  on dynamical gauge field configurations with $(L/a)^4$ hypercubic volumes. As for our project with $N_f=12$ flavors, we choose to generate ten flavor  configurations using tree-level improved Symanzik (L\"uscher-Weisz) gauge action \cite{Luscher:1984xn,Luscher:1985zq} and M\"obius domain wall fermions (MDWF) \cite{Brower:2012vk} with three levels of stout-smearing \cite{Morningstar:2003gk} ($\varrho=0.1$)   for the fermion action.\footnote{The good properties of this  action has been first demonstrated for simulations in QCD \cite{Kaneko:2013jla,Noaki:2015xpx} but is now also used in large scale simulations of a composite Higgs model \cite{Witzel:2019oej,Witzel:2018gxm,Appelquist:2020xua}} Our ensembles of gauge field configurations are generated using the hybrid Monte Carlo (HMC) update algorithm \cite{Duane:1987de} with five massless two-flavor fermion fields and trajectories of length $\tau=2$ in molecular dynamics time units (MDTU). Simulations with domain wall fermions are performed by creating a five dimensional Dirac operator where the fifth dimension, $L_s$, separates the chiral, physical modes of 4-d space-time. We choose $L_s=12$ for $\beta \ge 4.40$ and $L_s=16$ for $\beta \le 4.30$. We set the domain wall height $M_5=1$ and simulate with bare fermion mass $am_f=0$.

The finite extent of the fifth dimension leads to residual breaking of chiral symmetry which conventionally is parametrized by an additive mass term $am_\text{res}$. Although $am_\text{res}$ grows toward the strong coupling, we demonstrate in Sec.~\ref{Sec.Ls} that our choice of $L_s$ is sufficient .

We set periodic boundary conditions (BC) for the gauge field and antiperiodic BC for the fermion fields in all four directions, i.e~the fermion BC trigger a gap in the eigenvalues of the Dirac operator enabling simulations with zero input quark mass. To determine the step-scaling $\beta$ function we use hypercubic $(L/a)^4$ volumes with $L/a=8$, 10, 12, 14, 16, 20, 24, 28, and 32 and generate for each volume a set of 17 bare couplings, starting in the weak coupling limit with $\beta=7.00$ and choosing $4.02$ as the value of our strongest bare coupling. Simulations at $\beta=4.02$ are about the strongest bare coupling which can be achieved for our choice of actions. As we detail in Appendix \ref{Sec.Phase}, a bulk phase transition prevents investigating much stronger coupling. Typically our ensembles consist of 6-10k for $L/a\le 14$, 3-6k for $L=16$, 20, 24, and 2-4k for $L/a=28$, 32 thermalized MDTU and we perform measurements separated by 10 MDTU. The simulations are carried out using \texttt{GRID}\footnote{\url{https://github.com/paboyle/Grid}} \cite{Boyle:2015tjk} and we perform gradient flow measurements  using \texttt{Qlua}\footnote{\url{https://usqcd.lns.mit.edu/w/index.php/QLUA}} \cite{Pochinsky:2008zz}. The subsequent data analysis is performed using the $\Gamma$-method \cite{Wolff:2003sm} to estimate the integrated autocorrelation time and account for that as part of the statistical data analysis.

\section{Topological artifacts of the gradient flow}
\label{Sec.FlowTopo}

\begin{figure}[tb!]
  \includegraphics[width=0.99\columnwidth]{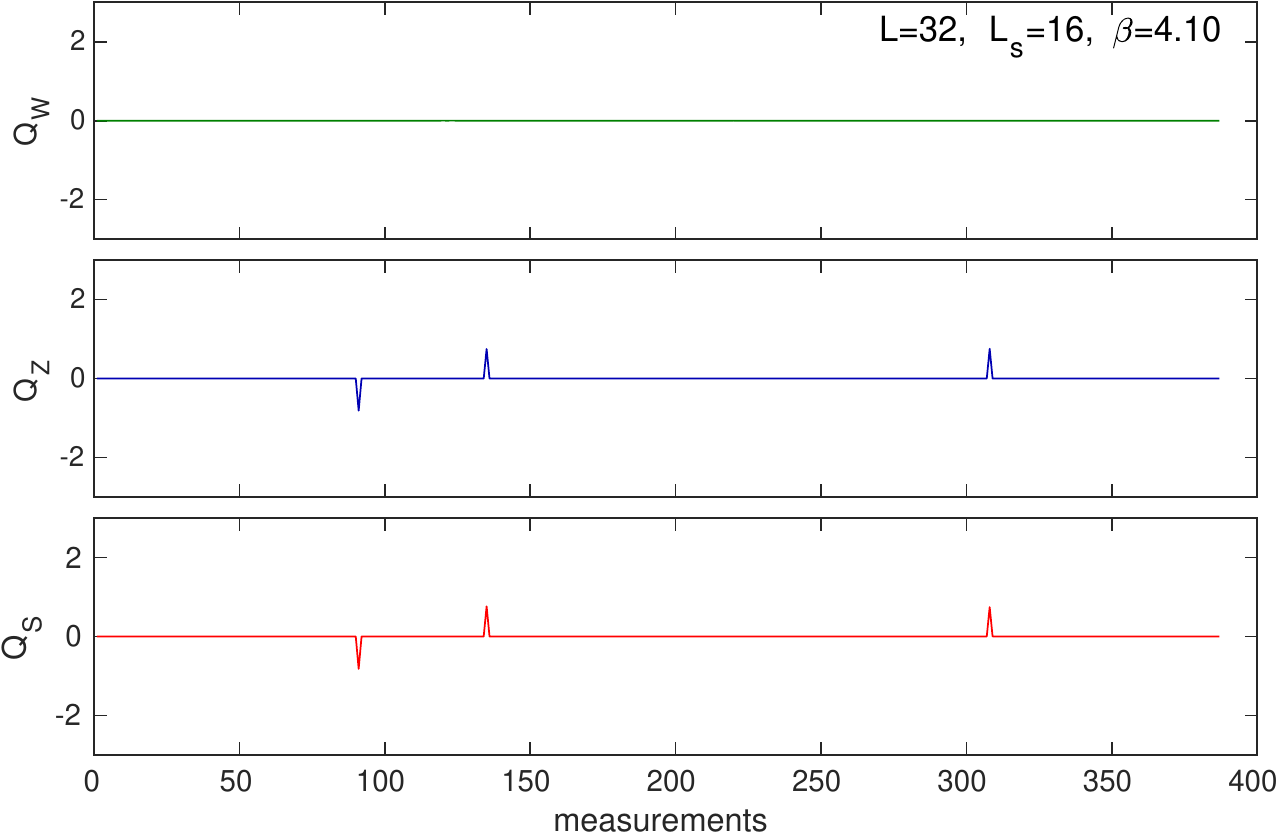} \\[5mm]
  \includegraphics[width=0.99\columnwidth]{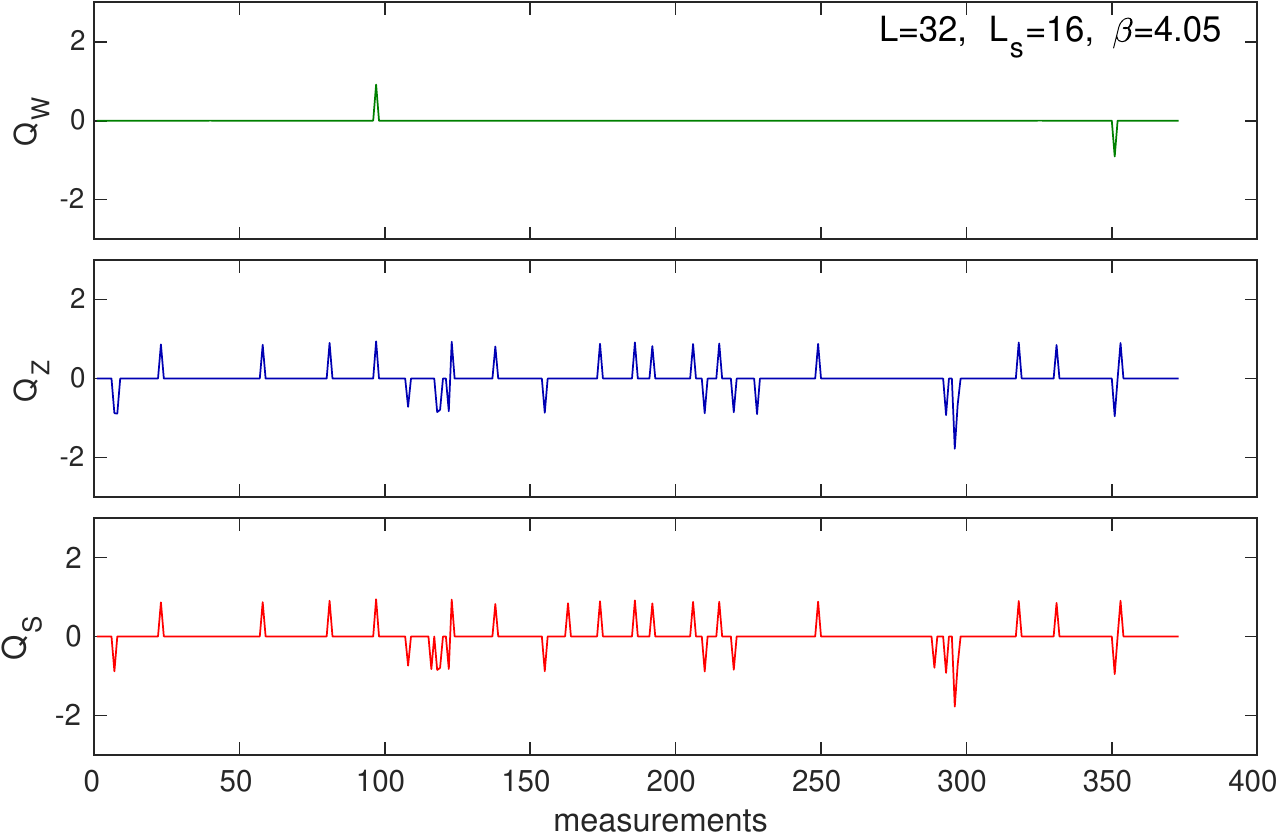} \\[5mm]
  \includegraphics[width=0.99\columnwidth]{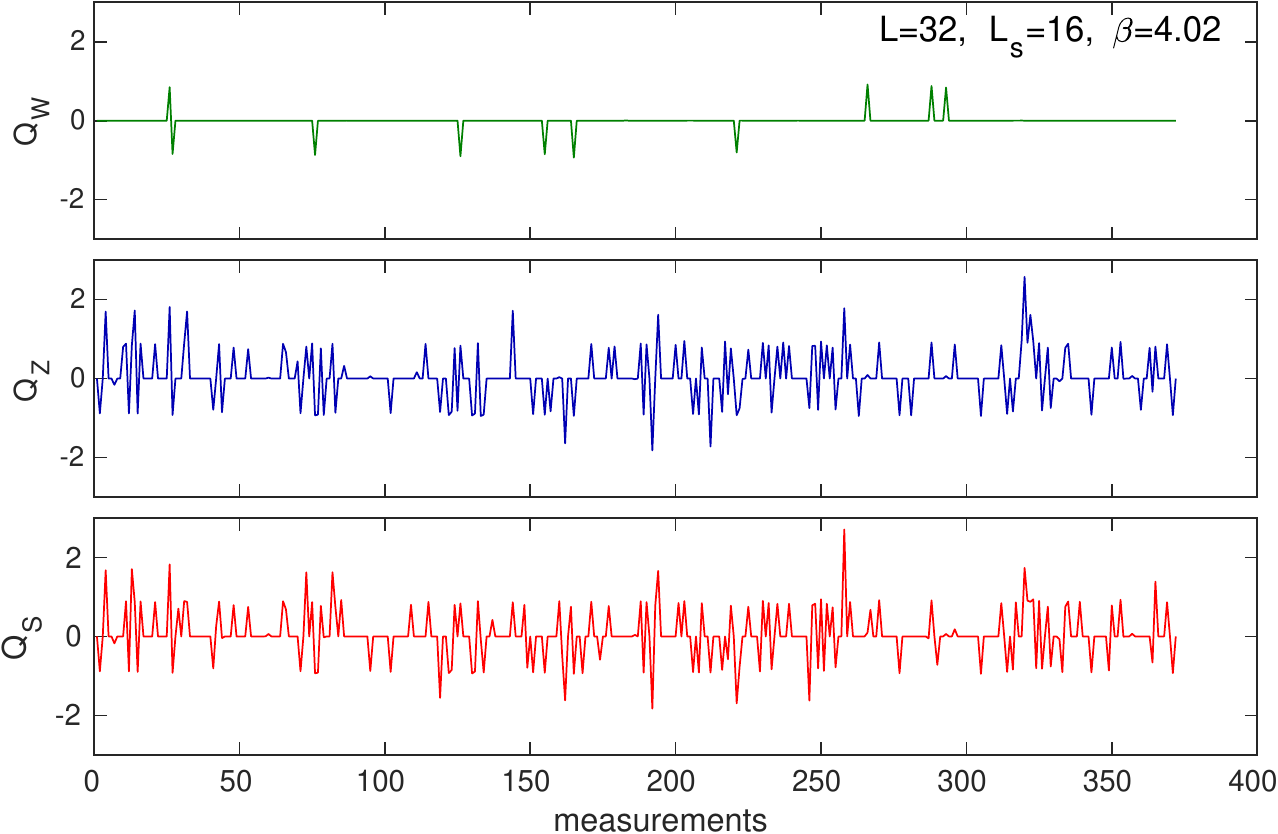}  
  \caption{MC history of topological charges $Q$ determined after gradient flow time $t/a^2=32$ on our $32^4$ ensembles in the strong coupling with $\beta\equiv 6/g_0^2=\{4.10$, 4.05, 4.02$\}$. For each bare coupling the three panel shows the determination with Wilson flow (W), Zeuthen flow (Z), and Symanzik flow, respectively.}
  \label{Fig.TopoCharge}
\end{figure}

Before presenting details of our gradient flow calculation, we discuss a so far not considered lattice artifact related to topology. In this section we provide a brief description of the issue and refer for further details to our companion work \cite{Hasenfratz:2020vta}.  

Our numerical simulations are performed using massless, chirally symmetric domain wall fermions. With strictly massless fermions \emph{all}\/ configurations would have net topological charge zero. Configuration with nonvanishing topological charge would have zero Boltzmann weight because unpaired instantons create zero modes in the eigenvalue spectrum of the Dirac operator.  If the topological charge is defined via the Dirac operator, $Q_\text{ferm}=0$.
On the lattice topology is not a conserved quantity and the geometric  definition of the charge
\begin{align}
Q_\text{geom} =\frac{1}{32\pi^2} \int dx \; \text{tr} ( F_{\mu\nu}(x){\tilde F_{\mu\nu}}(x) )
\end{align} 
 can differ from $Q_\text{ferm}$. Gradient flow removes the ultraviolet fluctuations of the gauge field and at large enough flow time $Q_\text{geom}$ is expected to be a good  estimate of the topological charge.

In our simulations we monitor $Q_\text{geom}$ and observe $Q_\text{geom}\approx 0$ on configurations at relatively weak gauge coupling.  However as the simulations are pushed into the strong coupling regime,  $Q_\text{geom} \ne 0$ configurations appear even at large flow time. We use the clover operator to estimate $F{\tilde F}$. In Fig.~\ref{Fig.TopoCharge} we show the Monte Carlo (MC) histories of the topological charge $Q_\text{geom}$ determined after  gradient flow time $t/a^2=32$ on our  $32^4$ ensembles at bare couplings $\beta=4.10$, 4.05, and 4.02. Each of the three  plots has three panels showing the results with  Wilson flow (W, green),  Zeuthen flow (Z, blue), and Symanzik flow (S, red). 
Even though the fermions cannot see unpaired instantons, apparently the gradient flow can promote vacuum fluctuations (dislocations) to instanton-like objects. This is an artifact of the gradient flow and not due to the action. In fact we have observed this effect in simulations with domain wall as well as with staggered fermions.
Different gradient flows find  different values for the topological charge on the same  configuration, further showing that $Q_\text{geom} \ne 0$ is an artifact of the flow.
The fraction of nonzero topological charge configurations  grows toward the strong coupling  i.e.~decreasing values of the bare coupling $\beta$. We also observe that Wilson flow has many fewer $Q_\text{geom}\ne 0$ configurations than Symanzik or Zeuthen flow.

In  Ref.~\cite{Hasenfratz:2020vta} we correlated   $Q_\text{geom}\ne 0$ with the value of the gradient flow coupling and found that both the renormalized coupling and the step-scaling function increases when  $Q_\text{geom}\ne 0$. Since the number of instantons is proportional to the square root of the volume, this nonperturbative artifact contributes a term to the step-scaling  function that  increases with the volume. This invalidates the perturbatively motivated $(a/L)^2 \to 0$ continuum limit extrapolation. The only way to control this artifact is to limit simulations to sufficiently weak coupling or use a gradient flow that suppresses $Q_\text{geom}$. 
As is evident from Fig.~\ref{Fig.TopoCharge}, Wilson flow  finds $Q_\text{geom}\ne 0$ only on a statistically negligible  number of configurations for $\beta > 4.02$,  while Symanzik and Zeuthen flows  may have large corrections. Although at $\beta=4.02$ even Wilson flow might be slightly affected, we choose Wilson flow for our preferred analysis. 

In Sec.~\ref{Sec.Alt} we revisit the effect  of topology  on the step-scaling function  when considering alternative flows.

\section{\texorpdfstring{Gradient flow $\beta$ function for\newline ten fundamental flavors}{Gradient flow beta function for ten fundamental flavors}}
\label{Sec.Nf10}
We have calculated renormalized couplings $g^2_c$ using three different gradient flows. However, we only consider Wilson flow to be acceptable at strong coupling and therefore obtain our preferred result using Wilson flow combined with the Symanzik operator. Due to too many configurations with nonzero topological charge, Zeuthen or Symanzik flows are solely used to estimate systematic effects.

\subsection{Preferred (n)WS analysis}

\begin{table*}[tb]
  \caption{Results of the interpolation fits for the five lattice volume pairs for our preferred nWS (top half) and WS (bottom half) analysis using renormalization schemes $c = 0.300$, $0.275$, and $0.250$. Since discretization effects are sufficiently small for nWS, we constrain the constant term $b_0 = 0$ in Eq.~(\ref{Eq.fit_form}) and perform fits with 13 degrees of freedom (d.o.f.). For WS the intercept $b_0$ is fitted and we have 12 d.o.f. In addition we list the $\chi^2/\text{d.o.f.}$~as well as the $p$-value.}
 \label{Tab.interpolations}
  \begin{tabular}{c@{~~~~}cc@{~~}c@{~~}cccccc}
    \hline \hline
    & analysis &    $c$ &$\chi^2$/d.o.f.&$p$-value& $b_4$     & $b_3$        & $b_2$     & $b_1$      & $b_0$ \\ \hline
    $ 8\to 16$& nWS & 0.300 & 0.712 & 0.753 &0.000147(52) &-0.00449(88) &0.0323(43) &-0.0158(60)  & --- \\
    $10\to 20$& nWS & 0.300 & 0.909 & 0.543 &0.000080(45) &-0.00388(84) &0.0294(46) &-0.0116(64)  & --- \\
    $12\to 24$& nWS & 0.300 & 0.862 & 0.593 &0.000122(52) &-0.00452(98) &0.0326(53) &-0.0155(74)  & --- \\
    $14\to 28$& nWS & 0.300 & 0.714 & 0.751 &0.000093(63) &-0.0041(12) &0.0351(66) &-0.026(10)  & --- \\
    $16\to 32$& nWS & 0.300 & 0.833 & 0.625 &-0.000125(91) &0.0002(17) &0.0118(87) &0.015(13) & --- \\
    \hline
    $ 8\to 16$& nWS & 0.275 & 0.836 & 0.621 &0.000193(53) &-0.00461(85) &0.0330(40) &-0.0180(53)  & --- \\
    $10\to 20$& nWS & 0.275 & 1.010 & 0.438 &0.000105(39) &-0.00429(72) &0.0314(38) &-0.0161(53)  & --- \\
    $12\to 24$& nWS & 0.275 & 1.075 & 0.376 &0.000121(41) &-0.00463(77) &0.0330(42) &-0.0171(59)  & --- \\
    $14\to 28$& nWS & 0.275 & 0.722 & 0.743 &0.000078(47) &-0.00393(89) &0.0330(50) &-0.0229(79)  & --- \\
    $16\to 32$& nWS & 0.275 & 0.740 & 0.724 &-0.000100(69) &-0.0004(13) &0.0138(67) &0.0108(98)  & --- \\
    \hline
    $ 8\to 16$& nWS & 0.250 & 1.238 & 0.244 &0.000350(58) &-0.00568(87) &0.0378(38) &-0.0230(48)  & --- \\
    $10\to 20$& nWS & 0.250 & 1.139 & 0.319 &0.000119(36) &-0.00436(63) &0.0318(32) &-0.0184(44)  & --- \\
    $12\to 24$& nWS & 0.250 & 1.523 & 0.100 &0.000131(33) &-0.00484(62) &0.0338(33) &-0.0192(47)  & --- \\
    $14\to 28$& nWS & 0.250 & 0.858 & 0.598 &0.000083(34) &-0.00409(65) &0.0329(37) &-0.0229(59) & --- \\ 
    $16\to 32$& nWS & 0.250 & 0.734 & 0.731 &-0.000071(51) &-0.00099(93) &0.0163(50) &0.0058(73) & --- \\ 
    \hline
    \hline 
    $ 8\to 16$&  WS & 0.300 & 0.614 & 0.832 &0.000095(46) &-0.0033(12) &0.030(10) &-0.168(36) &0.037(40)\\ 
    $10\to 20$&  WS & 0.300 & 0.767 & 0.686 &0.000109(53) &-0.0044(14) &0.039(12) &-0.137(40) &0.060(43)\\ 
    $12\to 24$&  WS & 0.300 & 0.908 & 0.538 &0.000071(76) &-0.0031(19) &0.024(16) &-0.049(51) &-0.012(51)\\ 
    $14\to 28$&  WS & 0.300 & 0.741 & 0.712 &0.00012(10) &-0.0047(25) &0.042(21) &-0.093(66) &0.036(67)\\
    $16\to 32$&  WS & 0.300 & 0.880 & 0.567 &-0.00005(16) &-0.0012(38) &0.022(30) &-0.047(95) &0.036(95)\\
    \hline    
    $ 8\to 16$&  WS & 0.275 & 0.627 & 0.821 &0.000105(40) &-0.0032(10) &0.0303(92) &-0.203(32) &0.046(37)\\
    $10\to 20$&  WS & 0.275 & 0.755 & 0.698 &0.000116(42) &-0.0044(11) &0.0394(96) &-0.157(33) &0.063(36)\\
    $12\to 24$&  WS & 0.275 & 1.111 & 0.345 &0.000100(56) &-0.0040(14) &0.032(12) &-0.088(40) &0.017(40)\\
    $14\to 28$&  WS & 0.275 & 0.724 & 0.729 &0.000106(72) &-0.0044(18) &0.040(15) &-0.097(48) &0.037(50)\\
    $16\to 32$&  WS & 0.275 & 0.754 & 0.699 &-0.00001(12) &-0.0021(28) &0.027(23) &-0.067(72) &0.048(72)\\
    \hline
    $ 8\to 16$&  WS & 0.250 & 0.740 & 0.713 &0.000144(35) &-0.00361(93) &0.0344(83) &-0.256(30) &0.067(35) \\
    $10\to 20$&  WS & 0.250 & 0.779 & 0.673 &0.000109(33) &-0.00400(88) &0.0367(78) &-0.178(27) &0.060(30) \\
    $12\to 24$&  WS & 0.250 & 1.426 & 0.145 &0.000122(41) &-0.0046(11)  &0.0378(92) &-0.125(31) &0.041(32) \\
    $14\to 28$&  WS & 0.250 & 0.749 & 0.704 &0.000117(49) &-0.0047(12)  &0.042(11) &-0.119(35) &0.052(38) \\
    $16\to 32$&  WS & 0.250 & 0.668 & 0.784 &0.000029(83) &-0.0030(20)  &0.033(16) &-0.092(53) &0.063(54) \\
    \hline \hline
  \end{tabular}
\end{table*}

\begin{figure*}[t]
  \begin{minipage}{0.49\textwidth}
   \flushright 
   \includegraphics[width=0.96\textwidth]{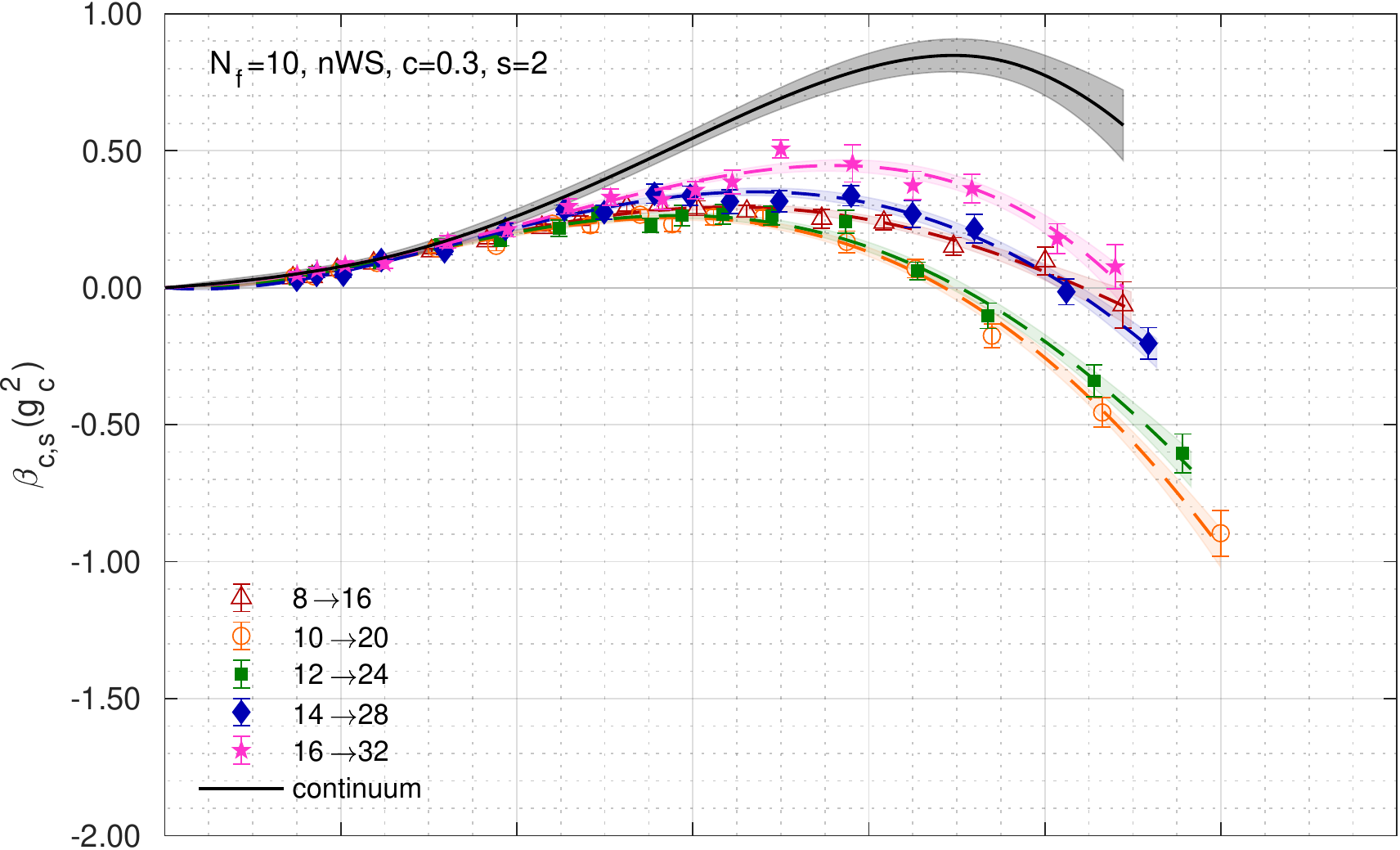}\\
   \includegraphics[width=0.932\textwidth]{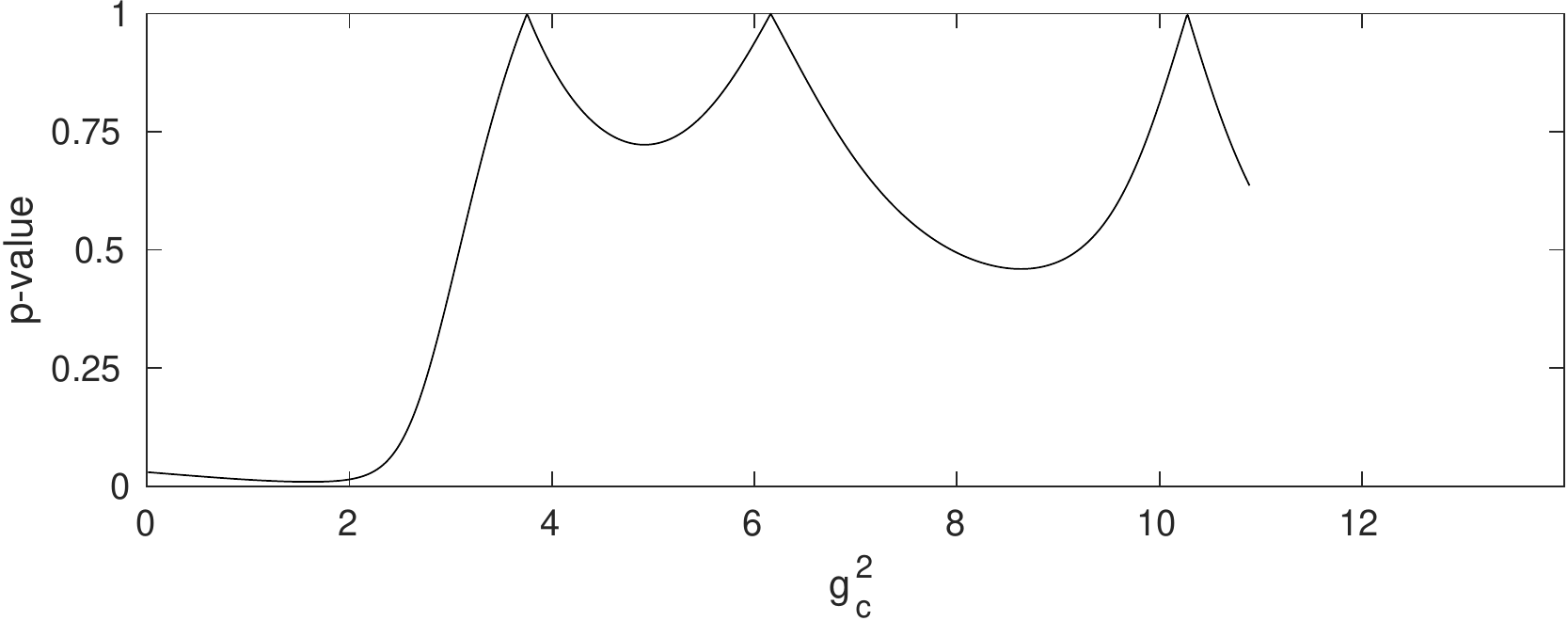}\\[3mm]
   \includegraphics[width=0.96\textwidth]{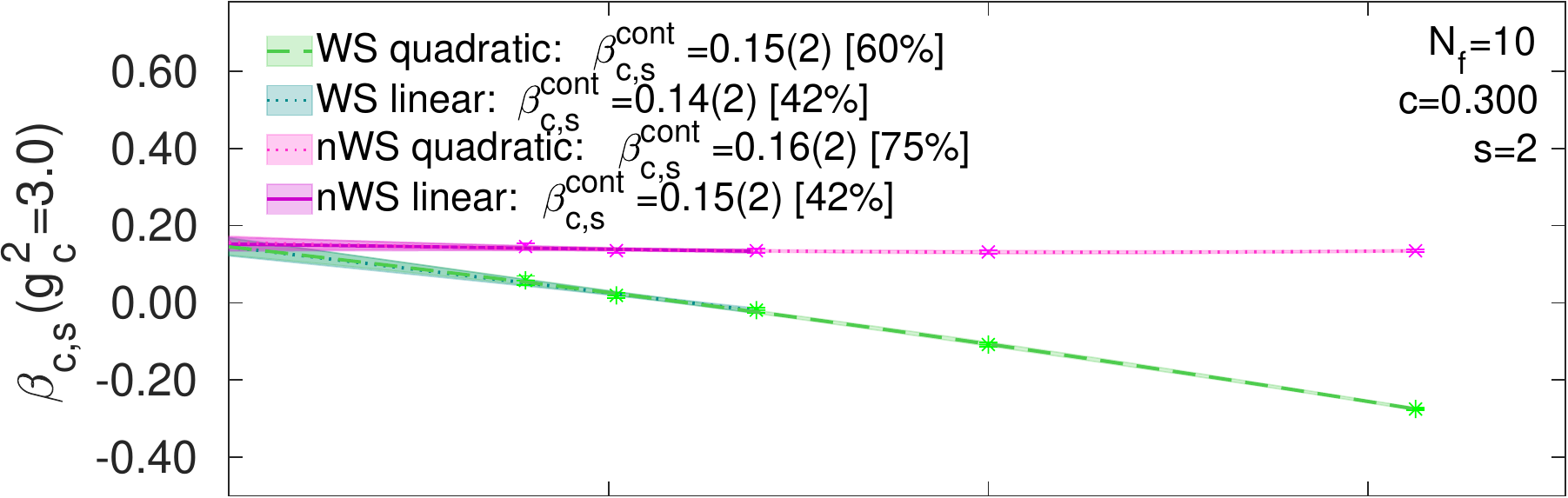}\\
   \includegraphics[width=0.96\textwidth]{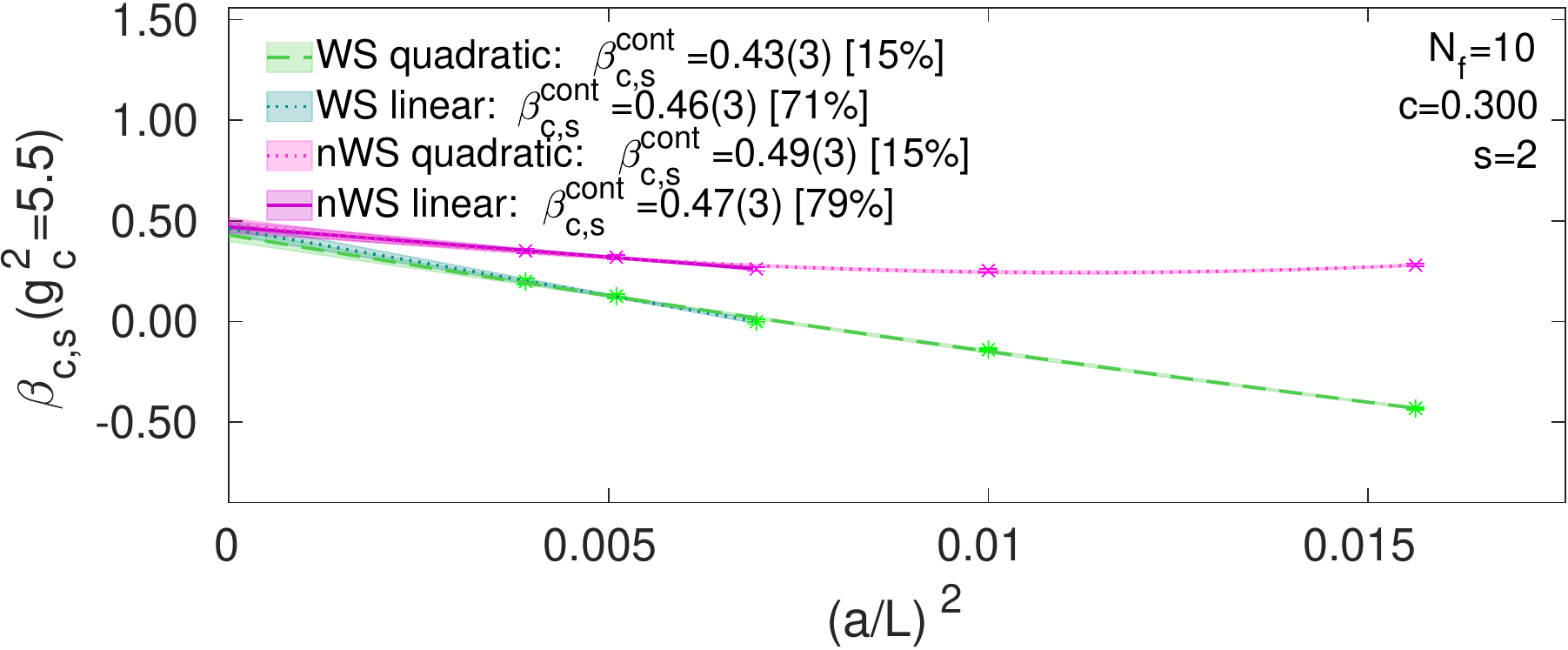}     
  \end{minipage}
  \begin{minipage}{0.49\textwidth}
    \flushright
    \includegraphics[width=0.96\textwidth]{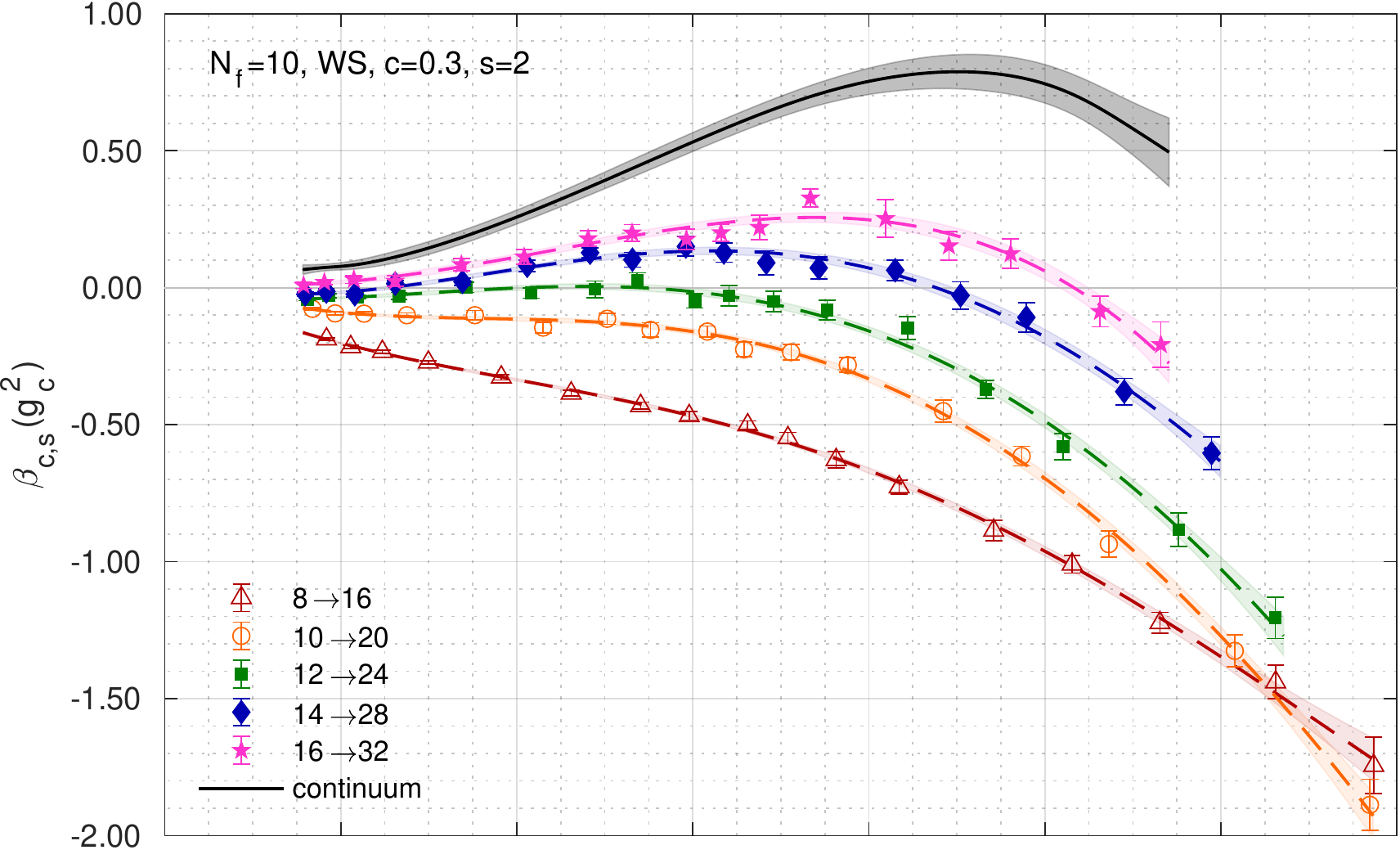}\\    
    \includegraphics[width=0.932\textwidth]{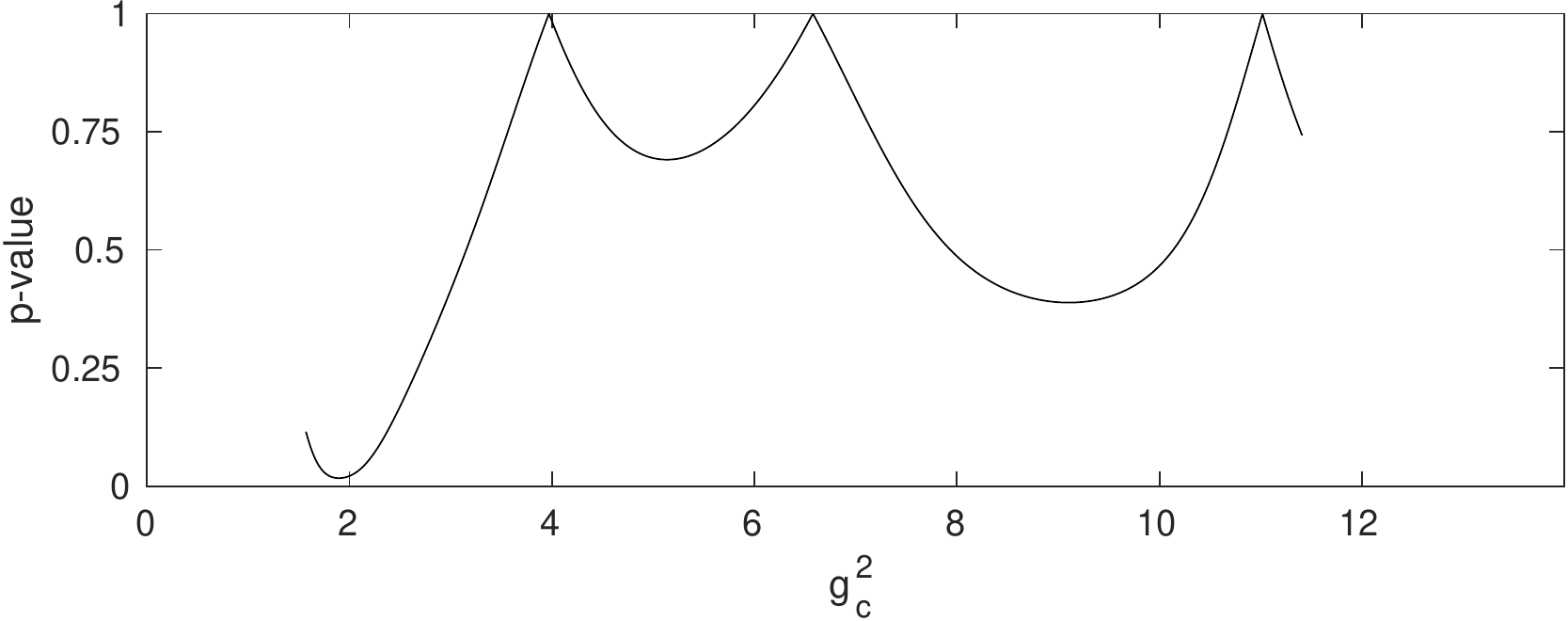}\\[3mm]
    \includegraphics[width=0.96\textwidth]{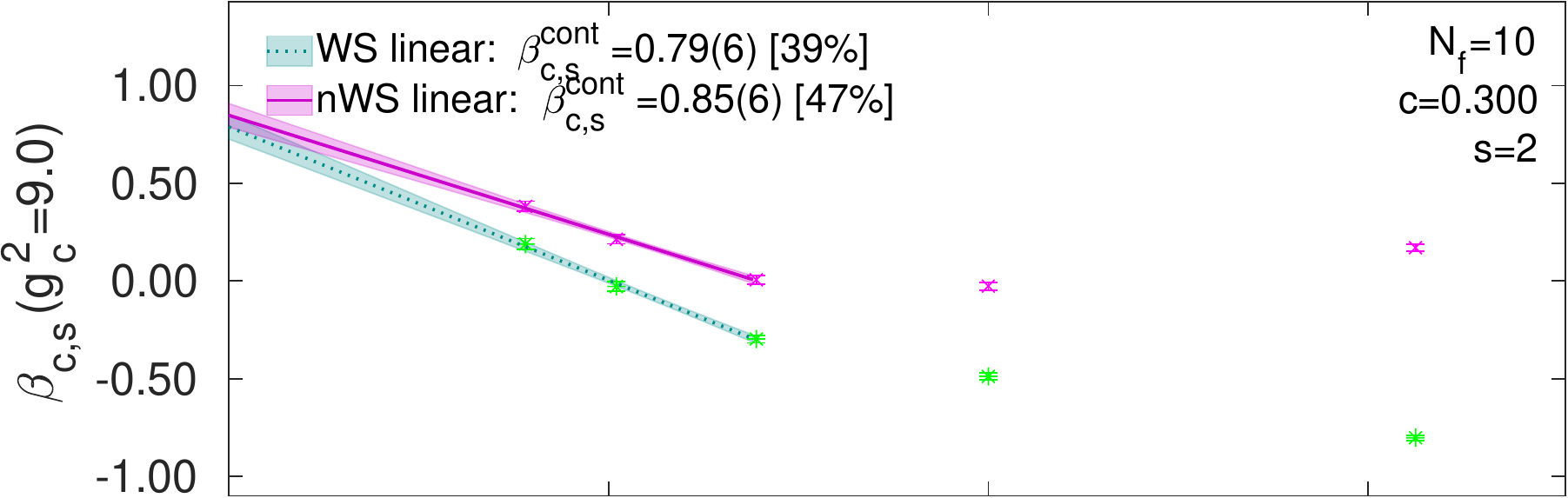}\\
    \includegraphics[width=0.96\textwidth]{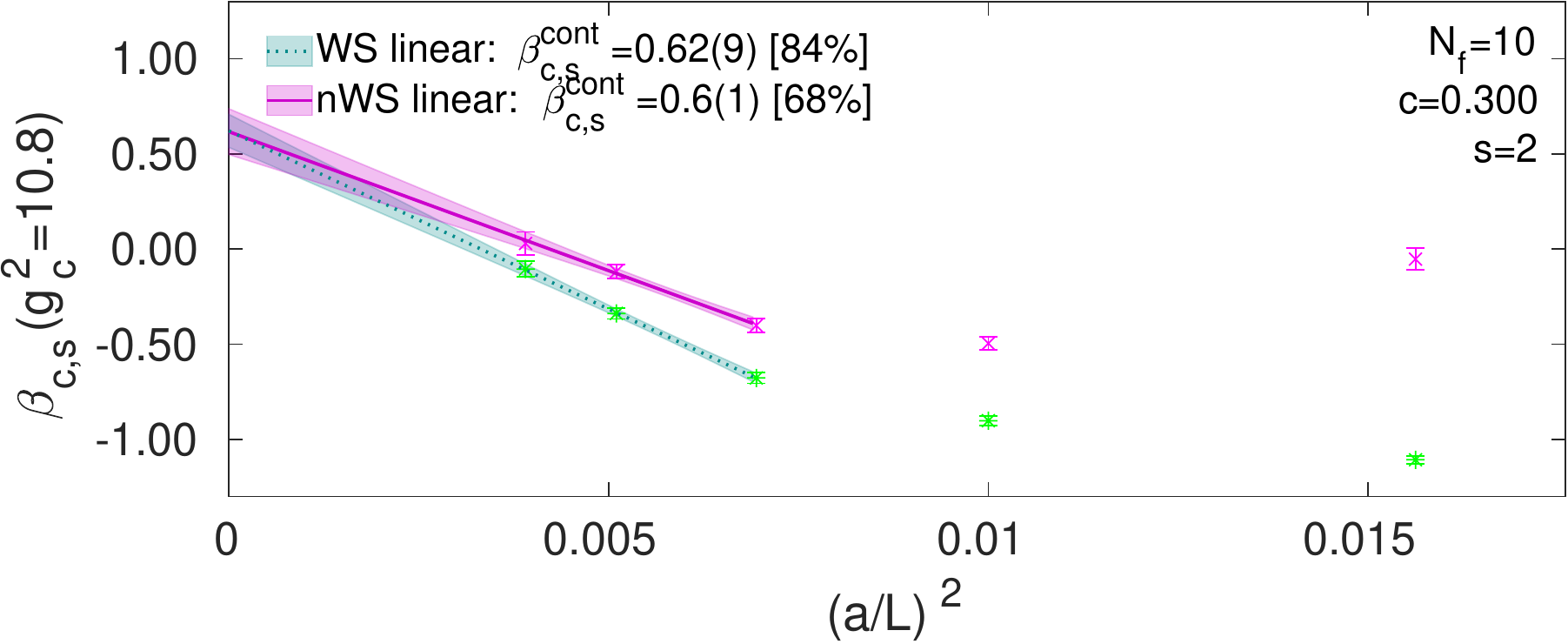}    
  \end{minipage}
  \caption{Discrete step-scaling $\beta$-function in the $c=0.300$ gradient flow scheme for our preferred nWS (left) and WS (right) data sets. The symbols in the top row show our results for the finite volume discrete $\beta$ function with scale change $s=2$. The dashed lines with shaded error bands in the same color of the data points show the interpolating fits. We take the continuum limit performing a linear fit (black line with gray error band) in $a^2/L^2$ to the three largest volume pairs (filled symbols). The $p$-values of the continuum extrapolation fit is shown in the plots in the second row. Further details of the continuum extrapolation at selected $g_c^2$ values are presented in the small panels at the bottom where the legend lists the extrapolated values in the continuum limit with $p$-values in brackets. Only statistical errors are shown.}
  \label{Fig.beta_c300}
\end{figure*}

\begin{figure*}[t]
  \begin{minipage}{0.49\textwidth}
   \flushright 
   \includegraphics[width=0.96\textwidth]{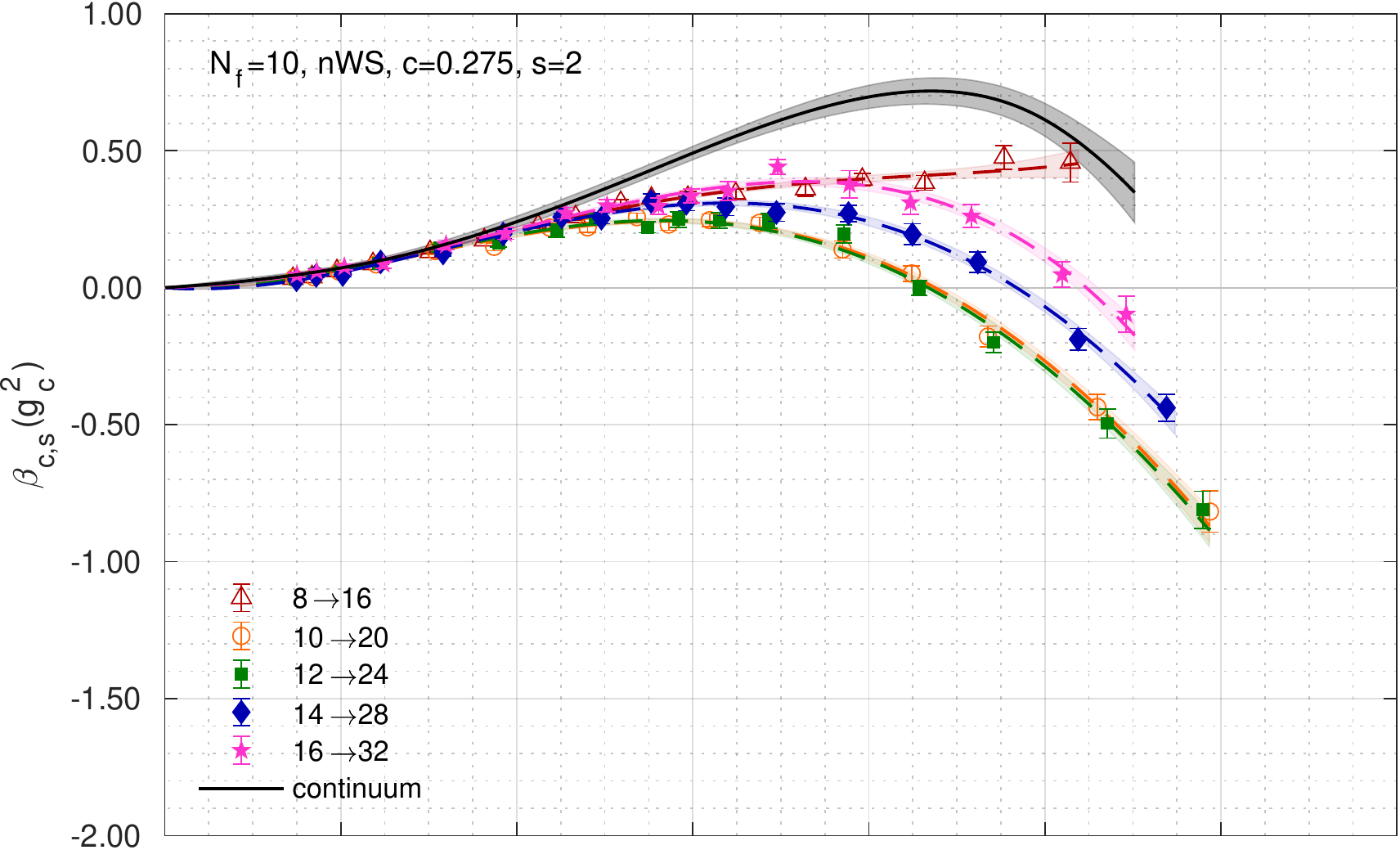}\\
   \includegraphics[width=0.932\textwidth]{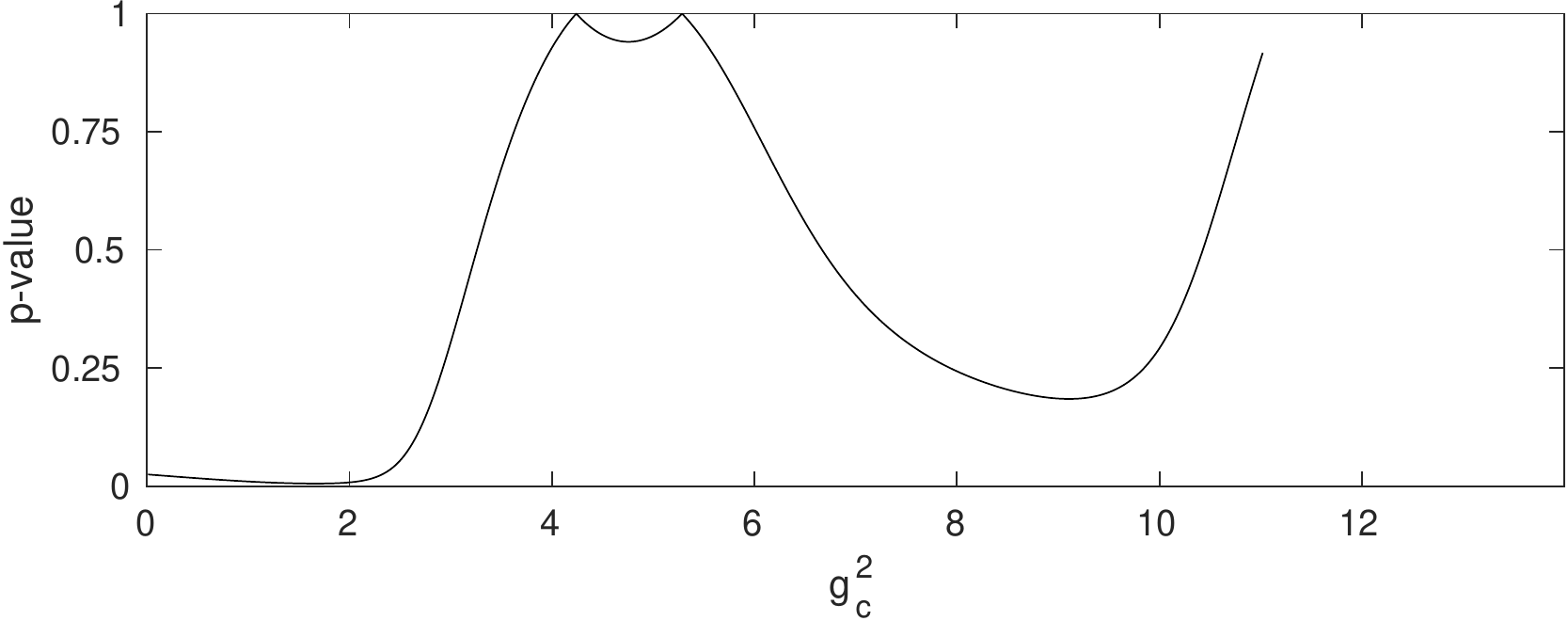}\\[3mm]
   \includegraphics[width=0.96\textwidth]{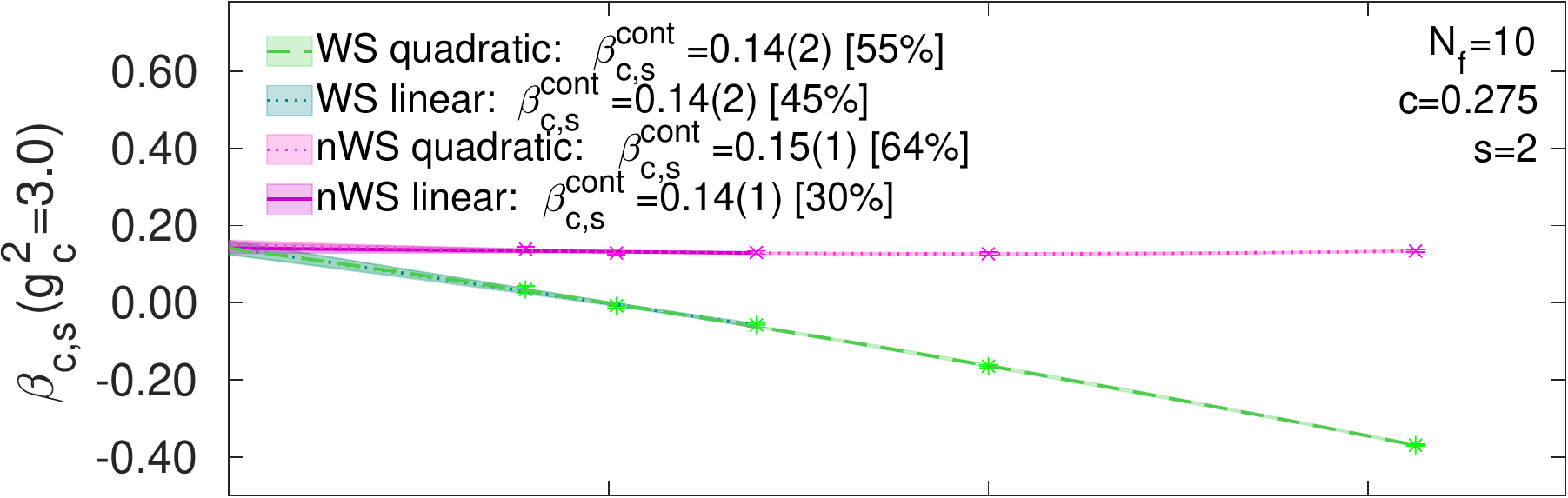}\\
   \includegraphics[width=0.96\textwidth]{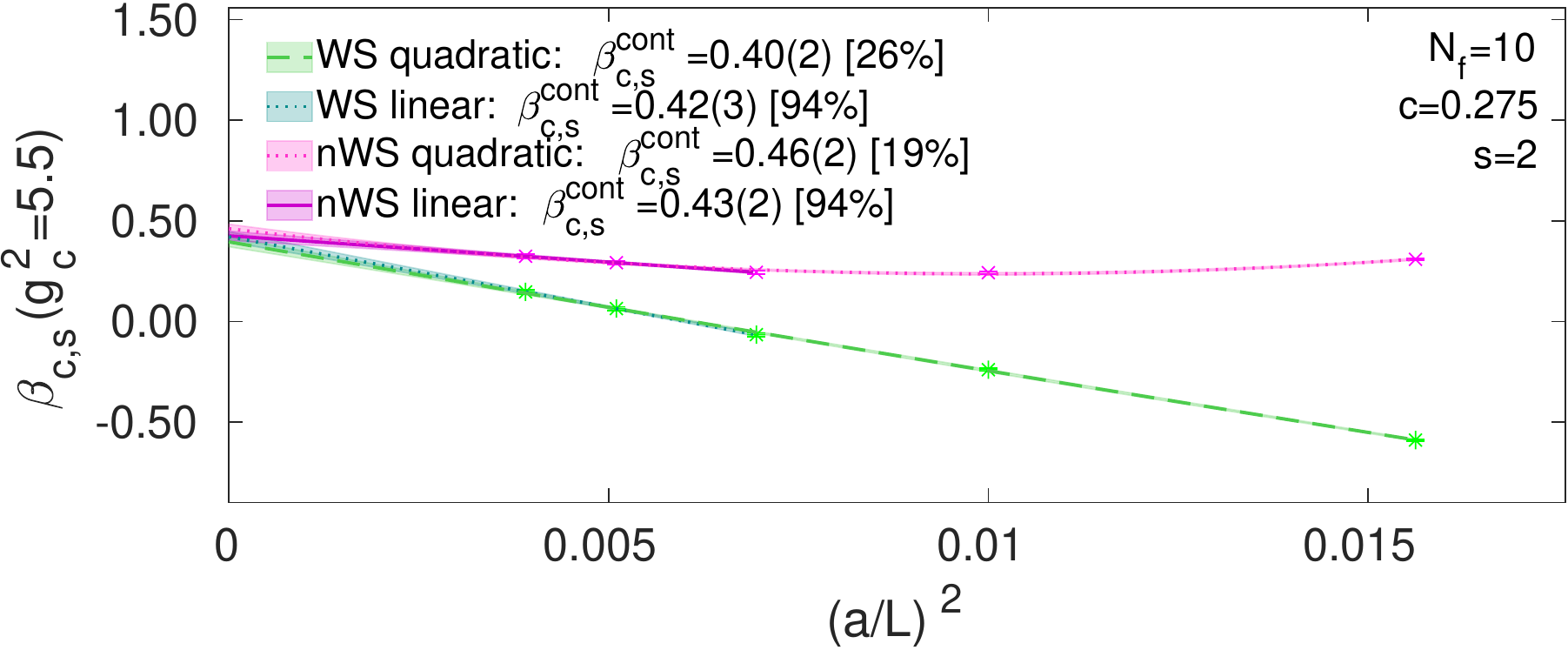}     
  \end{minipage}
  \begin{minipage}{0.49\textwidth}
    \flushright
    \includegraphics[width=0.96\textwidth]{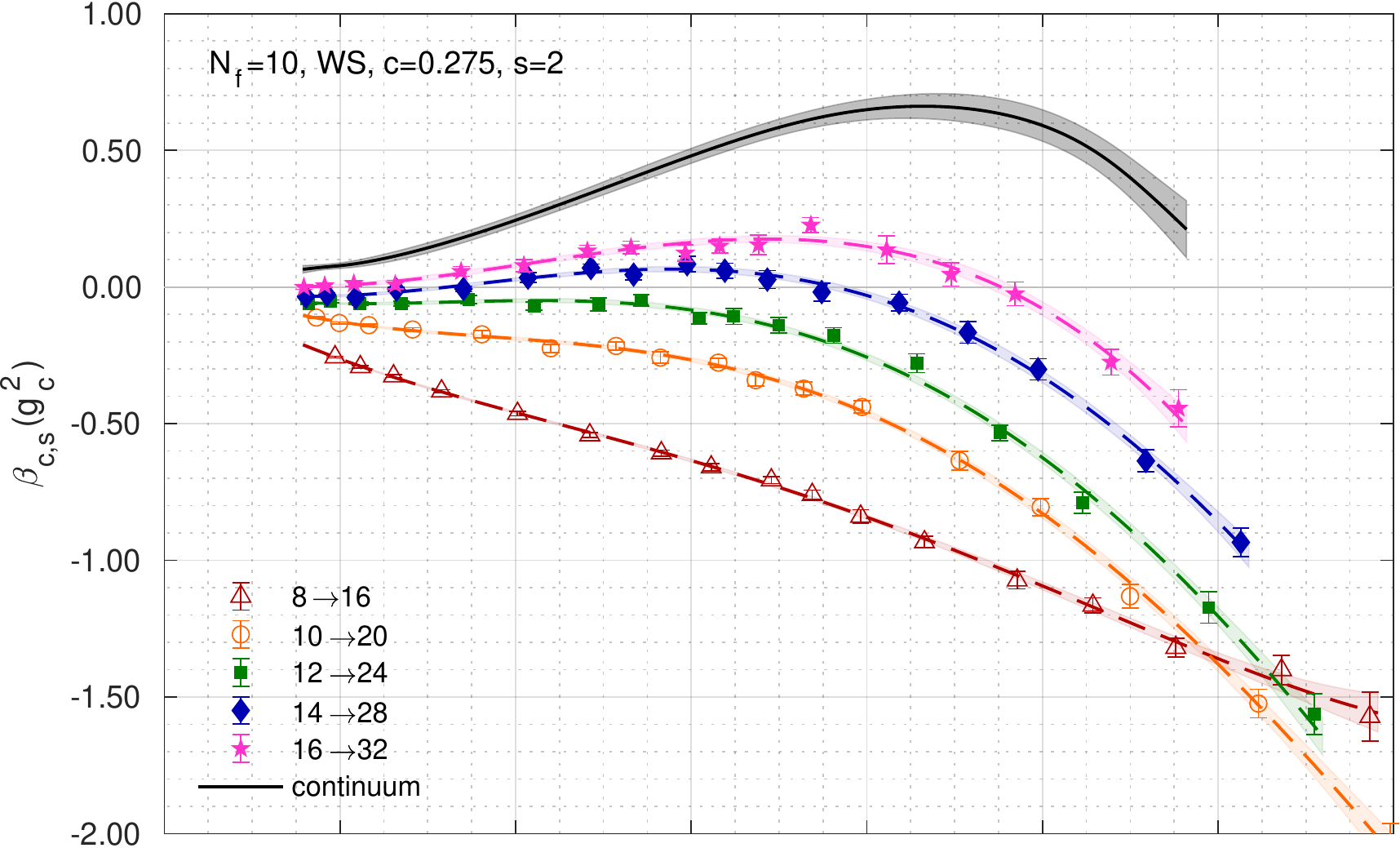}\\    
    \includegraphics[width=0.932\textwidth]{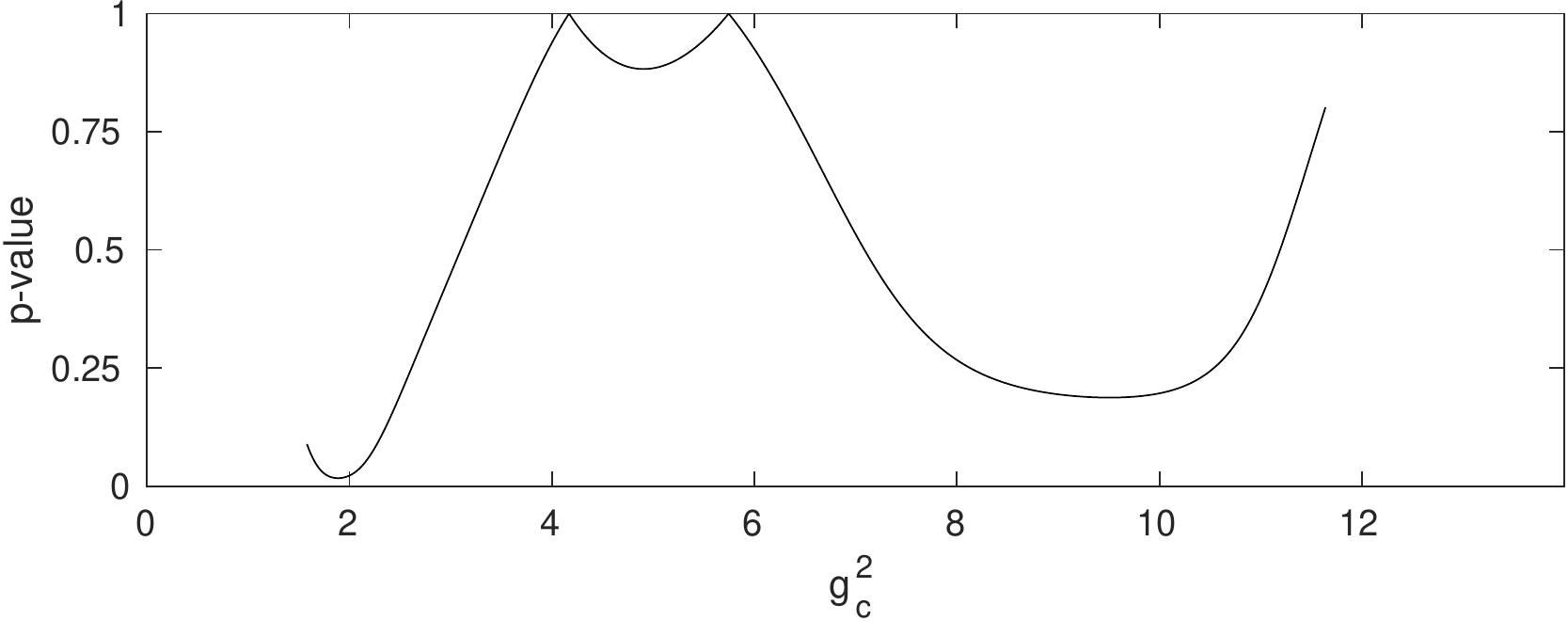}\\[3mm]
    \includegraphics[width=0.96\textwidth]{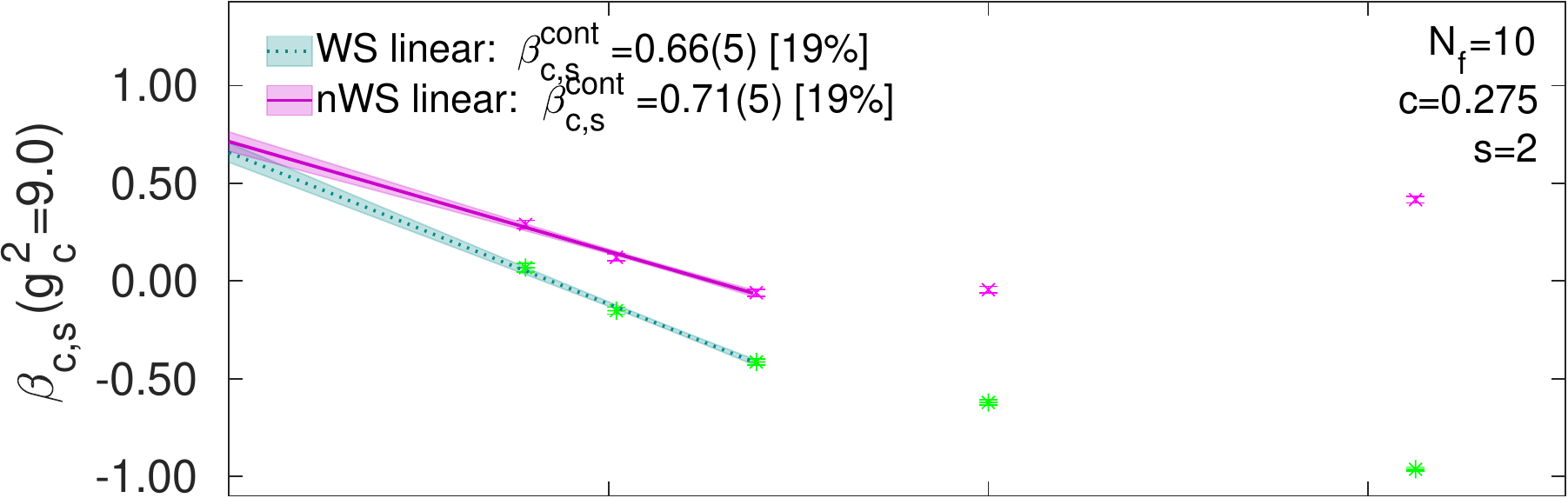}\\
    \includegraphics[width=0.96\textwidth]{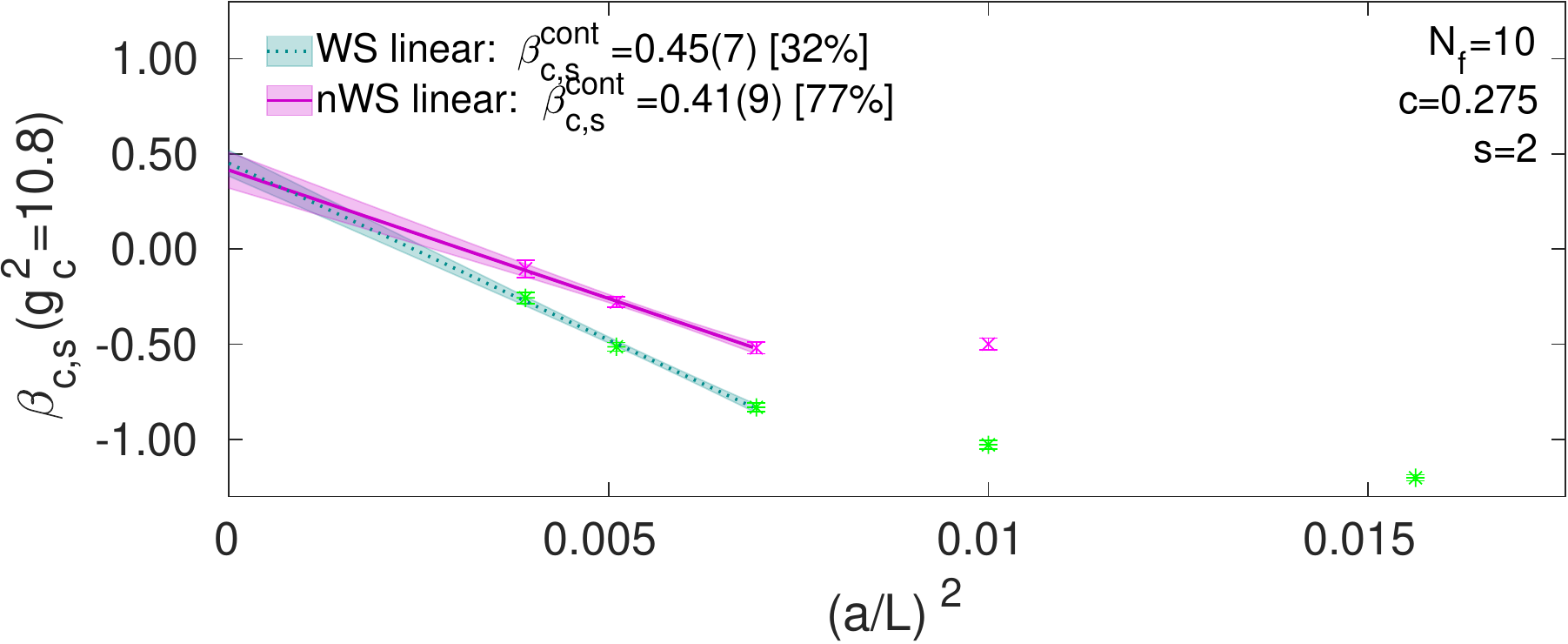}    
  \end{minipage}
  \caption{Discrete step-scaling $\beta$-function in the $c=0.275$ gradient flow scheme for our preferred nWS (left) and WS (right) data sets. The symbols in the top row show our results for the finite volume discrete $\beta$ function with scale change $s=2$. The dashed lines with shaded error bands in the same color of the data points show the interpolating fits. We take the continuum limit performing a linear fit (black line with gray error band) in $a^2/L^2$ to the three largest volume pairs (filled symbols). The $p$-values of the continuum extrapolation fit is shown in the plots in the second row. Further details of the continuum extrapolation at selected $g_c^2$ values are presented in the small panels at the bottom where the legend lists the extrapolated values in the continuum limit with $p$-values in brackets. Only statistical errors are shown.}
  \label{Fig.beta_c275}
\end{figure*}

\begin{figure*}[t]
  \begin{minipage}{0.49\textwidth}
   \flushright 
   \includegraphics[width=0.96\textwidth]{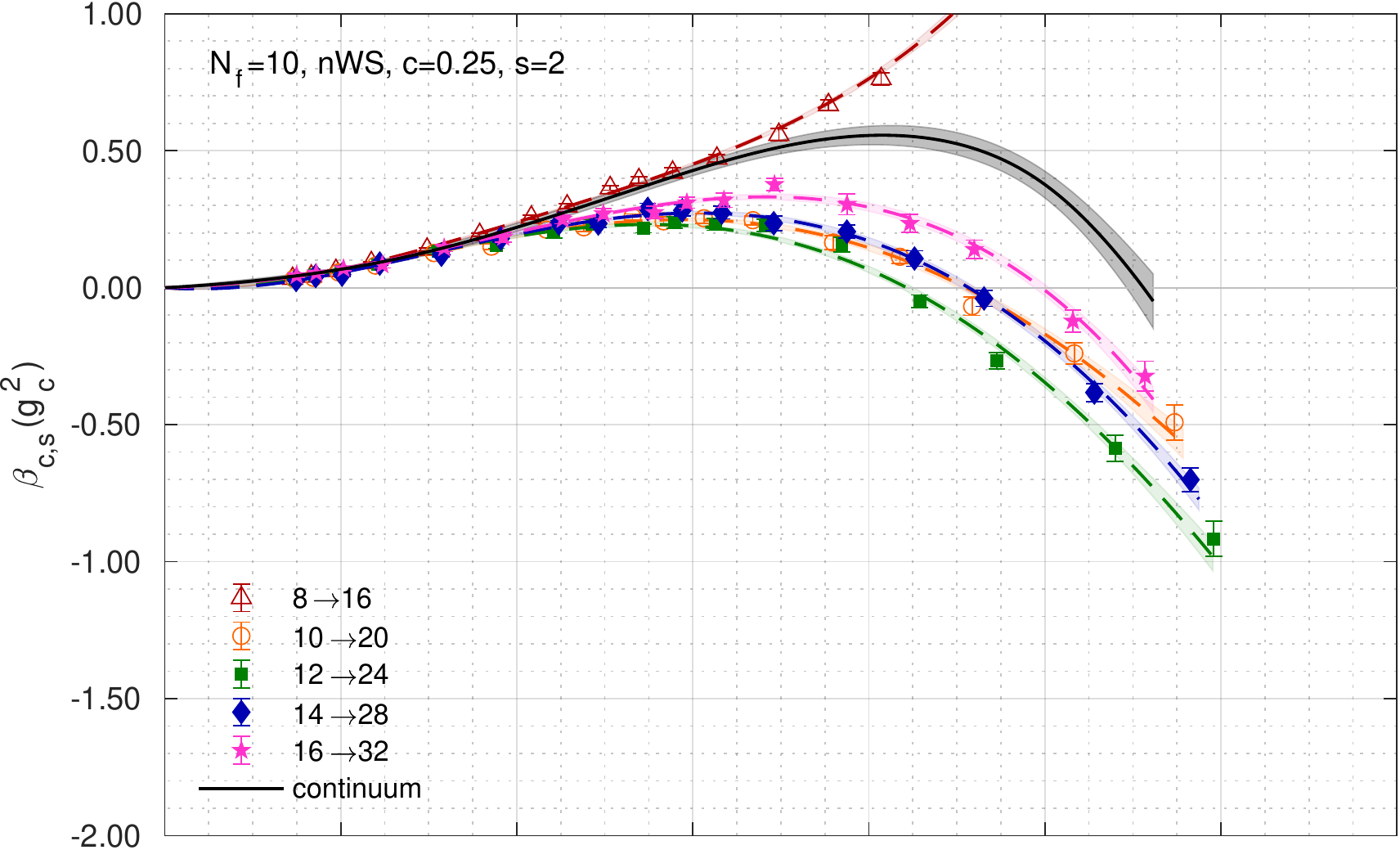}\\
   \includegraphics[width=0.932\textwidth]{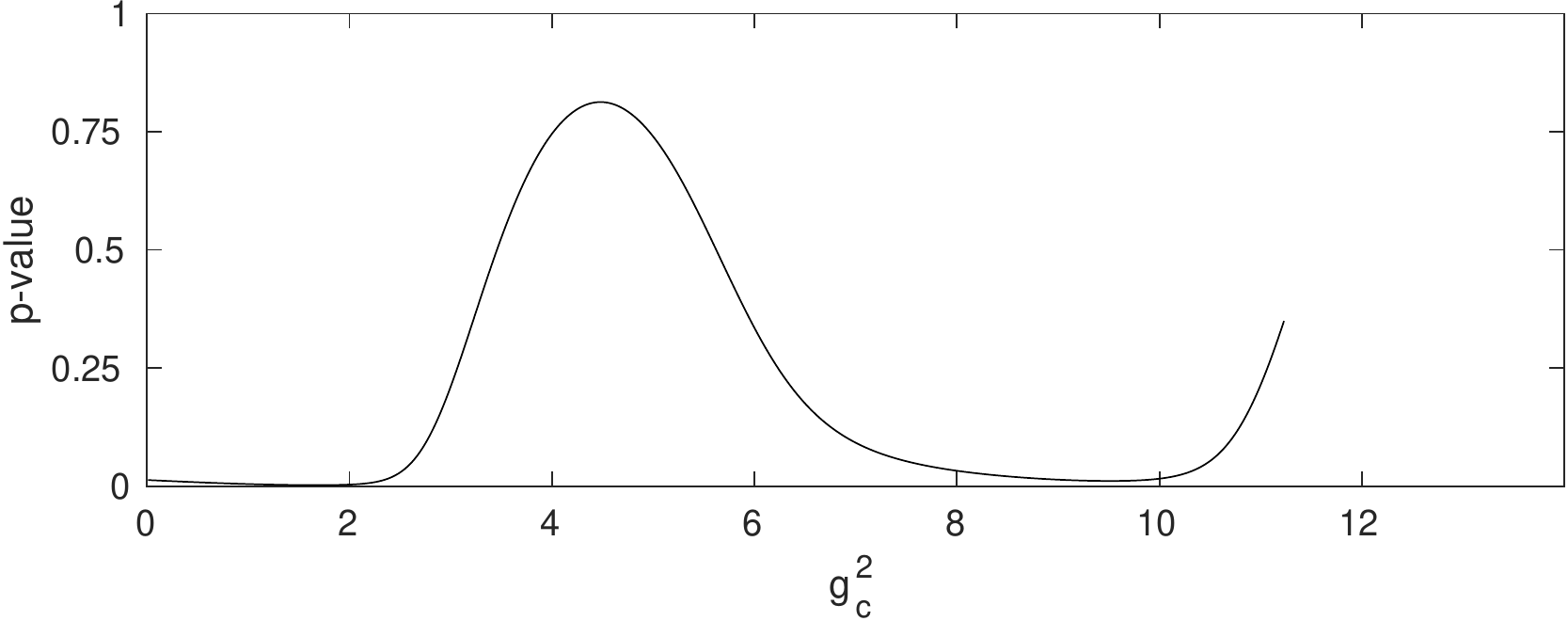}\\[3mm]
   \includegraphics[width=0.96\textwidth]{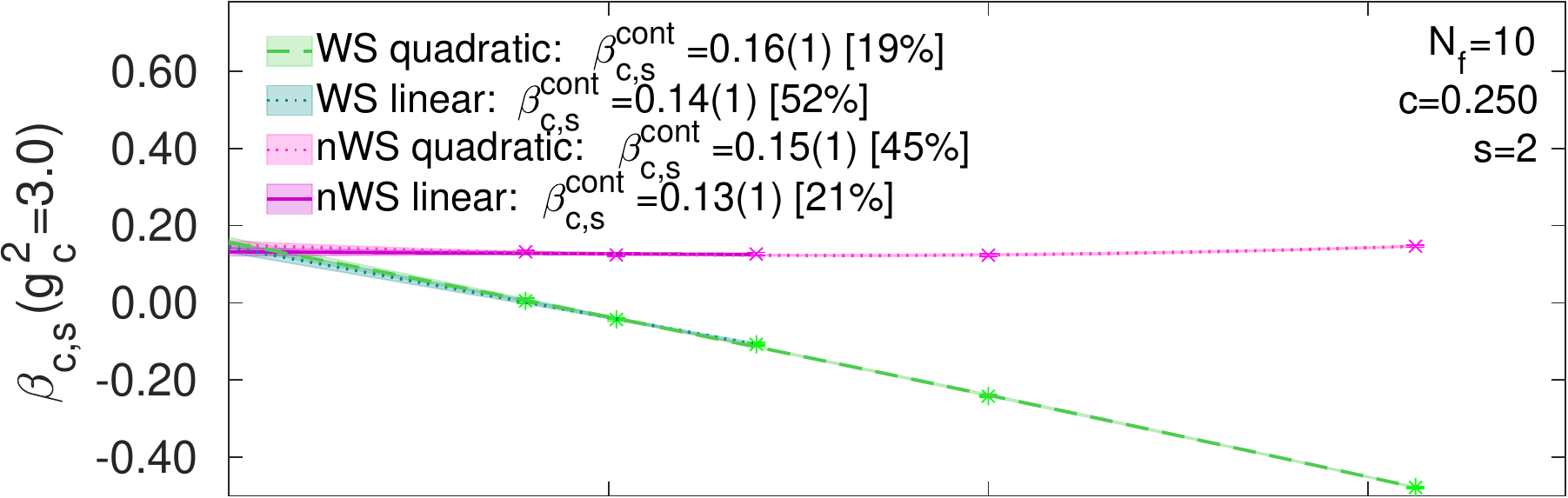}\\
   \includegraphics[width=0.96\textwidth]{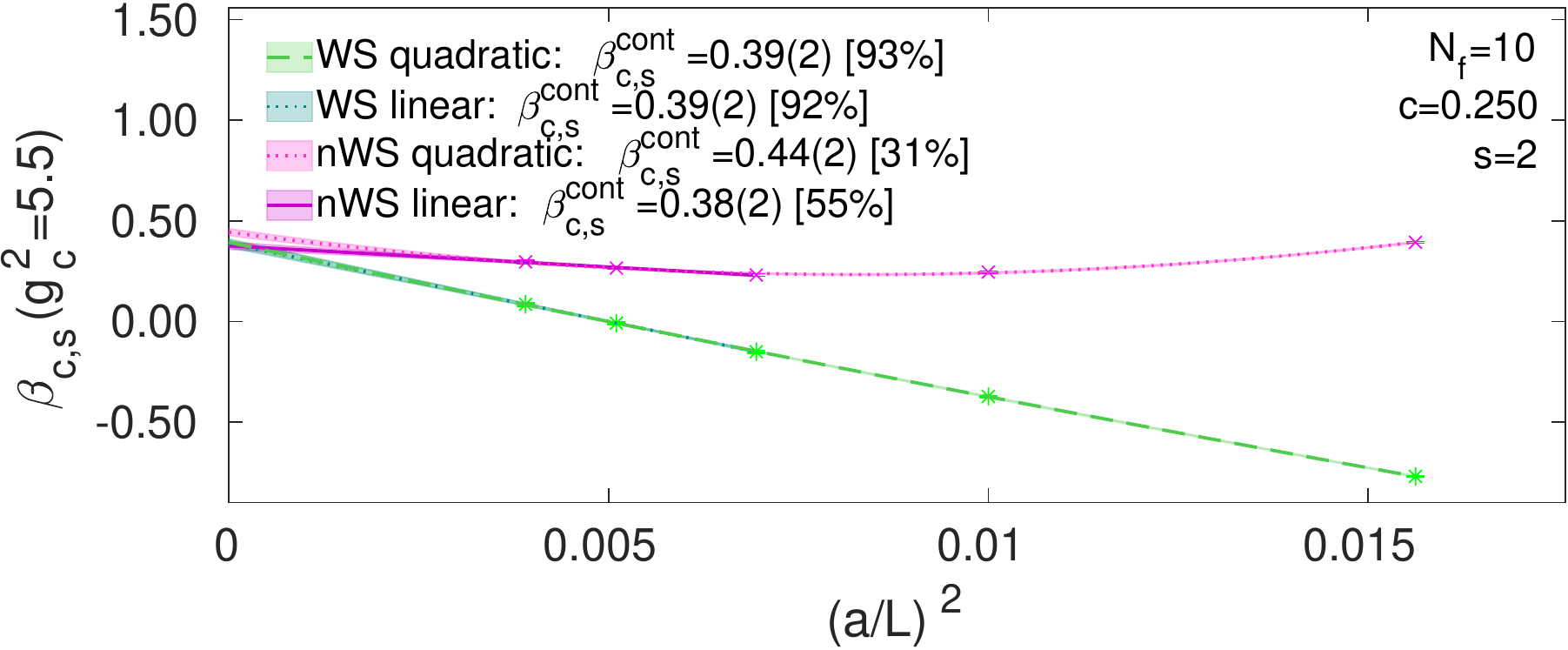}     
  \end{minipage}
  \begin{minipage}{0.49\textwidth}
    \flushright
    \includegraphics[width=0.96\textwidth]{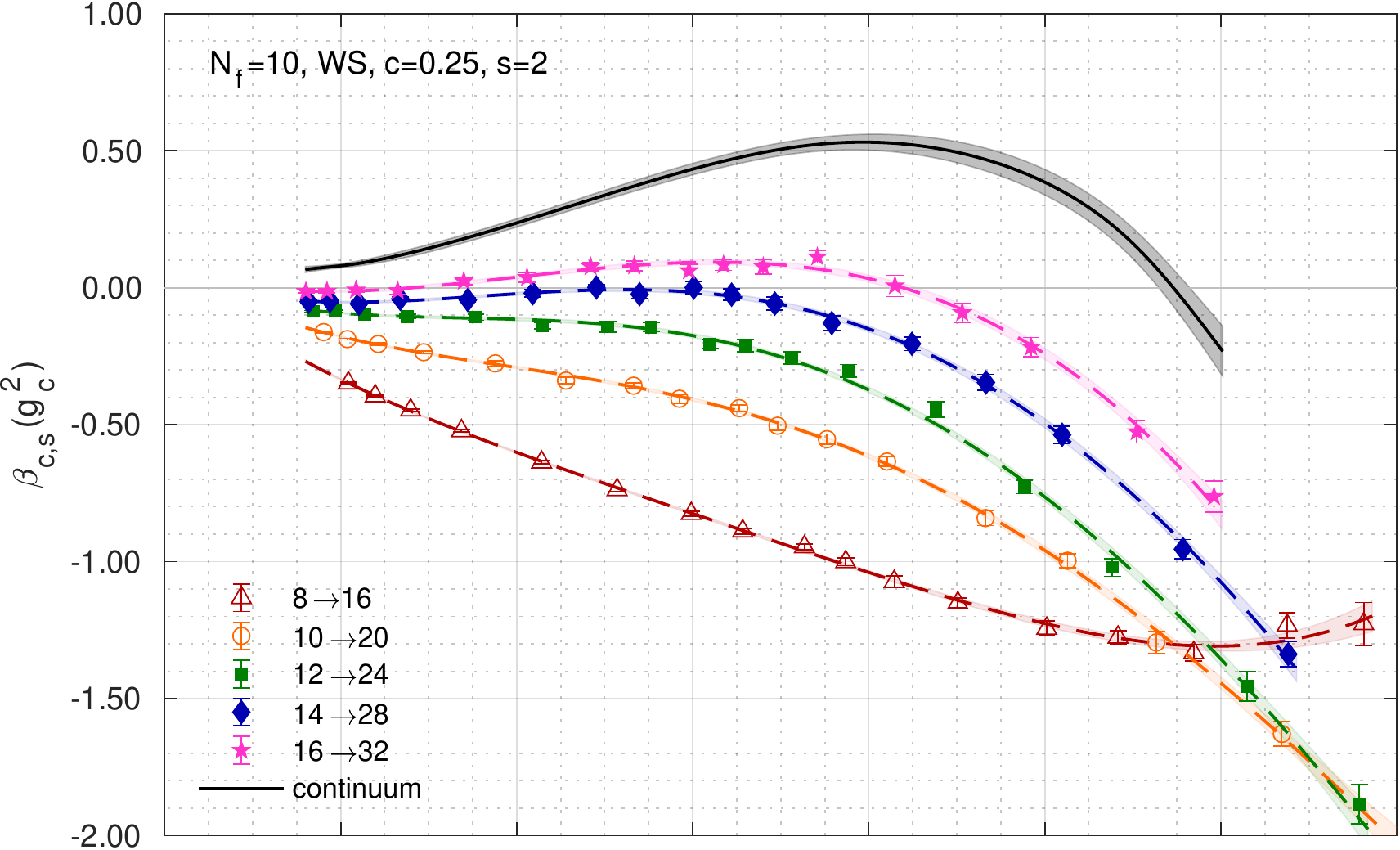}\\    
    \includegraphics[width=0.932\textwidth]{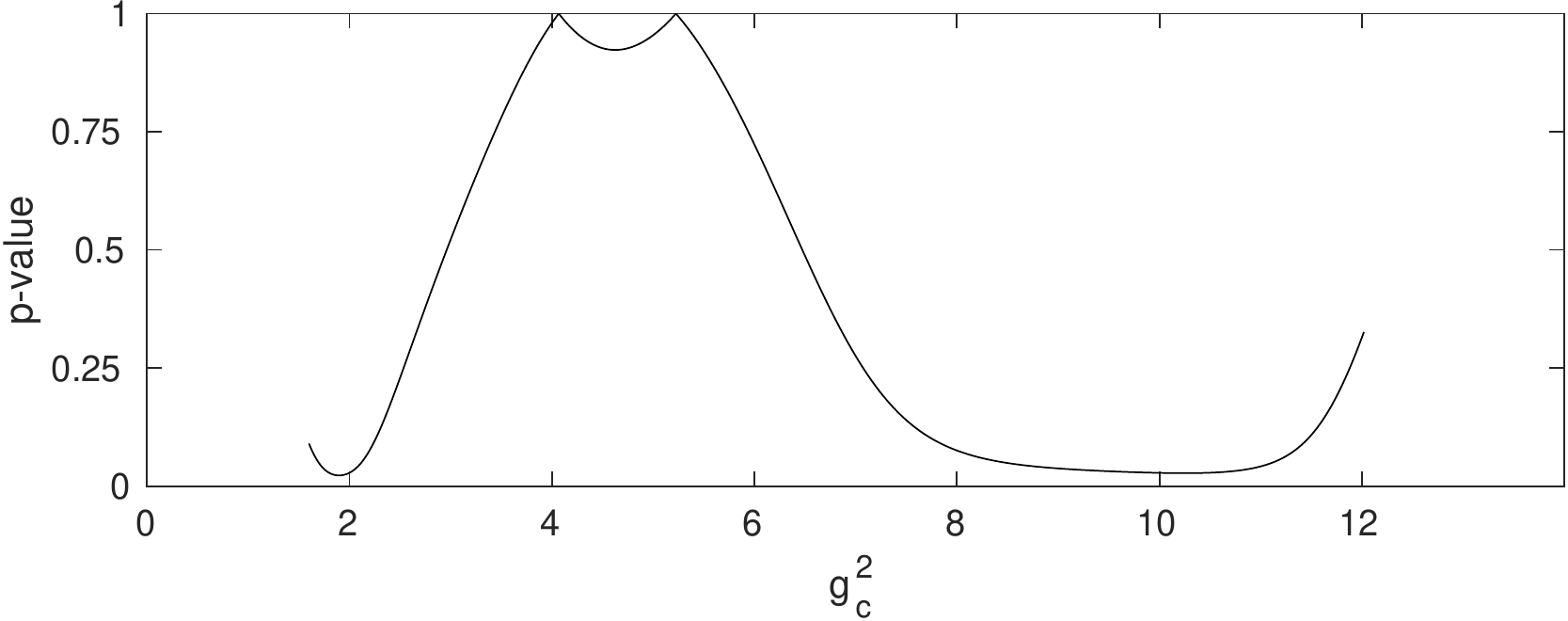}\\[3mm]
    \includegraphics[width=0.96\textwidth]{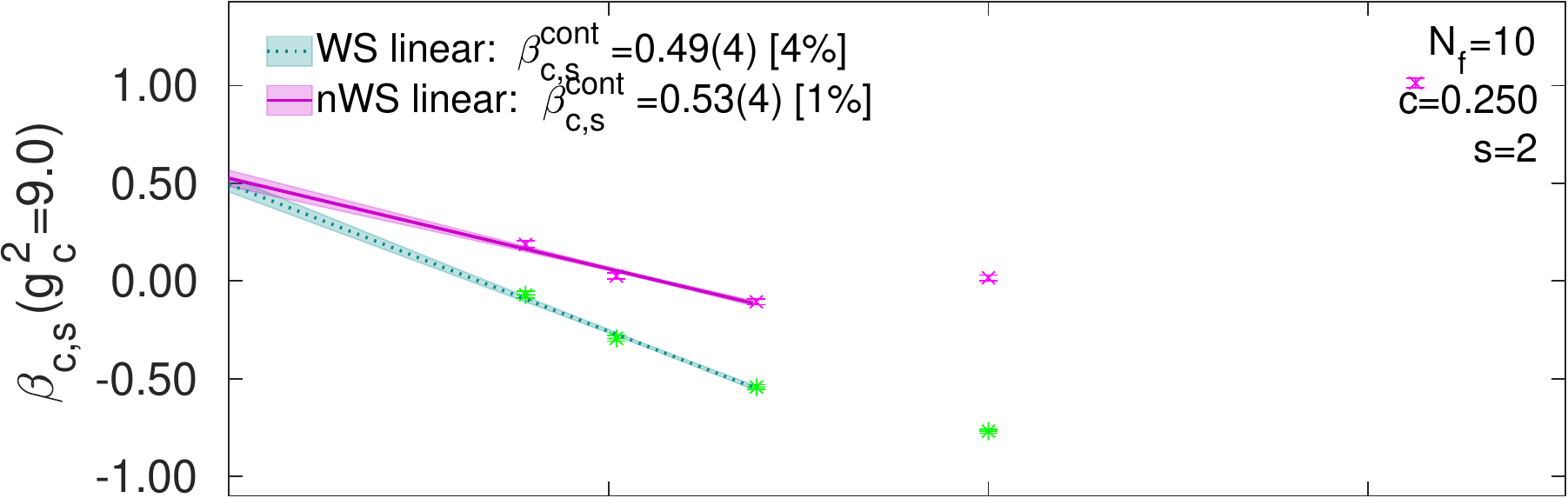}\\
    \includegraphics[width=0.96\textwidth]{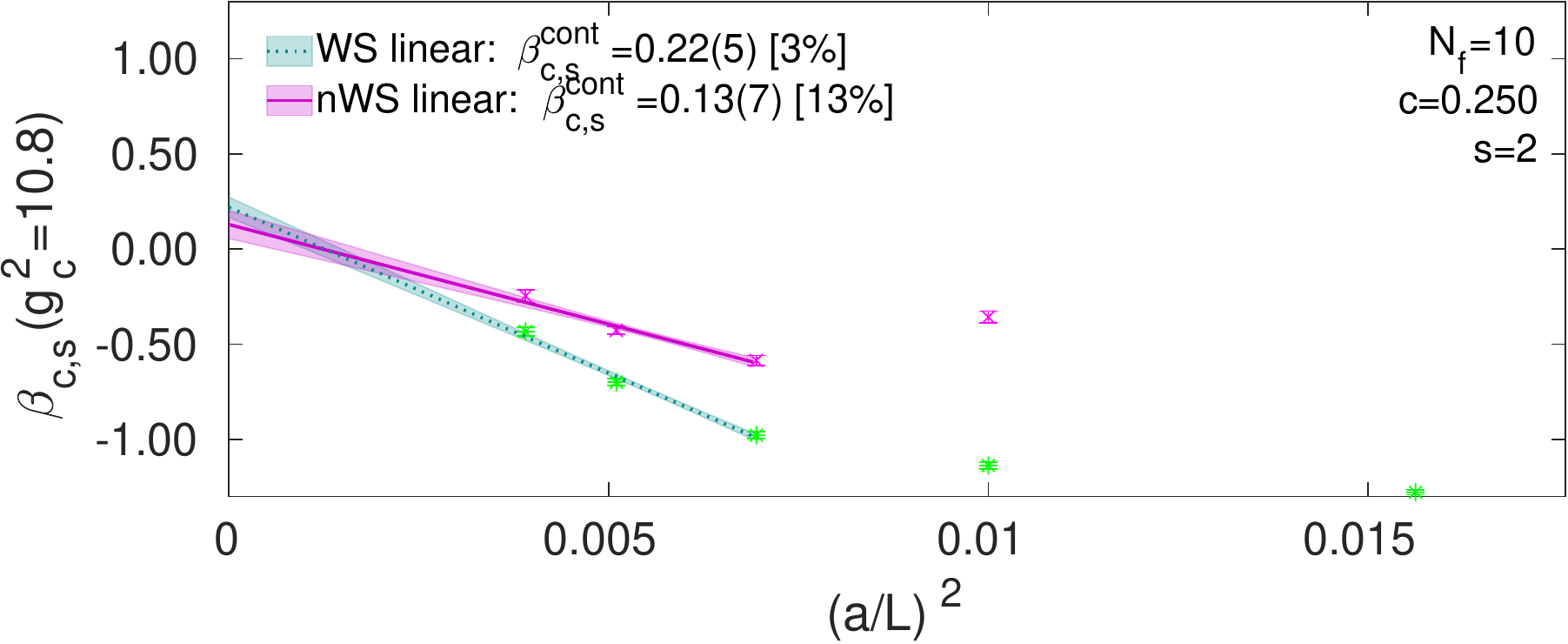}    
  \end{minipage}
  \caption{Discrete step-scaling $\beta$-function in the $c=0.250$ gradient flow scheme for our preferred nWS (left) and WS (right) data sets. The symbols in the top row show our results for the finite volume discrete $\beta$ function with scale change $s=2$. The dashed lines with shaded error bands in the same color of the data points show the interpolating fits. We take the continuum limit performing a linear fit (black line with gray error band) in $a^2/L^2$ to the three largest volume pairs (filled symbols). The $p$-values of the continuum extrapolation fit is shown in the plots in the second row. Further details of the continuum extrapolation at selected $g_c^2$ values are presented in the small panels at the bottom where the legend lists the extrapolated values in the continuum limit with $p$-values in brackets. Only statistical errors are shown.}
  \label{Fig.beta_c250}
\end{figure*}

 We present the values of the renormalized couplings $g_c^2$ according to Eq.~(\ref{Eq.gc2}) based on Wilson flow with the Symanzik operator in the renormalization schemes $c=0.300$, 0.275, and 0.250 in Appendix \ref{Sec.RenCouplings},  Table \ref{Tab.nWS_WS}.  
At weaker couplings, discretization effects can be reduced by applying perturbatively calculated tree-level normalization factors as included in Eq.~(\ref{Eq.gc2}). Since it is, however, not obvious if this perturbative correction is helpful at strong coupling, we carry out our preferred analysis with and without tree-level normalization and refer to the two analysis by nWS and WS, respectively. Our results show that for weaker couplings ($g_c^2\lesssim 6.0$) tree-level normalization clearly removes discretization effects but also for stronger coupling leads to a prediction of the continuum limit step-scaling function $\beta_{c,s}(g_c^2)$ which is consistent with the WS determination without tree-level normalization. We therefore present our results for the ten flavor step-scaling function in Figs.~\ref{Fig.beta_c300}--\ref{Fig.beta_c250} showing the nWS analysis on the left and the and WS analysis on the right. 

Our analysis proceeds in the following steps:
\begin{itemize}
\item We calculate the discrete $\beta_{c,s}(g_c^2;L)$ function defined in Eq.~(\ref{Eq.beta_cs}) for our five different volume pairs with scale change $s=2$. The outcome is shown by the colored data symbols in the top panels of Figs.~\ref{Fig.beta_c300}--\ref{Fig.beta_c250}. Since simulations at different bare coupling $\beta$ are statistically independent, also these data points are statistically independent.
\item Next we  interpolate the data for each pair of lattice volumes using a polynomial form motivated by the perturbative expansion
  \begin{align}
\beta_{c,s}(g_c^2;L) = \sum_{i=0}^{n} b_i g_c^{2i}.
\label{Eq.fit_form}
  \end{align}
We find that $n=4$ is sufficient to describe our data well  over the full  $g_c^2$ range, and the corresponding fits have all excellent $p$-values. Since discretization effects are sufficiently small at weak coupling for nWS, we constrain the intercept $b_0=0$ but fit $b_0$ for WS. We report the (correlated) fit coefficients in Table \ref{Tab.interpolations}. Some of the coefficients are not resolved  statistically.  This is not a concern,  as long as the interpolations describe our data well. This is evident from the $\chi^2/\text{d.o.f.}$~or $p$-values also given in Table \ref{Tab.interpolations} and visible by the color-shaded bands (with dashed lines indicating the central values) passing through the same color data points in the top row panels of Figs.~\ref{Fig.beta_c300}--\ref{Fig.beta_c250}. 
\item The interpolating fits provide us with finite volume discrete step-scaling functions $\beta_{c,s}(g_c^2; L)$ at continuous values of $g_c^2$. Due to the constraint $b_0=0$, nWS starts at zero  but WS begins with our first data point. The range of $g_c^2$ covered varies with the scheme $c$.
\item To obtain $\beta_{c,s}(g_c^2)$ in the continuum, we perform an infinite volume continuum limit extrapolation of the interpolated $\beta_{c,s}(g_c^2; L)$ functions at fixed $g^2_c$ values. Since in the strong coupling limit our smallest volume pair $8\to 16$,  and in some cases also $10\to 20$ pair shows large discretization effects, we predict the continuum limit by performing a linear fit in $a^2/L^2$ on the three largest volume pairs.  In Figs.~\ref{Fig.beta_c300}--\ref{Fig.beta_c250} we show the $8\to 16$ and $10\to 20$ volume pairs with open symbols and solely for illustrative purposes. Only volume pairs $12\to 24$, $14\to 28$, and $16\to 32$  displayed with filled symbols enter our final continuum limit result.
\item
  The quality of the continuum limit extrapolation is monitored by the $p$-value shown by the second row panels in Figs.~\ref{Fig.beta_c300}--\ref{Fig.beta_c250}. While overall the $p$-values are excellent for most of the $g_c^2$ range, the values rapidly drop for the weakest coupling.  Theoretically not expected, we can only guess that poor $p$-values at weak coupling are an artifact of very precise data. Since all  five volume pairs sit very close to each other, there is little doubt on the continuum limit. Moreover,  the weak coupling range is in good agreement with perturbative predictions which are certainly trustworthy for $g_c^2\lesssim 2.0$. More concerning is the drop of the $p$-value for $c=0.250$ in the range of $9 \lesssim g_c^2 \lesssim 11$. Here a low $p$-value around or even below 5\% gives rise to concerns that the three volumes included in the continuum extrapolation may not be large enough \cite{Fritzsch:2013je}.
\item  Further details of the continuum limit extrapolation are shown by the four panels at the bottom of Figs.~\ref{Fig.beta_c300}--\ref{Fig.beta_c250}. Selecting $g_c^2=3.0$, 5.5, 9.0, and 10.8 we show the continuum limit extrapolation vs.~$(a/L)^2$. At weaker coupling ($g_c^2=3.0$ and 5.5), our preferred linear fit for the three largest volume pairs is shown together with a quadratic fit using all five volume pairs. At stronger coupling ($g_c^2=9.0$ and 10.8), the smallest volume pairs exhibit large cutoff effects and quadratic fits  to all five volume pairs have very low $p$-values introducing significant corrections for the smaller volume pairs. This may indicate the need for even higher order terms or nonperturbative corrections to reliably describe all five data points. Performing a linear continuum limit extrapolation  using only two data points corresponding to our largest volume pairs  predicts a step-scaling function with larger uncertainties but still in agreement with our preferred result at the $1\sigma$ level.  Hence we show only a linear fit to the three largest data points. In all cases we observe excellent agreement between fits to the nWS and WS data. 
\end{itemize}

Taking a closer look at the nWS analysis shown on the left of Figs.~\ref{Fig.beta_c300}--\ref{Fig.beta_c250}, it is noteworthy to point out that discretization effects 
are barely, if at all, visible for $g_c^2\lesssim 4.5$. This range coincides with the value of $g^2$ where the perturbative predictions of the $\beta$-function at 3-, 4, and 5-loop order are also in good agreement (see e.g.~Fig.~\ref{Fig.beta_Nf10_final}).

\subsection{Alternative flow/operators}
\label{Sec.Alt}

\begin{figure}[t]
  \includegraphics[width=0.95\columnwidth]{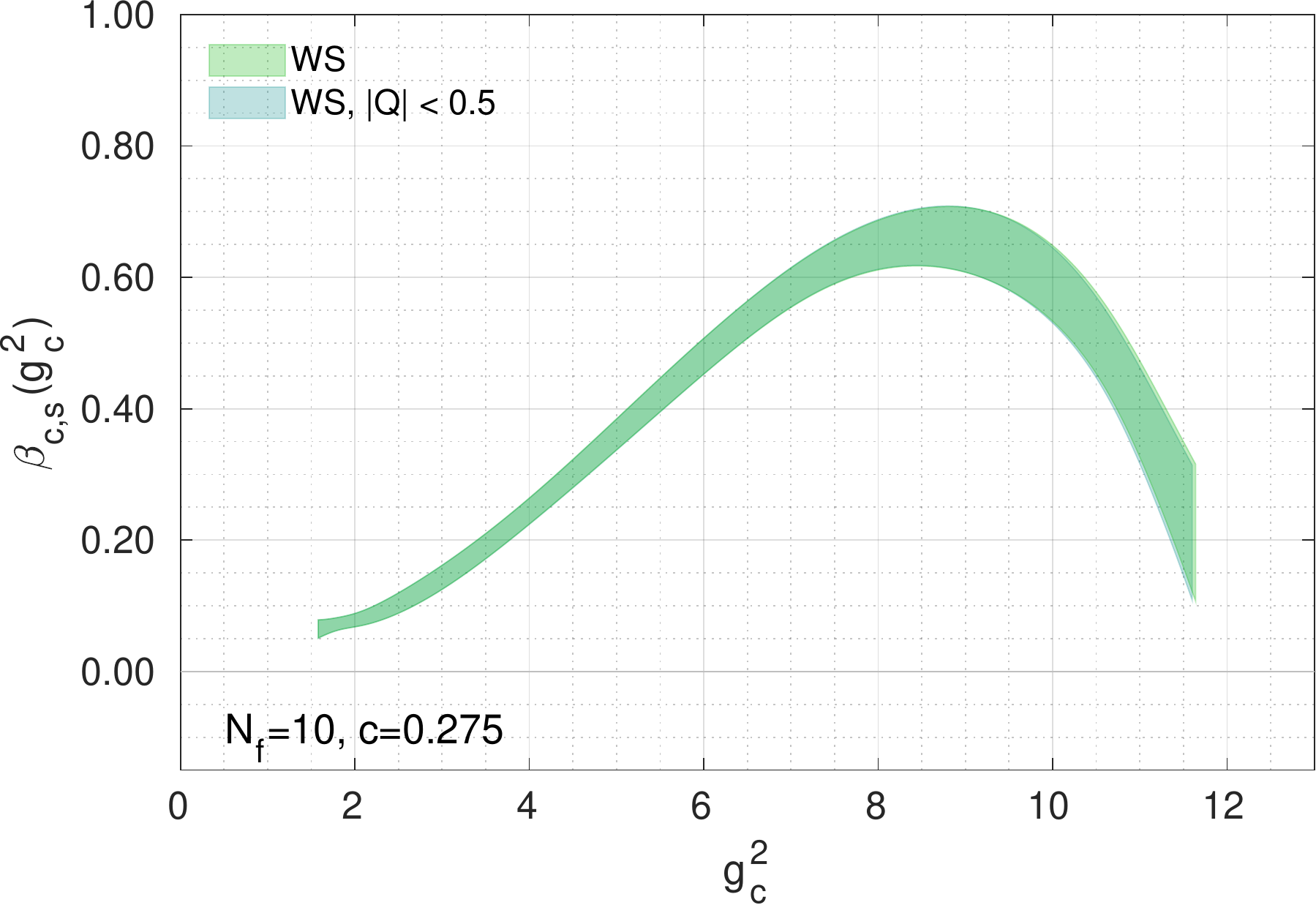}  \\[3mm]
  \includegraphics[width=0.95\columnwidth]{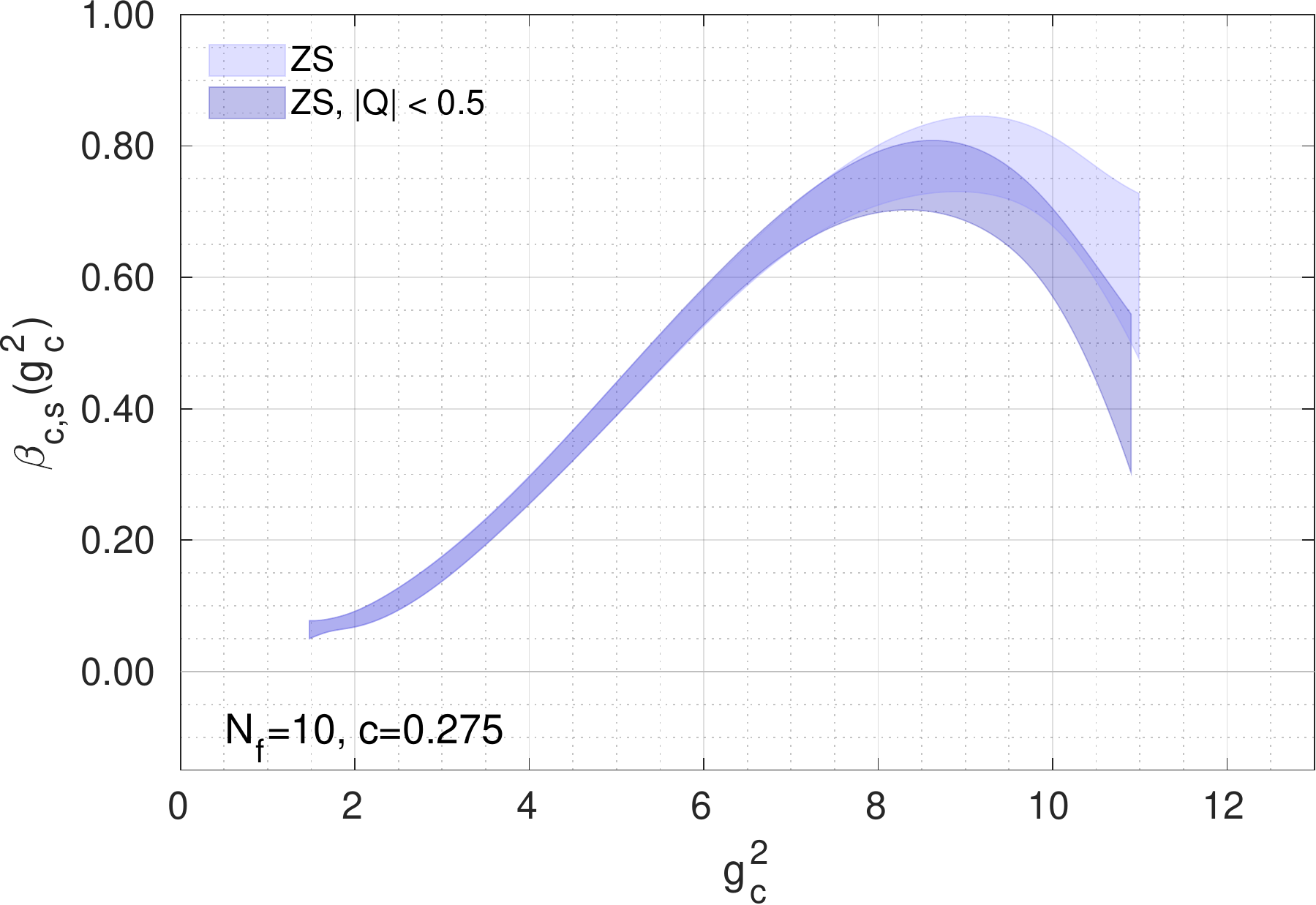}  \\[3mm]
  \includegraphics[width=0.95\columnwidth]{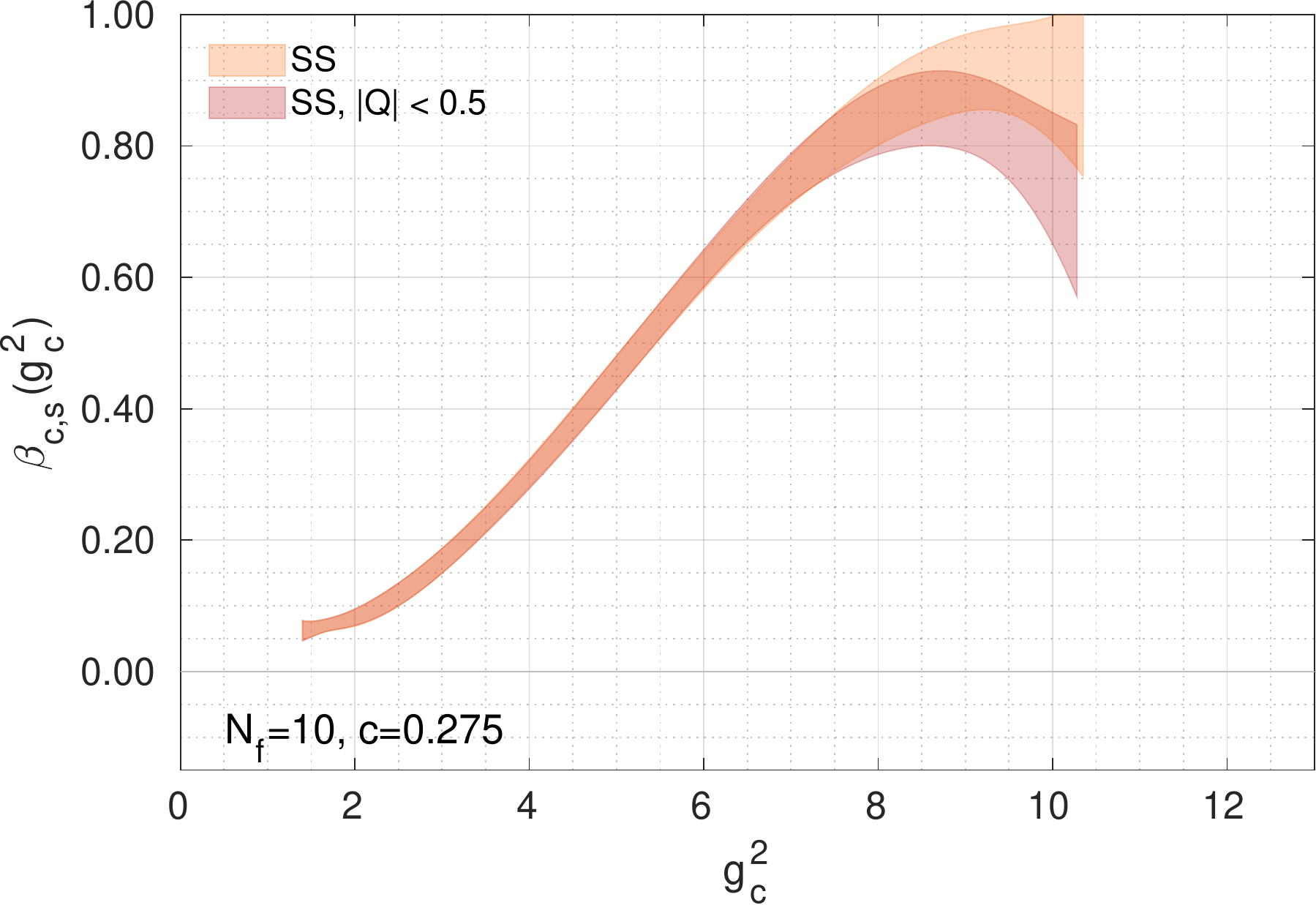}
  \caption{Impact of nonzero topological charge $Q$ on the continuum limit of the $\beta$ functions for scheme $c=0.275$. Performing the analysis with the Symanzik operator we show  Wilson (top), Zeuthen (middle), and Symanzik (bottom) flow.} 
  \label{Fig.betaTopoCharge}
\end{figure}

\begin{figure*}[p]
  \includegraphics[height=0.304\textheight]{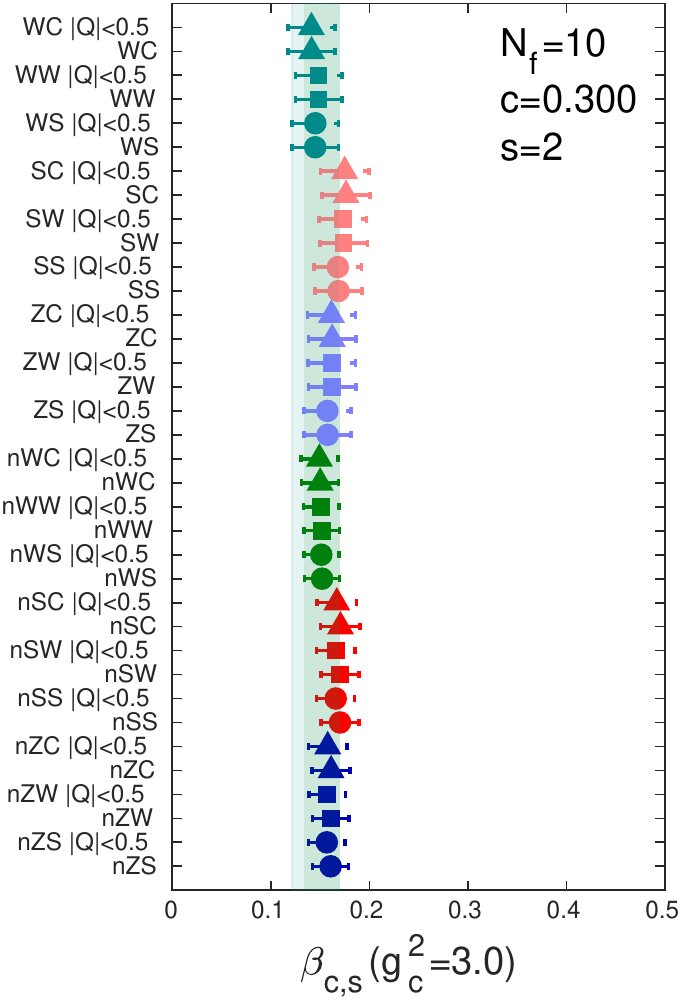}
  \includegraphics[height=0.304\textheight]{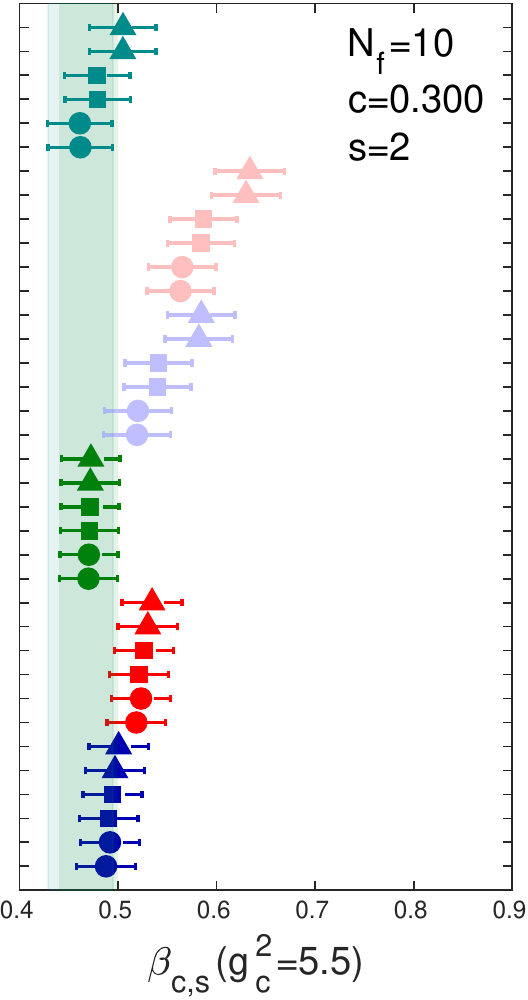}
  \includegraphics[height=0.304\textheight]{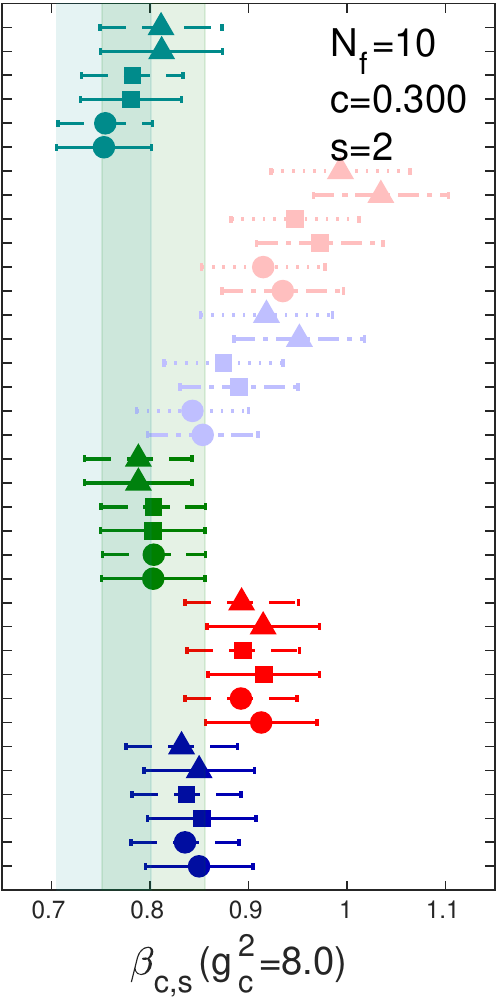}
  \includegraphics[height=0.304\textheight]{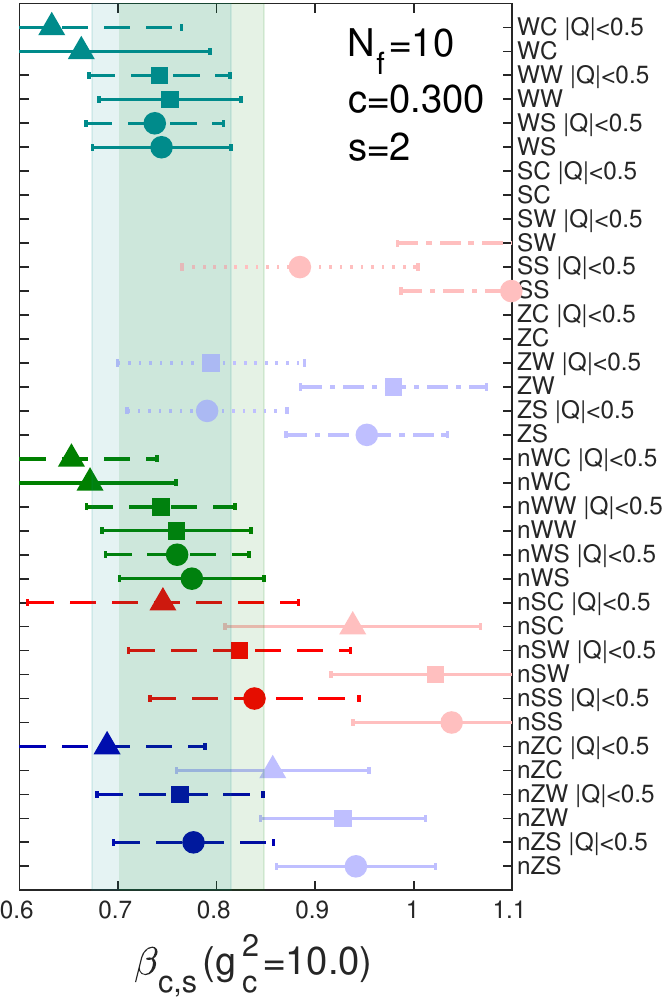}\\  
  \includegraphics[height=0.304\textheight]{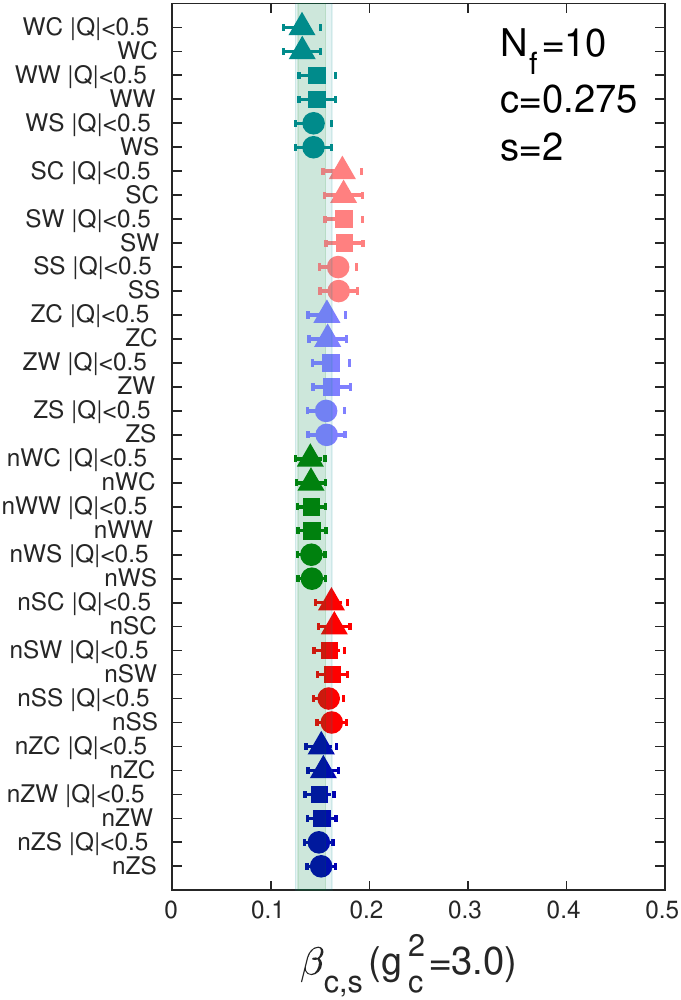}
  \includegraphics[height=0.304\textheight]{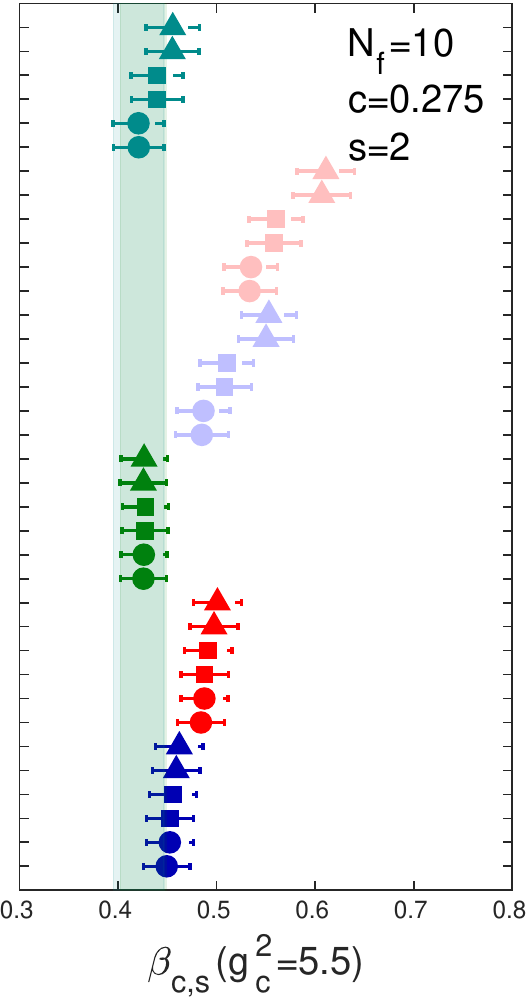}
  \includegraphics[height=0.304\textheight]{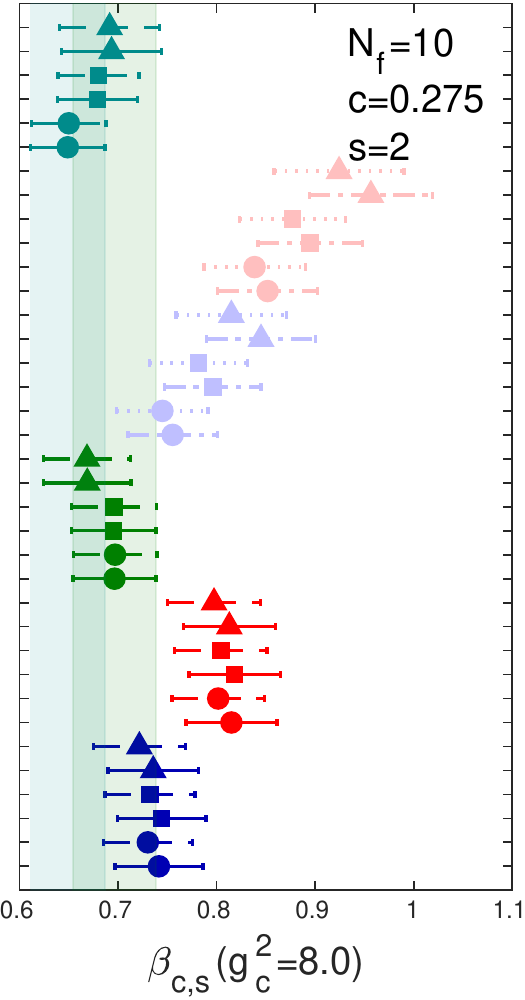}
  \includegraphics[height=0.304\textheight]{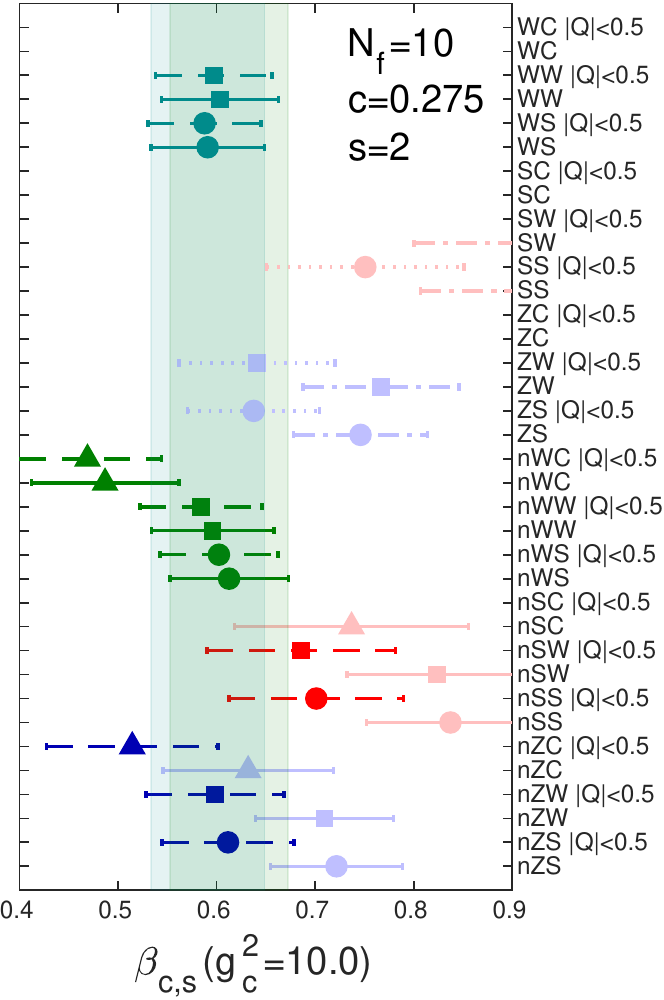}\\
  \includegraphics[height=0.304\textheight]{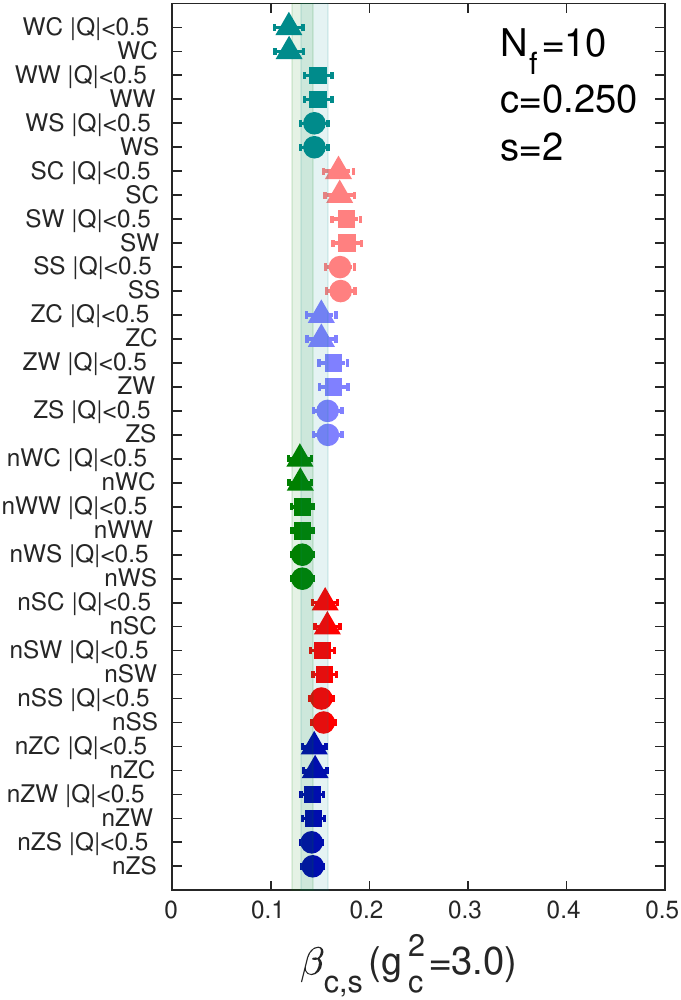}
  \includegraphics[height=0.304\textheight]{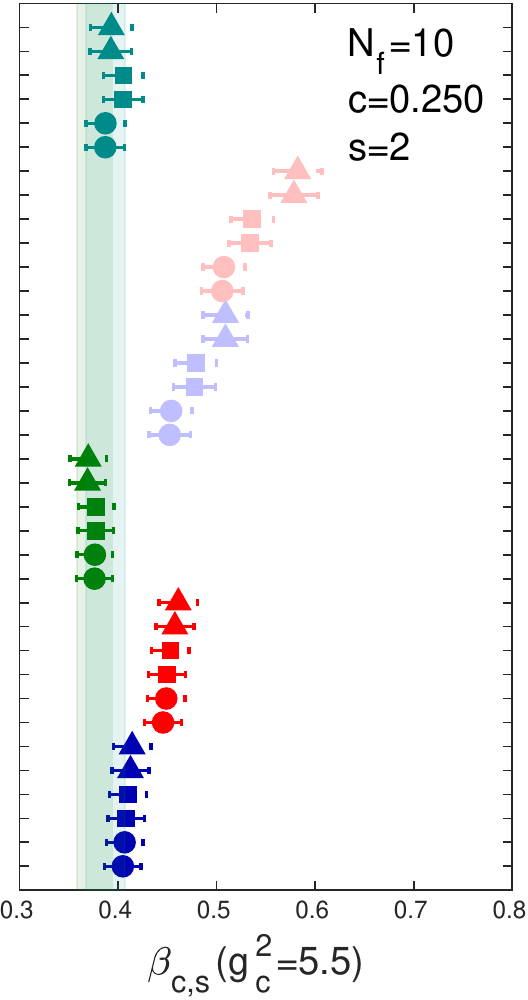}
  \includegraphics[height=0.304\textheight]{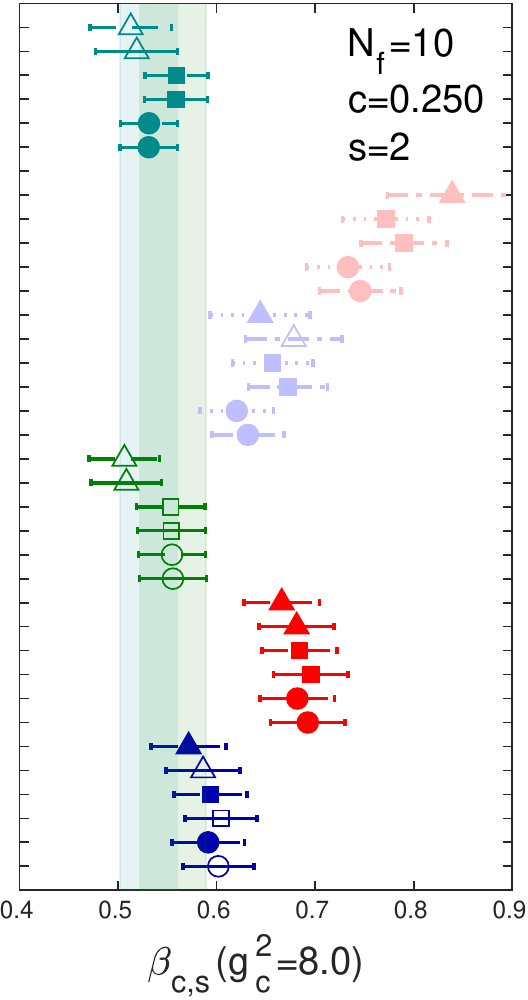}
  \includegraphics[height=0.304\textheight]{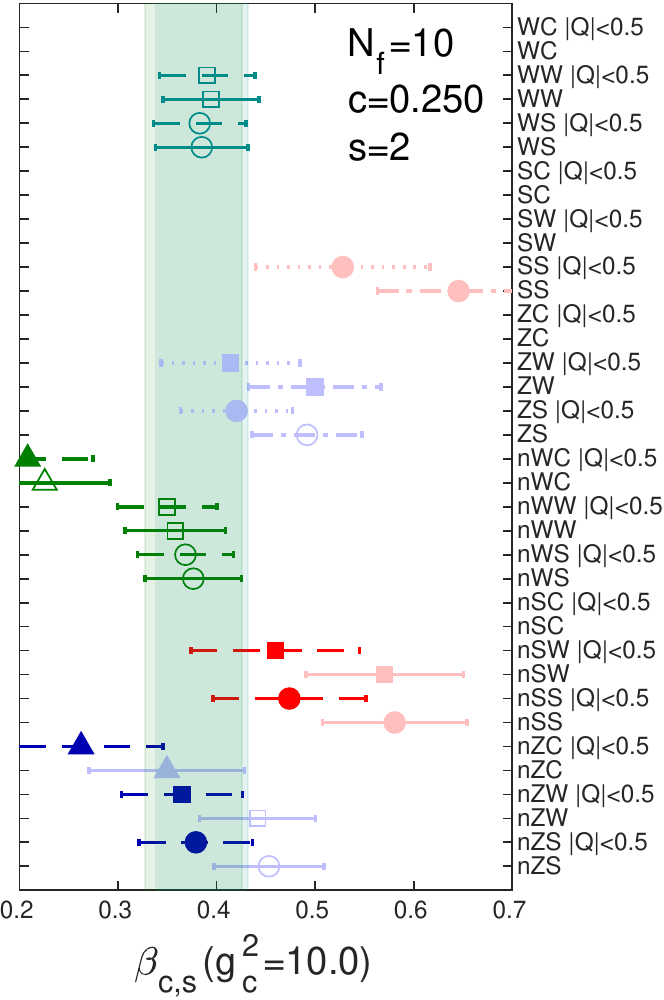}  
  \caption{Systematic effects on $\beta_{c,s}(g_c^2)$ due to tree-level improvement, different flows and operators as well as nonzero topological charge. The continuum limit is in all cases obtained by linear extrapolation to the three largest volume pairs. The columns show our continuum limit results at selective $g_c^2$ values of 3.0, 5.5, 8.0, and 10.0; the rows correspond to renormalization schemes $c=0.300$, 0.275, 0.250. Open symbols indicate extrapolations with a $p$-value below 5\% and light shaded data points are considered not reliable. The vertical shaded bands highlight our preferred (n)WS analysis.}
  \label{Fig.beta_sys}
\end{figure*}

Considering different operators and/or gradient flows can help to understand systematic effects. Compared to our $N_f=12$ calculation, our simulations for ten flavors extend to much stronger coupling. This has two consequences: on the one hand we established that Zeuthen and Symanzik flows exhibit novel lattice artifacts showing up as nonzero topological charge, on the other hand we know that discretization effects of different operators and gradient flows can also differ significantly. Hence care must be taken when considering alternative flows/operators in order to avoid that a particularly poor choice dominates the result.

To understand the effect of nonzero topological charge, we repeat our analysis for each flow/operator combination with a filtering option where we select only configurations with $|Q| < 0.5$,\footnote{From now on we simply use $Q$ in place $Q_\text{geom}$.} i.e.~we discard all configurations with topological charge $|Q|>0.5$ at flow time $t=(cL)^2/8$. 
Of course such an ad hoc measure is not theoretically sound. It gives, however, insight into the effect due to nonzero $Q$. For all three flows (Wilson, Zeuthen, Symanzik) we use the Symanzik operator in $c=0.275$ scheme to calculate the continuum limit of the step-scaling function. The three plots in Fig.~\ref{Fig.betaTopoCharge} show the overlay for each flow with and without filtering the topological charge. 
Further details of the Zeuthen flow analysis are presented in Figs.~\ref{Fig.beta_alt_ZS} and \ref{Fig.beta_alt_ZS_topo}  in Appendix \ref{Sec.AltFlow}.    We would like to stress that for our preferred analysis based on Wilson flow filtering on $Q$ has no effect, but filtering lowers  the predicted step-scaling function in the strong coupling regime both for Zeuthen and Symanzik flow.  

In total we consider nine flow/operator combinations: Wilson (W), Symanzik (S), and Zeuthen (Z) flow each with three operators Wilson plaquette (W), Symanzik (S), or clover (C) and refer to the analysis by two capital letters indicating the flow and the operator. In addition we analyze the data without and with tree-level normalization, prefixing in the latter case 'n' to our short hand notation. Further, we perform the analysis with and without topological filtering. Choosing four selected values of $g_c^2=3.0,$ 5.5, 8.0, and 10.0 to cover the full range of $g_c^2$, we show the predicted values of $\beta_{c,s}(g_c^2)$ in Fig.~\ref{Fig.beta_sys}.  The first row uses $c=0.300$, the second row $c=0.275$, and the last row $c=0.250$. In all cases we show continuum limit results obtained  from a linear fit in $a^2/L^2$ to our three largest volume pairs. The shape of the symbol indicates the operator (square: Wilson plaquette, circle: Symanzik, triangle: clover) and the color distinguishes the flow (green: Wilson, blue: Zeuthen, red: Symanzik). At the strongest coupling ($g_c^2=10.0$) shown, the effect of nonzero topological charge is most significant for Zeuthen and Symanzik flow. Hence the determination without filtering is shown in pale colors. In addition operators of the same flow exhibiting large discretization effects are displayed in pale colors.

Focusing on determinations we consider reliable and  displayed in vivid colors, we observe a very good consistency with our preferred (n)WS determinations indicated by the vertical green bands. Most determinations agree better than at the $1\sigma$ level and only the tree-level improved determinations with Symanzik flow (red) differ by more than $2\sigma$. Moreover, Fig.~\ref{Fig.beta_sys} shows that systematic effects significantly grow at very strong coupling making a nonperturbative determination more and more challenging. The overall good consistency nevertheless bolsters confidence in our result.

\subsection{\texorpdfstring{Effect of finite $L_s$}{Effect of finite Ls}}
\label{Sec.Ls}
\begin{figure}[tb]
  \includegraphics[width=0.99\columnwidth]{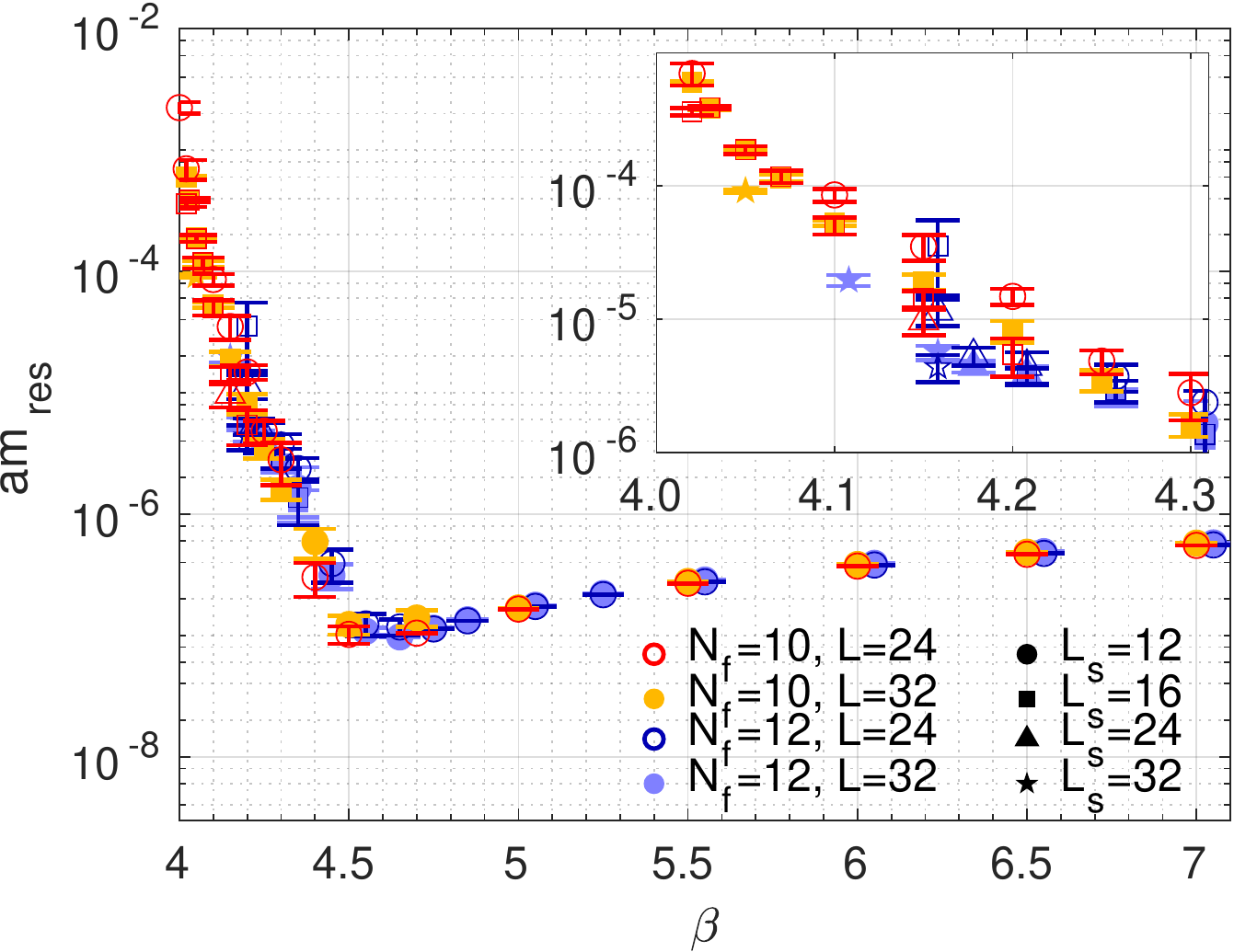}
  \caption{Residual chiral symmetry breaking, measured in terms of the residual mass $am_\text{res}$,  as function of the bare gauge coupling $\beta$ using $(L/a)^4$ volumes with $L/a=24$ and 32 for systems with ten (red, orange) or twelve (dark blue, light blue) flavors. No dependence on the number of flavors or volumes is resolved. Only statistical errors are shown and $N_f=12$ data have a small horizontal offset.} 
  \label{Fig.mres_vs_beta}
\end{figure}

\begin{figure}[tb]
  \includegraphics[width=0.99\columnwidth]{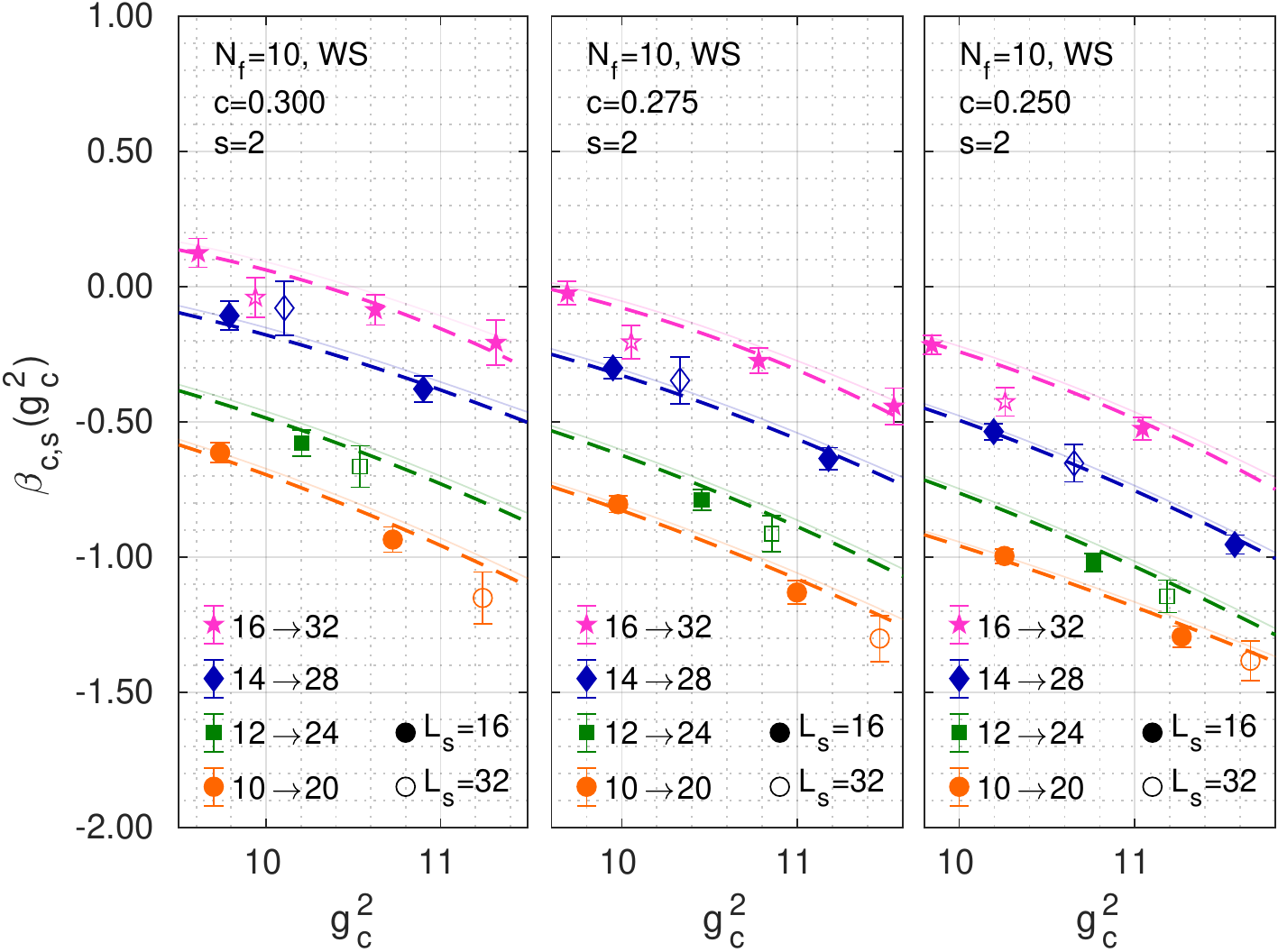}
  \caption{Effect of changing $L_s=16$ to 32 at $\beta=4.05$. The filled symbols show our preferred WS analysis using $L_s=16$ data and the dashed lines with shaded error band the corresponding interpolation for a section in $g_c^2$. Overlayed with open symbols are the $L_s=32$ data points at $\beta=4.05$. Increasing $L_s$ increases $g_c^2$ but slightly decreases $\beta_{c,s}(g_c^2;L)$ and the data points effectively slides along the interpolated curve to the lower right.} 
  \label{Fig.CheckLs}
\end{figure}

\begin{table}[tb]
  \caption{Renormalized coupling $g_c^2(L;\beta)$ and $\beta_{c,s}(g_c^2;L,\beta)$ determined at $\beta=4.05$ for our preferred WS analysis using ensembles with $L_s=16$ and 32.}
  \label{Tab.Ls}
  \begin{tabular}{ccccccc}
    \hline\hline
    $L_s$ &  $c$  &  $L$ &   $g_c^2(L;\beta)$ &  $s\cdot L$ &  $g_c^2(sL;\beta)$ & $\beta_{c,s}(g_c^2; L,\beta)$  \\
    \hline
   16 & 0.250 & 10 & 11.266(47)  & 20 & 9.471(27) & -1.295(39)\\
   32 & 0.250 & 10 & 11.658(56)  & 20 & 9.739(84) & -1.384(73)\\
   16 & 0.250 & 12 & 10.766(30)  & 24 & 9.350(34) & -1.021(33)\\
   32 & 0.250 & 12 & 11.184(71)  & 24 & 9.596(45) & -1.146(60)\\
   16 & 0.250 & 14 & 10.196(27)  & 28 & 9.452(33) & -0.537(30)\\
   32 & 0.250 & 14 & 10.654(41)  & 28 & 9.747(87) & -0.654(69)\\
   16 & 0.250 & 16 & 9.844(21)  & 32 & 9.543(44) & -0.217(35)\\
   32 & 0.250 & 16 & 10.261(43)  & 32 & 9.669(59) & -0.427(52)\\
   \hline
   16 & 0.275 & 10 & 10.997(50)  & 20 & 9.429(33) & -1.132(43)\\
   32 & 0.275 & 10 & 11.469(63)  & 20 & 9.664(99) & -1.302(85)\\
   16 & 0.275 & 12 & 10.456(31)  & 24 & 9.362(45) & -0.789(39)\\
   32 & 0.275 & 12 & 10.854(69)  & 24 & 9.587(61) & -0.914(67)\\   
   16 & 0.275 & 14 & 9.950(29)  & 28 & 9.532(45) & -0.301(39)\\
   32 & 0.275 & 14 & 10.332(41)  & 28 & 9.85(11) & -0.348(86)\\
   16 & 0.275 & 16 & 9.689(22)  & 32 & 9.654(56) & -0.025(43)\\
   32 & 0.275 & 16 & 10.054(46)  & 32 & 9.768(72) & -0.206(61)\\
   \hline
   16 & 0.300 & 10 & 10.726(51)  & 20 & 9.429(41) & -0.936(47)\\
   32 & 0.300 & 10 & 11.242(68)  & 20 & 9.65(11) & -1.152(96)\\
   16 & 0.300 & 12 & 10.204(30)  & 24 & 9.400(59) & -0.580(48)\\
   32 & 0.300 & 12 & 10.538(66)  & 24 & 9.615(84) & -0.666(77)\\
   16 & 0.300 & 14 & 9.790(32)  & 28 & 9.639(66) & -0.108(53)\\
   32 & 0.300 & 14 & 10.105(45)  & 28 & 9.99(14) & -0.08(10)\\
   16 & 0.300 & 16 & 9.611(26)  & 32 & 9.782(70) & 0.124(54)\\
   32 & 0.300 & 16 & 9.939(53)  & 32 & 9.882(85) & -0.041(73)\\
    \hline\hline
  \end{tabular}
\end{table}

As mentioned in Sec.~\ref{Sec.Numerical}, domain wall fermions exhibit a small residual chiral symmetry breaking because for practical simulations the extent $L_s$ of the fifth dimension has to be finite. The residual chiral symmetry breaking is conventionally expressed in terms of an additive mass term $am_\text{res}$. As part of our simulations we monitor $am_\text{res}$ by numerically determining it from the ratio of the midpoint-pseudoscalar over the pseudoscalar-pseudoscalar correlator. In Fig.~\ref{Fig.mres_vs_beta} we show $am_\text{res}$ as a function of the bare coupling $\beta$ for our simulations with ten and twelve dynamical flavors. Starting at $\beta = 4.5$ we observe a rapid growth of $am_\text{res}$ as $\beta$ is decreased. While for our $N_f=12$ simulations we force $a m_\text{res} < 5\cdot 10^{-6}$ \cite{Hasenfratz:2019dpr}, this is not viable for $N_f=10$ where we intend to explore much stronger couplings. Instead we use $L_s=16$ for all simulations with bare coupling $\beta \le 4.30$ and generated additional ensembles at $\beta=4.05$ using $L_s=32$ to verify that $L_s=16$ is indeed sufficient. 

Choosing $\beta=4.05$ has the advantage that Wilson flow still suppresses topological charges sufficiently well (see Fig.~\ref{Fig.TopoCharge}) and the effect of varying $L_s$ does not get obscured by nonzero topological charge contributions. Repeating our preferred WS analysis on the $L_s=32$ ensembles, we list in Table \ref{Tab.Ls} the values for $g_c^2(L;\beta)$ and $\beta_{c,s}(g_c^2;L,\beta)$ for our four largest volume pairs using the three renormalization schemes $c=0.300$, 0.275 and 0.250. Even though increasing $L_s$ from 16 to 32 leads to statistically resolved changes in $g_c^2(L;\beta)$, the effect on $\beta_{c,s}(g_c^2;L,\beta)$ is much less significant. The  difference is even further attenuated when one compares $\beta_{c,s}(g_c^2;L,\beta)$ vs.~$g_c^2(L;\beta)$ for $L_s=32$ to  our default $L_s=16$ analysis as shown in Fig.~\ref{Fig.CheckLs}. Overlaying the $L_s=32$ data point with an open symbol on the segments of our preferred WS analysis, the $L_s=32$ data points ``slide'' along the interpolated band to the right and are largely consistent with $L_s=16$ analysis. Since simulations with $L_s=32$ are about five times more expensive than $L_s=16$ simulations, all $L_s=32$ ensembles have considerably less statistics and  consequently larger statistical uncertainties.  Additional aspects of simulations with $L_s=32$ are presented in Ref.~\cite{Hasenfratz:2020vta}.

\section{Conclusion\label{Sec.Summary}}

\begin{figure}[t]
  \includegraphics[width=0.98\columnwidth]{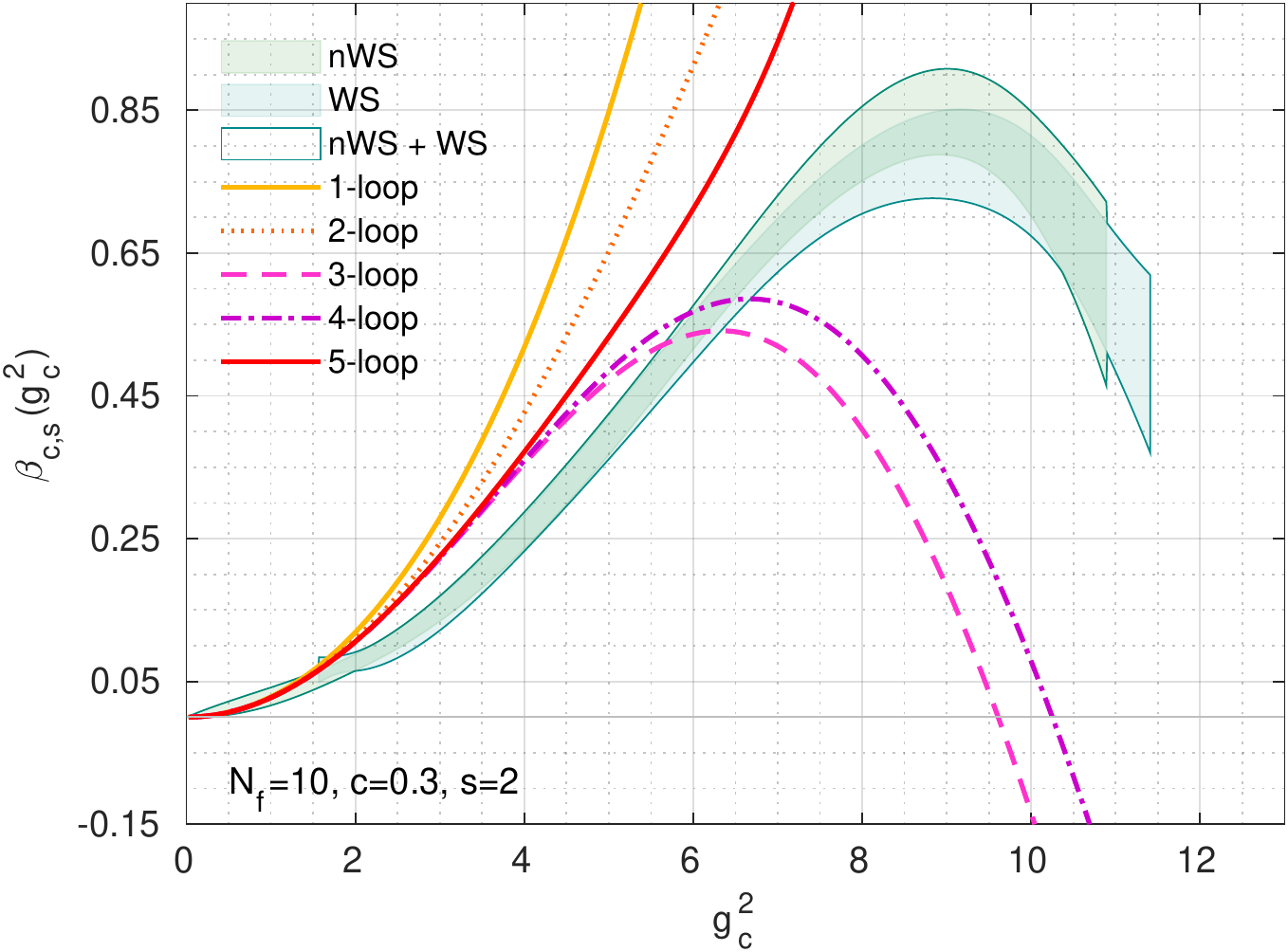}
  \includegraphics[width=0.98\columnwidth]{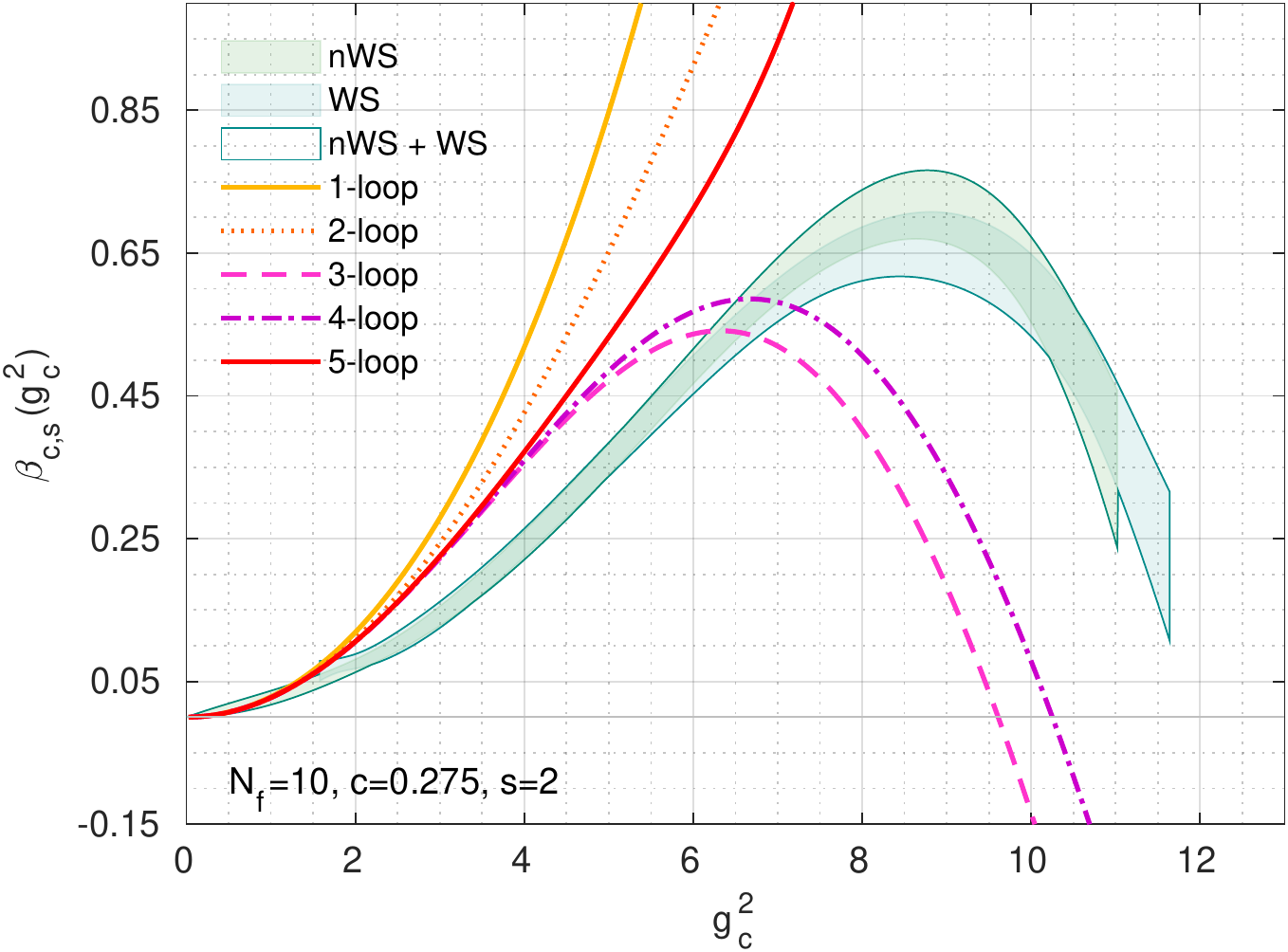} 
  \includegraphics[width=0.98\columnwidth]{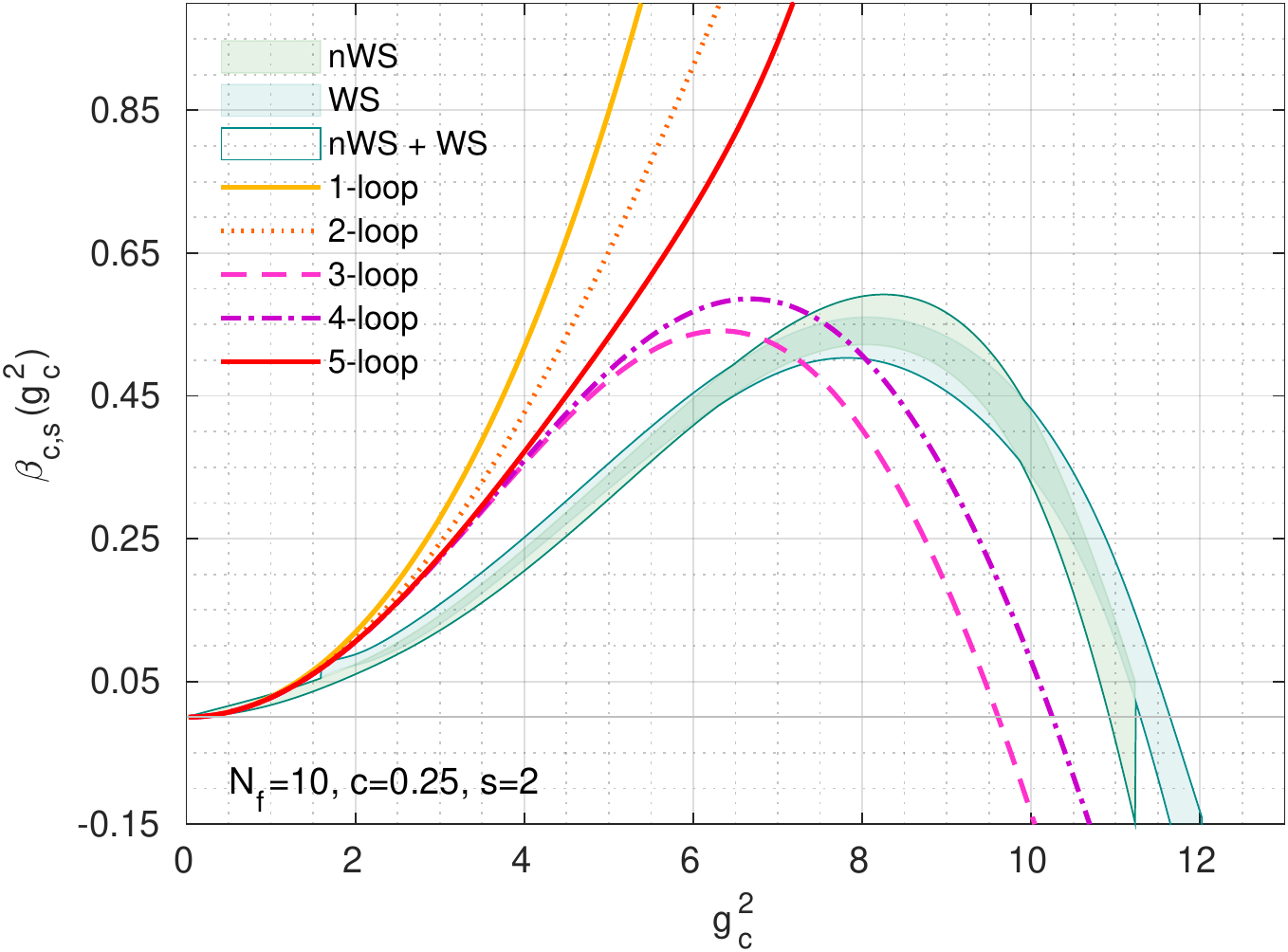}
  \caption{Continuum extrapolations of our preferred (n)WS data set for $c = 0.300$, 0.275, and 0.250 in comparison to perturbative 1--5-loop $\MSbar$ predictions.}
  \label{Fig.gettingfinal}
\end{figure}

Using gauge field configurations generated with stout-smeared M\"obius domain wall fermions and Symanzik gauge action, we have calculated the gradient flow step-scaling function for SU(3) with ten dynamical flavors. Our simulations explore the range of strong coupling so far not investigated in lattice calculations. 
Pursuing simulations in the range for $g_c^2 \gtrsim 8.0$, we observe that the gradient flow occasionally promotes vacuum fluctuations (dislocations) to instanton-like objects. This is a lattice artifact that has not been  described previously.   The effect is more pronounced for some gradient flows than for others  but always causes the gradient flow coupling to increase and run faster. 
Since Wilson flow does the best job in suppressing such dislocations compared to Zeuthen or Symanzik flow, we choose Wilson flow with Symanzik operator for our preferred analysis. Further we consider performing our analysis with and without tree-level normalization to reduce cutoff effects. 
Although justified only in the weak coupling limit, we find that our result with tree-level normalization is consistent to our unimproved result throughout the full range covered in $g_c^2$. Hence we quote, as shown in Fig.~\ref{Fig.gettingfinal}, the envelope covering our nWS and WS prediction as our final result.\footnote{ASCII files containing the data corresponding to our final results (envelope of nWS and WS) are uploaded as Supplemental Material.} 
The two determinations mostly overlap with each other. Thus this choice may only account for some of the systematic effects. Using alternative flow/operator combinations to obtain a better estimate of systematic effects is however troublesome because of lattice artifacts induced by nonzero topological charge in the strong coupling regime. Discretization effects of some flow-operator combinations also grow substantially at strong coupling.  Moreover, we studied the effect due to the finite extent of $L_s$ which results  in a small  chiral symmetry breaking. Increasing $L_s$ from 16 to 32 at $\beta=4.05$ we observe changes in $g_c^2(L;\beta)$ which however mostly cancel in the difference $\beta_{c,s}(g_c^2;L,\beta)$. In relation to our preferred analysis based on $L_s=16$ ensembles, the $L_s=32$ data are largely consistent with the  interpolated $L_s=16$ result. This suggests that the overall effect due to the finite value of $L_s$ is negligible compared to other effects. Another possible systematic effect may enter when predicting the continuum limit. We extrapolate the three largest volume pairs using a linear Ansatz in $(a/L)^2$.  This form is motivated perturbatively because for our actions the irrelevant operators enter at $\mathcal{O}(a^2)$ at the Gaussian FP. At a strongly coupled IRFP, the leading irrelevant exponent could be different. We are however not able to resolve a nontrivial exponent. Eventually different, nontrivial exponents may explain the difference between different operators and flows.

For all three values of the renormalization scheme $c$ considered, we observe that the step-scaling $\beta$ function exhibits a maximum in the range $8 \lesssim g_c^2 \lesssim 9$ and rapidly decreases for stronger coupling pointing to an IRFP  at $g_c^2\gtrsim  11$. Although our analysis shows the $\beta$ function changes sign in the $c=0.250$ scheme, this result has to be taken with caution because poor $p$-values for the continuum extrapolation in that range may signal significant finite volume effects. Nevertheless, our results strongly suggest that the RG $\beta$ function for $N_f=10$ exhibits an IRFP i.e.~the SU(3) gauge-fermion system with ten flavors is likely conformal. This observation is fully in agreement with the spectrum analysis of the composite Higgs model featuring SU(3) with four light and six heavy flavors. There hyperscaling of ratios has been demonstrated \cite{Witzel:2019oej,Witzel:2018gxm,Appelquist:2020xua} which also implies that $N_f=10$ is most likely conformal. Further, our data for the step-scaling $\beta$-function resolve a dependence on the renormalization scheme $c$.

As already shown in Fig.~\ref{Fig.beta_Nf10_final}, our result is in excellent agreement with the determination by Chiu \cite{Chiu:2016uui,Chiu:2017kza,Chiu:2018edw} for $g_c^2\lesssim 5.8$ but significantly differ from his prediction for stronger coupling. At stronger coupling our result is however compatible with LatHC \cite{Fodor:2017gtj,Fodor:2018tdg,Fodor:2019ypi} who reached $g_c^2\sim 8.0$. 
Although the nonperturbative gradient flow results correspond to a different renormalization scheme than  the perturbative $\MSbar$   calculation, it is nevertheless instructive to compare to perturbative predictions \cite{Baikov:2016tgj,Ryttov:2010iz,Ryttov:2016ner,Ryttov:2016hal}. 
Up to $g_c^2\sim 4$, predictions at three, four, and five loop order are close. While 3- and 4-loop predict an IRFP around $g^2\sim 10$, the 5-loop \MSbar result does not have a fixed point. Reference \cite{Ryttov:2016ner} suggested to improve the convergence of the perturbative series by using Pad\'e approximation, and Ref.~\cite{DiPietro:2020jne} has considered Borel re-summation to extend the perturbative range. Both references conclude that  $N_f=10$ is so strongly coupled that these analytic calculations do not give a reliable result. Thus $N_f=10$ demonstrates the difficulties to obtain a perturbative prediction on the RG $\beta$-function and highlights the need for nonperturbative lattice calculations. 

\begin{acknowledgments}
  We are very grateful to Peter Boyle, Guido Cossu, Antonin Portelli, and Azusa Yamaguchi who develop the \texttt{Grid} software library providing the basis of this work and who assisted us in installing and running \texttt{Grid} on different architectures and computing centers. A.H.~and O.W.~acknowledge support by DOE grant No.~DE-SC0010005 and C.R. by DOE Grant No.~DE-SC0015845. A.H.~would like to acknowledge the Mainz Institute for Theoretical Physics (MITP) of the Cluster of Excellence PRISMA+ (Project ID 39083149) for enabling us to complete a portion of this work. O.W.~acknowledges partial support by the Munich Institute for Astro- and Particle Physics (MIAPP) which is funded by the Deutsche Forschungsgemeinschaft (DFG, German Research Foundation) under Germany's Excellence Strategy – EXC-2094 – 390783311. 

Computations for this work were carried out in part on facilities of the USQCD Collaboration, which are funded by the Office of Science of the U.S.~Department of Energy, the RMACC Summit supercomputer \cite{UCsummit}, which is supported by the National Science Foundation (awards No.~ACI-1532235 and No.~ACI-1532236), the University of Colorado Boulder, and Colorado State University, and on computers at the MGHPCC, in part funded by the?National Science Foundation (award No.~OCI-1229059). This work used the Extreme Science and Engineering Discovery Environment (XSEDE), which is supported by National Science Foundation grant number ACI-1548562 \cite{xsede} through allocation TG-PHY180005 on the XSEDE resource \texttt{stampede2}.  This research also used resources of the National Energy Research Scientific Computing Center (NERSC), a U.S. Department of Energy Office of Science User Facility operated under Contract No. DE-AC02-05CH11231.  We thank  Fermilab,  Jefferson Lab, NERSC, the University of Colorado Boulder, TACC, the NSF, and the U.S.~DOE for providing the facilities essential for the completion of this work. 
\end{acknowledgments}
\clearpage
\appendix
\section{Bulk phase structure} \label{Sec.Phase}
\begin{figure}[htb]
  \includegraphics[width=0.99\columnwidth]{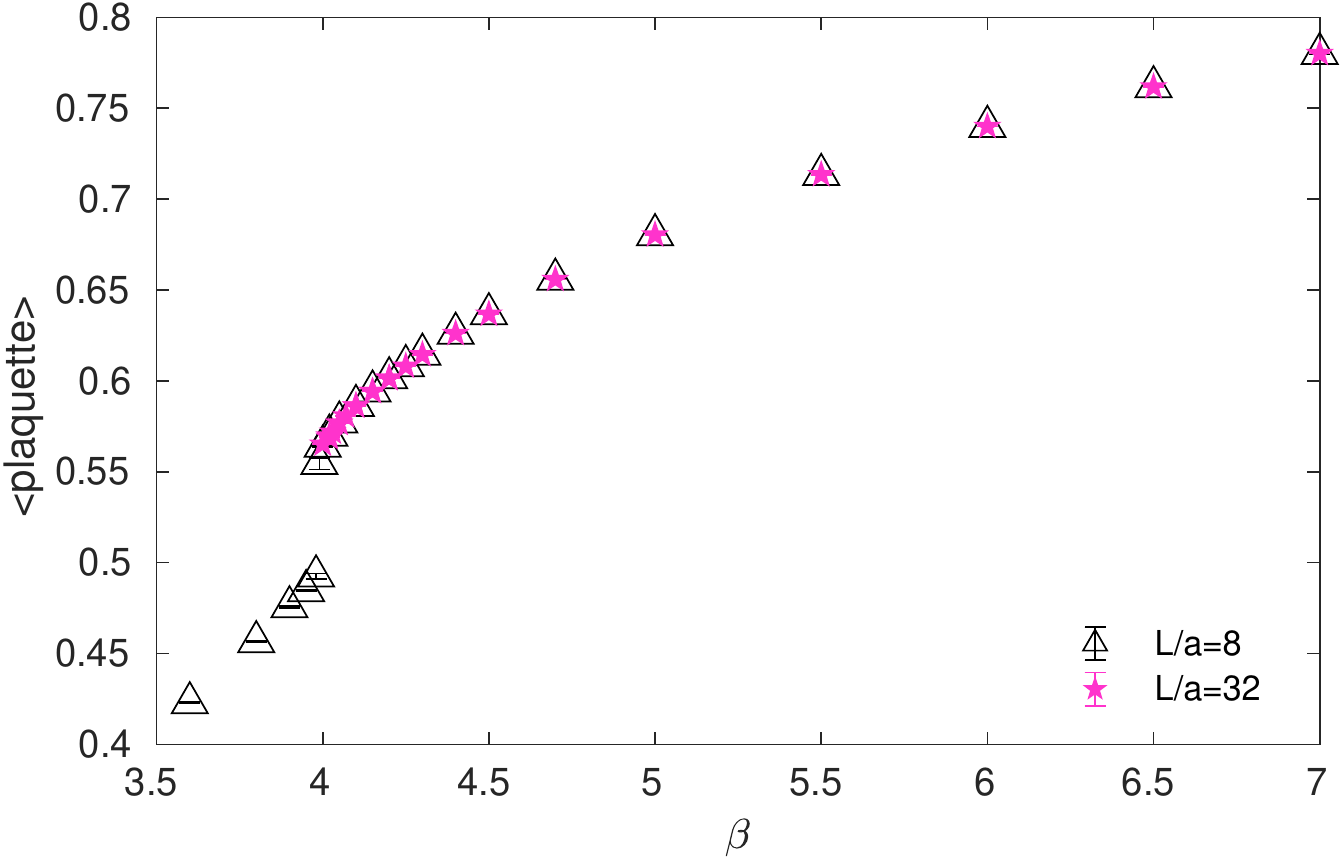}
  \caption{Average value of the plaquette as function of the bare gauge coupling $\beta$. Around $\beta\sim 4.00$ we find a bulk phase transition on the smaller $L/a=8$ lattices. Larger volumes predict the same values for the plaquette suggesting that the observed first order transition is a bulk transition and not related to finite volume.}
  \label{Fig.plaq}
\end{figure}

We have scanned  the range of bare coupling $\beta$ using small $L/a=8$ volumes to identify any possible bulk phase transition of our system. We identified a first order phase transition around $\beta\sim 4.00$ where the plaquette, as shown in Fig.~\ref{Fig.plaq}, is discontinuous. This motivates the choice of $\beta=4.02$ as the strongest coupling in our step-scaling analysis. 

\begin{figure*}[htb]  
  \includegraphics[width=0.99\columnwidth]{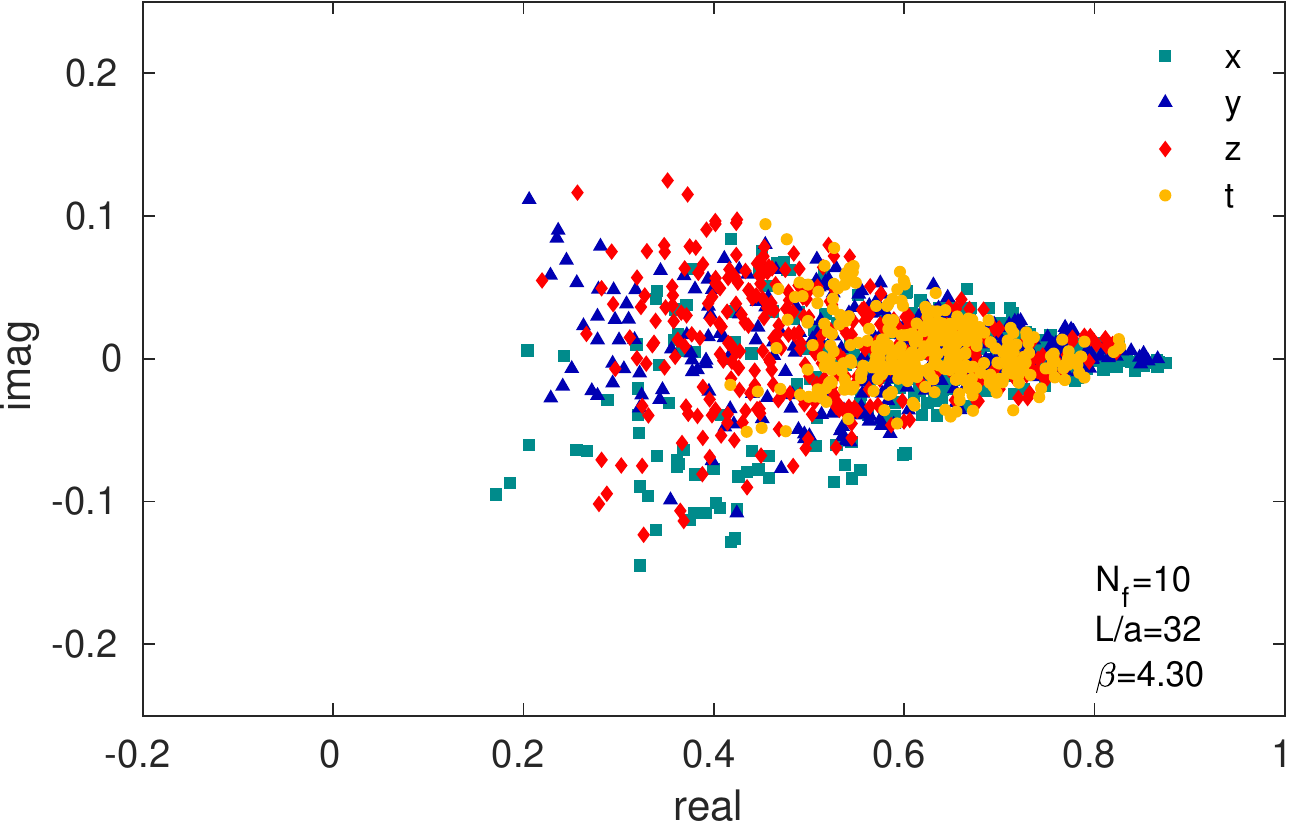}
  \includegraphics[width=0.99\columnwidth]{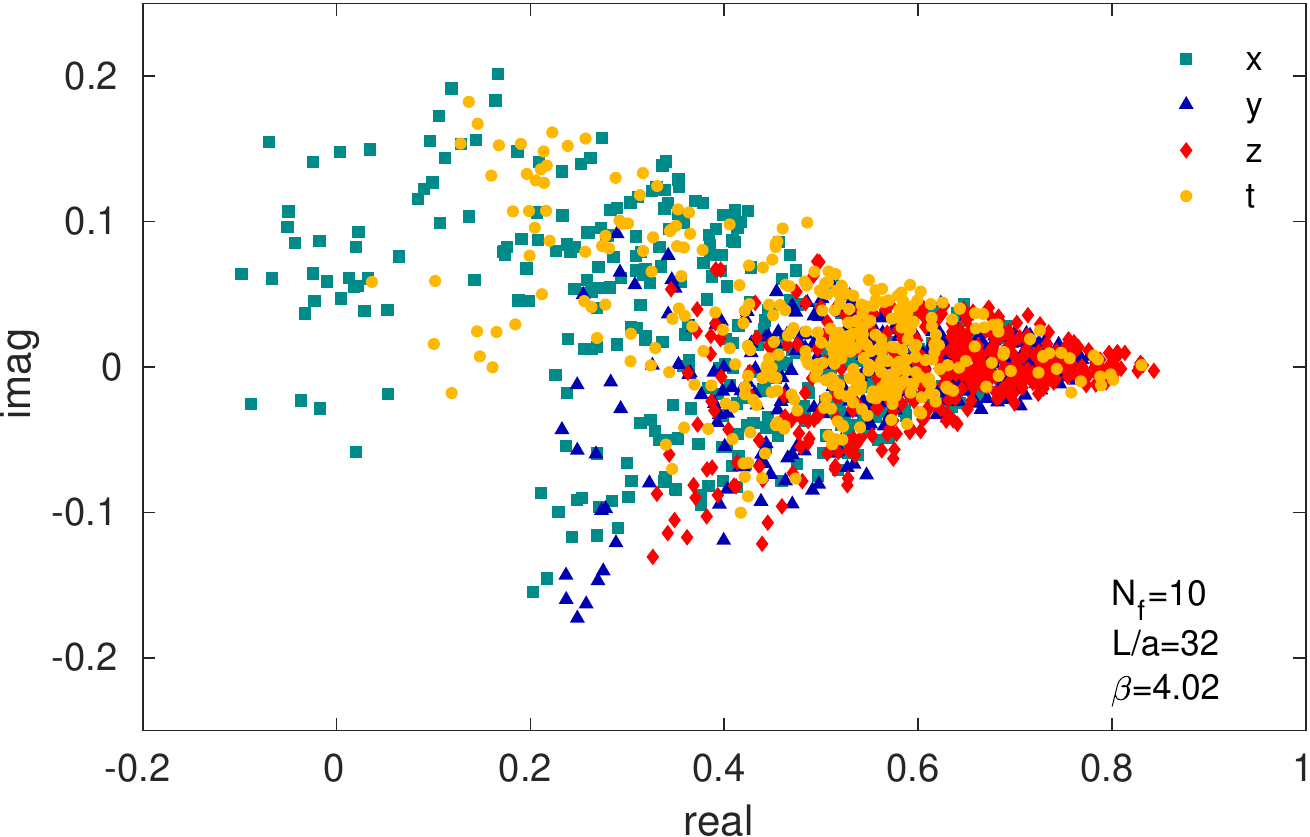}
  \caption{Real and imaginary part of the Polyakov loop in all four space-time directions at flow time $t/a^2=32$ on $L/a=32$ lattices with $\beta=4.30$ and 4.02, respectively.}
  \label{Fig.polylp}
\end{figure*}

In addition to the average plaquette, we also monitor the Polyakov loop. In Fig.~\ref{Fig.polylp}  we show the scatter plot of the real and imaginary parts of the Polyakov loop in all four space-time directions for simulations at $\beta=4.30$ and $4.02$ on lattices with $L/a=32$. Due to our choice of antiperiodic boundary conditions for the fermions in all four directions, the Polyakov loop in a deconfined regime is expected to fluctuate around a positive real value. Were the system confining, the Polyakov loop would fluctuate around zero. We measure the Polyakov loop at our maximal flow time, $t=L^2/32$ to reduce statistical fluctuations. Even at the strongest bare coupling $\beta=4.02$ the expectation value of the Polyakov loop is positive. While occasionally it wanders toward zero in one direction or an other, we never observe even two directions at once close to zero. Our simulations are clearly in the deconfined phase. 

\setlength{\LTcapwidth}{\textwidth}
\section{\texorpdfstring{Renormalized couplings $g_c^2$}{Renormalized couplings gc2}}
\label{Sec.RenCouplings}

\begin{longtable*}{cccccccccccc}
  \caption{Details of our preferred analysis based on Wilson flow and Symanzik operator. For each ensemble specified by the spatial extent $L/a$ and bare gauge coupling $\beta$ we list $N$, the number of measurements, as well as the renormalized couplings $g_c^2$ for the analysis with (nWS) and without tree-level improvement (WS) for the three renormalization schemes $c=0.300$, 0.275 and 0.250. In addition the integrated autocorrelation times determined using the $\Gamma$-method \cite{Wolff:2003sm} are listed in units of 10 MDTU.} \label{Tab.nWS_WS}\\
  
  \hline\hline
      &         &     & \multicolumn{3}{c}{$c=0.300$}&\multicolumn{3}{c}{$c=0.275$}&\multicolumn{3}{c}{$c=0.250$}\\
  $L/a$ & $\beta$ & $N$ & $g_c^2$(nWS) & $g_c^2$(WS) & $\tau_\text{int}$& $g_c^2$(nWS) &  $g_c^2$(WS)  & $\tau_\text{int}$ &$g_c^2$(
nWS)  &$g_c^2$(WS)  & $\tau_\text{int}$\\
  \hline
  \endfirsthead

  \hline
      &         &     & \multicolumn{3}{c}{$c=0.300$}&\multicolumn{3}{c}{$c=0.275$}&\multicolumn{3}{c}{$c=0.250$}\\
  $L/a$ & $\beta$ & $N$ & $g_c^2$(nWS) &  $g_c^2$(WS) & $\tau_\text{int}$& $g_c^2$(nWS) &$g_c^2$(WS) & $\tau_\text{int}$ & $g_c^2$(nWS) & $g_c^2$(WS) & $\tau_\text{int}$\\
  \hline
  \endhead

  \hline
  \endfoot

  \hline \hline
  \endlastfoot

8 & 4.02 & 1005 & 10.89(10)  & 13.74(13) & 2.5(6) & 10.290(84)  & 13.73(11) & 2.5(6) & 9.485(64)  & 13.623(92) & 2.5(6)\\ 
 8 & 4.03 & 1001 & 10.004(59)  & 12.621(75) & 1.4(3) & 9.536(49)  & 12.723(65) & 1.4(3) & 8.879(37)  & 12.752(54) & 1.4(3)\\ 
 8 & 4.05 & 1003 & 8.962(37)  & 11.307(47) & 1.0(2) & 8.633(31)  & 11.518(41) & 1.0(2) & 8.140(24)  & 11.691(34) & 1.0(2)\\ 
 8 & 4.07 & 1024 & 8.171(28)  & 10.309(36) & 1.0(2) & 7.925(24)  & 10.574(31) & 1.0(2) & 7.541(19)  & 10.831(27) & 0.9(1)\\ 
 8 & 4.10 & 721 & 7.465(24)  & 9.418(30) & 0.8(1) & 7.280(20)  & 9.712(27) & 0.7(1) & 6.977(16)  & 10.021(23) & 0.7(1)\\ 
 8 & 4.15 & 640 & 6.614(23)  & 8.345(29) & 0.9(2) & 6.488(19)  & 8.656(26) & 0.9(2) & 6.274(15)  & 9.010(22) & 0.9(2)\\ 
 8 & 4.20 & 561 & 6.046(17)  & 7.628(21) & 0.7(1) & 5.944(14)  & 7.930(19) & 0.7(1) & 5.771(12)  & 8.288(17) & 0.7(1)\\ 
 8 & 4.25 & 730 & 5.612(12)  & 7.080(15) & 0.7(1) & 5.530(10)  & 7.378(14) & 0.7(1) & 5.3872(85)  & 7.737(12) & 0.66(10)\\ 
 8 & 4.30 & 740 & 5.2494(94)  & 6.623(12) & 0.49(5) & 5.1810(79)  & 6.912(11) & 0.49(5) & 5.0609(65)  & 7.2687(93) & 0.50(5)\\ 
 8 & 4.40 & 750 & 4.7230(94)  & 5.959(12) & 0.62(9) & 4.6681(78)  & 6.228(10) & 0.61(9) & 4.5722(61)  & 6.5668(88) & 0.57(7)\\ 
 8 & 4.50 & 749 & 4.2858(78)  & 5.4072(98) & 0.59(9) & 4.2419(64)  & 5.6595(85) & 0.58(8) & 4.1654(51)  & 5.9825(73) & 0.58(8)\\ 
 8 & 4.70 & 720 & 3.6604(62)  & 4.6182(79) & 0.55(7) & 3.6316(49)  & 4.8452(66) & 0.51(5) & 3.5786(39)  & 5.1397(55) & 0.49(5)\\ 
 8 & 5.00 & 746 & 3.0310(45)  & 3.8241(57) & 0.46(5) & 3.0147(37)  & 4.0222(50) & 0.47(5) & 2.9812(30)  & 4.2818(44) & 0.48(5)\\ 
 8 & 5.50 & 716 & 2.3744(36)  & 2.9956(46) & 0.50(5) & 2.3663(30)  & 3.1571(40) & 0.50(4) & 2.3476(24)  & 3.3717(34) & 0.49(4)\\ 
 8 & 6.00 & 653 & 1.9601(31)  & 2.4729(39) & 0.50(8) & 1.9566(26)  & 2.6104(35) & 0.55(7) & 1.9454(20)  & 2.7941(28) & 0.51(6)\\ 
 8 & 6.50 & 716 & 1.6756(23)  & 2.1140(29) & 0.47(4) & 1.6732(18)  & 2.2324(25) & 0.45(3) & 1.6652(14)  & 2.3917(21) & 0.44(3)\\ 
 8 & 7.00 & 694 & 1.4571(21)  & 1.8384(26) & 0.50(5) & 1.4573(17)  & 1.9443(22) & 0.49(5) & 1.4530(13)  & 2.0869(19) & 0.49(5)\\ 
 \hline 
10 & 4.02 & 1002 & 12.00(10)  & 13.69(12) & 2.9(7) & 11.874(93)  & 13.96(11) & 3.1(8) & 11.473(81)  & 14.091(99) & 3.2(8)\\ 
 10 & 4.03 & 1002 & 10.651(53)  & 12.157(61) & 1.6(3) & 10.595(49)  & 12.459(57) & 1.5(3) & 10.334(42)  & 12.692(51) & 1.5(3)\\ 
 10 & 4.05 & 1002 & 9.398(45)  & 10.726(51) & 1.6(3) & 9.353(42)  & 10.997(50) & 1.8(4) & 9.173(39)  & 11.266(47) & 1.9(4)\\ 
 10 & 4.07 & 1005 & 8.530(28)  & 9.736(32) & 1.2(2) & 8.487(26)  & 9.979(30) & 1.2(2) & 8.352(22)  & 10.257(27) & 1.2(2)\\ 
 10 & 4.10 & 1002 & 7.749(23)  & 8.844(26) & 1.0(2) & 7.702(20)  & 9.056(23) & 0.9(2) & 7.593(17)  & 9.326(20) & 0.9(1)\\ 
 10 & 4.15 & 921 & 6.799(15)  & 7.760(17) & 0.8(1) & 6.758(13)  & 7.946(16) & 0.8(1) & 6.682(11)  & 8.206(14) & 0.8(1)\\ 
 10 & 4.20 & 994 & 6.235(11)  & 7.116(13) & 0.61(8) & 6.1924(96)  & 7.281(11) & 0.58(7) & 6.1262(80)  & 7.5240(98) & 0.55(7)\\ 
 10 & 4.25 & 936 & 5.767(11)  & 6.582(13) & 0.69(10) & 5.7268(90)  & 6.734(11) & 0.64(9) & 5.6690(75)  & 6.9624(93) & 0.65(8)\\ 
 10 & 4.30 & 941 & 5.402(11)  & 6.165(12) & 0.8(1) & 5.3654(86)  & 6.309(10) & 0.7(1) & 5.3141(68)  & 6.5265(83) & 0.64(9)\\ 
 10 & 4.40 & 767 & 4.836(10)  & 5.520(12) & 0.8(1) & 4.8048(83)  & 5.6498(98) & 0.7(1) & 4.7617(66)  & 5.8482(81) & 0.7(1)\\ 
 10 & 4.50 & 783 & 4.4037(77)  & 5.0264(88) & 0.58(8) & 4.3750(62)  & 5.1444(73) & 0.55(8) & 4.3370(50)  & 5.3266(61) & 0.54(8)\\ 
 10 & 4.70 & 569 & 3.7667(75)  & 4.2993(85) & 0.57(8) & 3.7428(61)  & 4.4011(72) & 0.56(8) & 3.7123(47)  & 4.5593(57) & 0.50(4)\\ 
 10 & 5.00 & 821 & 3.0877(57)  & 3.5242(65) & 0.7(1) & 3.0755(46)  & 3.6164(54) & 0.7(1) & 3.0585(35)  & 3.7564(43) & 0.63(9)\\ 
 10 & 5.50 & 599 & 2.4106(43)  & 2.7515(49) & 0.50(6) & 2.4044(35)  & 2.8273(41) & 0.49(6) & 2.3951(28)  & 2.9415(34) & 0.48(6)\\ 
 10 & 6.00 & 736 & 1.9846(33)  & 2.2652(37) & 0.57(8) & 1.9809(27)  & 2.3292(31) & 0.57(7) & 1.9751(21)  & 2.4257(26) & 0.56(8)\\ 
 10 & 6.50 & 612 & 1.6967(34)  & 1.9366(39) & 0.7(1) & 1.6944(27)  & 1.9924(32) & 0.6(1) & 1.6901(21)  & 2.0757(26) & 0.63(10)\\ 
 10 & 7.00 & 735 & 1.4722(21)  & 1.6803(24) & 0.49(4) & 1.4720(17)  & 1.7309(20) & 0.48(4) & 1.4703(14)  & 1.8058(17) & 0.47(4)\\ 
 \hline 
12 & 4.02 & 1022 & 11.565(82)  & 12.617(89) & 4(1) & 11.794(84)  & 13.095(94) & 4(1) & 11.918(83)  & 13.573(94) & 5(1)\\ 
 12 & 4.03 & 1021 & 10.560(59)  & 11.520(64) & 2.8(7) & 10.711(59)  & 11.893(65) & 2.9(7) & 10.798(57)  & 12.298(65) & 2.9(7)\\ 
 12 & 4.05 & 1009 & 9.353(27)  & 10.204(30) & 1.1(2) & 9.417(28)  & 10.456(31) & 1.2(2) & 9.453(26)  & 10.766(30) & 1.2(2)\\ 
 12 & 4.07 & 1001 & 8.553(22)  & 9.330(24) & 1.1(2) & 8.571(20)  & 9.517(22) & 1.0(2) & 8.579(18)  & 9.770(21) & 1.0(2)\\ 
 12 & 4.10 & 770 & 7.737(19)  & 8.440(21) & 0.8(1) & 7.717(17)  & 8.568(18) & 0.8(1) & 7.693(14)  & 8.761(16) & 0.8(1)\\ 
 12 & 4.15 & 766 & 6.898(16)  & 7.525(17) & 0.8(1) & 6.863(13)  & 7.621(14) & 0.7(1) & 6.824(11)  & 7.771(12) & 0.7(1)\\ 
 12 & 4.20 & 743 & 6.342(15)  & 6.918(16) & 0.8(1) & 6.303(12)  & 6.998(13) & 0.7(1) & 6.2559(95)  & 7.124(11) & 0.7(1)\\ 
 12 & 4.25 & 727 & 5.877(14)  & 6.411(15) & 0.9(2) & 5.839(12)  & 6.483(13) & 0.9(2) & 5.7938(92)  & 6.598(10) & 0.8(1)\\ 
 12 & 4.30 & 687 & 5.523(13)  & 6.025(14) & 0.8(1) & 5.485(10)  & 6.091(11) & 0.7(1) & 5.4403(78)  & 6.1955(89) & 0.65(10)\\ 
 12 & 4.40 & 393 & 4.922(14)  & 5.369(15) & 0.7(1) & 4.891(11)  & 5.431(12) & 0.6(1) & 4.8543(86)  & 5.5281(98) & 0.6(1)\\ 
 12 & 4.50 & 416 & 4.482(12)  & 4.889(13) & 0.7(2) & 4.454(10)  & 4.946(11) & 0.7(1) & 4.4207(82)  & 5.0344(93) & 0.7(1)\\ 
 12 & 4.70 & 525 & 3.8157(97)  & 4.163(11) & 0.8(2) & 3.7940(81)  & 4.2127(90) & 0.8(2) & 3.7682(65)  & 4.2913(74) & 0.8(1)\\ 
 12 & 5.00 & 414 & 3.1372(83)  & 3.4225(90) & 0.8(2) & 3.1223(65)  & 3.4669(73) & 0.7(1) & 3.1044(52)  & 3.5353(59) & 0.7(1)\\ 
 12 & 5.50 & 524 & 2.4403(45)  & 2.6622(49) & 0.49(4) & 2.4316(37)  & 2.6999(41) & 0.47(4) & 2.4209(30)  & 2.7570(34) & 0.45(4)\\ 
 12 & 6.00 & 377 & 2.0120(57)  & 2.1950(62) & 0.8(2) & 2.0057(46)  & 2.2271(51) & 0.8(2) & 1.9979(37)  & 2.2753(42) & 0.7(2)\\ 
 12 & 6.50 & 515 & 1.7077(45)  & 1.8630(49) & 1.0(2) & 1.7044(34)  & 1.8926(38) & 0.8(2) & 1.6999(25)  & 1.9358(28) & 0.7(1)\\ 
 12 & 7.00 & 386 & 1.4846(38)  & 1.6195(41) & 0.7(1) & 1.4833(31)  & 1.6470(34) & 0.6(1) & 1.4807(25)  & 1.6863(28) & 0.6(1)\\ 
 \hline 
14 & 4.02 & 1001 & 11.176(44)  & 11.893(47) & 1.9(4) & 11.382(43)  & 12.259(46) & 2.0(4) & 11.654(44)  & 12.765(48) & 2.1(5)\\ 
 14 & 4.03 & 961 & 10.244(35)  & 10.901(37) & 1.5(3) & 10.377(34)  & 11.178(36) & 1.6(3) & 10.562(33)  & 11.569(36) & 1.6(3)\\ 
 14 & 4.05 & 961 & 9.199(30)  & 9.790(32) & 1.7(3) & 9.238(26)  & 9.950(29) & 1.6(3) & 9.309(24)  & 10.196(27) & 1.5(3)\\ 
 14 & 4.07 & 961 & 8.496(21)  & 9.042(23) & 1.2(2) & 8.495(18)  & 9.150(20) & 1.2(2) & 8.518(16)  & 9.330(18) & 1.2(2)\\ 
 14 & 4.10 & 961 & 7.799(16)  & 8.300(17) & 0.9(2) & 7.769(13)  & 8.368(15) & 0.9(1) & 7.750(12)  & 8.489(13) & 0.9(1)\\ 
 14 & 4.15 & 963 & 6.982(18)  & 7.431(19) & 1.3(2) & 6.950(14)  & 7.486(15) & 1.1(2) & 6.919(11)  & 7.579(13) & 1.0(2)\\ 
 14 & 4.20 & 961 & 6.422(18)  & 6.834(19) & 1.5(3) & 6.375(14)  & 6.867(15) & 1.4(3) & 6.329(11)  & 6.932(12) & 1.3(2)\\ 
 14 & 4.25 & 963 & 5.970(12)  & 6.353(12) & 0.9(1) & 5.9264(94)  & 6.383(10) & 0.8(1) & 5.8802(73)  & 6.4406(80) & 0.7(1)\\ 
 14 & 4.30 & 962 & 5.564(12)  & 5.921(13) & 1.0(2) & 5.5274(95)  & 5.954(10) & 0.9(2) & 5.4876(75)  & 6.0107(82) & 0.8(1)\\ 
 14 & 4.40 & 963 & 4.993(10)  & 5.314(11) & 0.8(1) & 4.9619(81)  & 5.3444(87) & 0.8(1) & 4.9259(64)  & 5.3954(70) & 0.7(1)\\ 
 14 & 4.50 & 963 & 4.5411(96)  & 4.833(10) & 0.9(1) & 4.5108(77)  & 4.8585(83) & 0.8(1) & 4.4769(59)  & 4.9036(65) & 0.8(1)\\ 
 14 & 4.70 & 962 & 3.8704(78)  & 4.1189(83) & 0.8(1) & 3.8454(62)  & 4.1419(67) & 0.8(1) & 3.8176(47)  & 4.1815(51) & 0.65(8)\\ 
 14 & 5.00 & 963 & 3.1796(65)  & 3.3837(69) & 1.0(2) & 3.1630(50)  & 3.4068(54) & 0.8(1) & 3.1441(38)  & 3.4438(41) & 0.70(10)\\ 
 14 & 5.50 & 963 & 2.4607(48)  & 2.6187(51) & 1.0(2) & 2.4532(38)  & 2.6423(41) & 0.9(1) & 2.4434(28)  & 2.6763(31) & 0.7(1)\\ 
 14 & 6.00 & 963 & 2.0298(39)  & 2.1601(41) & 0.9(1) & 2.0232(31)  & 2.1792(33) & 0.8(1) & 2.0153(24)  & 2.2073(26) & 0.7(1)\\ 
 14 & 6.50 & 963 & 1.7257(30)  & 1.8365(32) & 0.7(1) & 1.7212(23)  & 1.8539(25) & 0.63(8) & 1.7152(18)  & 1.8787(20) & 0.57(7)\\ 
 14 & 7.00 & 963 & 1.5008(26)  & 1.5971(27) & 0.7(1) & 1.4976(21)  & 1.6131(23) & 0.7(1) & 1.4933(17)  & 1.6357(18) & 0.67(9)\\ 
 \hline 
16 & 4.02 & 595 & 10.799(56)  & 11.317(59) & 2.6(8) & 10.923(52)  & 11.549(55) & 2.6(8) & 11.138(53)  & 11.921(57) & 2.8(9)\\ 
 16 & 4.03 & 631 & 10.139(40)  & 10.626(42) & 1.6(4) & 10.196(35)  & 10.780(37) & 1.6(4) & 10.318(33)  & 11.043(35) & 1.6(4)\\ 
 16 & 4.05 & 601 & 9.170(25)  & 9.611(26) & 1.0(2) & 9.164(21)  & 9.689(22) & 1.0(2) & 9.197(19)  & 9.844(21) & 1.1(2)\\ 
 16 & 4.07 & 632 & 8.501(25)  & 8.909(26) & 1.1(2) & 8.473(21)  & 8.959(22) & 1.0(2) & 8.466(18)  & 9.061(20) & 1.1(2)\\ 
 16 & 4.10 & 379 & 7.814(41)  & 8.190(43) & 2.2(7) & 7.779(33)  & 8.225(35) & 2.0(6) & 7.752(27)  & 8.296(29) & 2.0(6)\\ 
 16 & 4.15 & 519 & 7.000(18)  & 7.336(19) & 0.9(2) & 6.963(15)  & 7.363(15) & 0.9(2) & 6.928(12)  & 7.415(13) & 0.8(2)\\ 
 16 & 4.20 & 200 & 6.448(33)  & 6.757(35) & 1.1(4) & 6.399(27)  & 6.766(28) & 1.1(4) & 6.354(21)  & 6.800(22) & 0.9(3)\\ 
 16 & 4.25 & 611 & 6.031(23)  & 6.320(24) & 1.7(4) & 5.982(18)  & 6.324(19) & 1.6(4) & 5.933(14)  & 6.349(15) & 1.5(3)\\ 
 16 & 4.30 & 605 & 5.656(17)  & 5.928(18) & 1.2(3) & 5.611(13)  & 5.933(14) & 1.1(2) & 5.5658(100)  & 5.957(11) & 0.9(2)\\ 
 16 & 4.40 & 485 & 5.067(17)  & 5.310(18) & 1.2(3) & 5.026(13)  & 5.314(14) & 1.1(3) & 4.985(10)  & 5.335(11) & 1.0(2)\\ 
 16 & 4.50 & 428 & 4.589(15)  & 4.810(15) & 1.0(3) & 4.556(12)  & 4.818(12) & 1.0(2) & 4.5212(89)  & 4.8389(95) & 0.9(2)\\ 
 16 & 4.70 & 551 & 3.897(11)  & 4.084(12) & 1.1(3) & 3.8726(93)  & 4.0945(98) & 1.1(2) & 3.8464(72)  & 4.1167(77) & 1.0(2)\\ 
 16 & 5.00 & 446 & 3.216(12)  & 3.371(13) & 1.4(4) & 3.1965(94)  & 3.3798(100) & 1.2(3) & 3.1751(70)  & 3.3982(75) & 1.0(3)\\ 
 16 & 5.50 & 551 & 2.4979(64)  & 2.6178(67) & 1.0(2) & 2.4866(53)  & 2.6291(56) & 0.9(2) & 2.4732(45)  & 2.6470(48) & 1.0(2)\\ 
 16 & 6.00 & 307 & 2.0508(63)  & 2.1492(66) & 0.7(2) & 2.0415(52)  & 2.1585(55) & 0.7(2) & 2.0315(41)  & 2.1742(44) & 0.7(1)\\ 
 16 & 6.50 & 323 & 1.7313(60)  & 1.8145(63) & 1.2(4) & 1.7277(47)  & 1.8267(49) & 1.1(3) & 1.7226(35)  & 1.8437(37) & 0.9(2)\\ 
 16 & 7.00 & 352 & 1.5057(49)  & 1.5780(52) & 1.0(3) & 1.5029(38)  & 1.5890(40) & 0.9(2) & 1.4991(30)  & 1.6044(32) & 0.8(2)\\ 
 \hline 
20 & 4.02 & 477 & 10.754(56)  & 11.076(58) & 2.7(9) & 10.741(45)  & 11.123(46) & 2.4(7) & 10.792(38)  & 11.258(39) & 2.2(7)\\ 
 20 & 4.03 & 465 & 10.020(52)  & 10.319(54) & 2.3(7) & 9.991(43)  & 10.346(44) & 2.2(7) & 10.002(35)  & 10.434(37) & 2.1(6)\\ 
 20 & 4.05 & 490 & 9.155(40)  & 9.429(41) & 2.2(7) & 9.105(32)  & 9.429(33) & 2.0(6) & 9.079(25)  & 9.471(27) & 1.7(5)\\ 
 20 & 4.07 & 502 & 8.626(36)  & 8.884(37) & 2.0(6) & 8.559(29)  & 8.863(30) & 1.8(5) & 8.508(24)  & 8.875(25) & 1.7(4)\\ 
 20 & 4.10 & 301 & 7.981(49)  & 8.219(51) & 2.5(9) & 7.895(39)  & 8.176(40) & 2.3(9) & 7.822(30)  & 8.160(32) & 2.1(7)\\ 
 20 & 4.15 & 353 & 7.155(32)  & 7.369(33) & 1.7(5) & 7.086(25)  & 7.337(26) & 1.6(5) & 7.023(20)  & 7.327(21) & 1.4(4)\\ 
 20 & 4.20 & 333 & 6.593(37)  & 6.790(39) & 1.9(6) & 6.534(30)  & 6.766(31) & 1.8(6) & 6.477(23)  & 6.757(24) & 1.6(5)\\ 
 20 & 4.25 & 302 & 6.088(34)  & 6.270(36) & 2.4(9) & 6.047(28)  & 6.261(29) & 2.3(8) & 6.005(21)  & 6.264(22) & 1.9(7)\\ 
 20 & 4.30 & 346 & 5.771(25)  & 5.943(25) & 1.4(4) & 5.722(19)  & 5.926(20) & 1.2(3) & 5.672(14)  & 5.917(15) & 1.0(2)\\ 
 20 & 4.40 & 284 & 5.152(33)  & 5.306(34) & 2.3(8) & 5.110(26)  & 5.292(26) & 2.0(7) & 5.067(20)  & 5.286(21) & 1.8(6)\\ 
 20 & 4.50 & 304 & 4.728(26)  & 4.869(27) & 1.8(6) & 4.679(19)  & 4.845(20) & 1.5(5) & 4.631(14)  & 4.831(14) & 1.2(3)\\ 
 20 & 4.70 & 255 & 3.978(24)  & 4.097(25) & 2.0(7) & 3.950(19)  & 4.090(19) & 1.8(6) & 3.920(14)  & 4.089(15) & 1.5(5)\\ 
 20 & 5.00 & 316 & 3.287(22)  & 3.385(23) & 3(1) & 3.261(17)  & 3.376(18) & 2.5(9) & 3.234(13)  & 3.374(13) & 2.1(7)\\ 
 20 & 5.50 & 368 & 2.536(10)  & 2.612(11) & 1.4(4) & 2.5222(83)  & 2.6118(86) & 1.3(4) & 2.5068(62)  & 2.6151(65) & 1.1(3)\\ 
 20 & 6.00 & 411 & 2.0719(78)  & 2.1339(80) & 1.3(4) & 2.0625(61)  & 2.1357(63) & 1.2(3) & 2.0525(46)  & 2.1411(48) & 1.0(2)\\ 
 20 & 6.50 & 381 & 1.7539(59)  & 1.8064(61) & 1.4(4) & 1.7476(47)  & 1.8097(48) & 1.2(3) & 1.7405(35)  & 1.8157(37) & 1.0(2)\\ 
 20 & 7.00 & 279 & 1.5276(53)  & 1.5733(55) & 0.9(2) & 1.5219(43)  & 1.5760(44) & 0.8(2) & 1.5157(34)  & 1.5812(36) & 0.7(2)\\ 
 \hline 
24 & 4.02 & 401 & 10.727(54)  & 10.947(55) & 2.5(8) & 10.670(40)  & 10.929(41) & 1.9(6) & 10.647(31)  & 10.960(32) & 1.6(5)\\ 
 24 & 4.03 & 388 & 10.088(56)  & 10.295(58) & 2.4(8) & 10.024(44)  & 10.267(46) & 2.3(7) & 9.986(34)  & 10.280(35) & 2.0(7)\\ 
 24 & 4.05 & 510 & 9.211(58)  & 9.400(59) & 4(1) & 9.140(44)  & 9.362(45) & 3(1) & 9.083(33)  & 9.350(34) & 2.9(9)\\ 
 24 & 4.07 & 501 & 8.638(38)  & 8.815(39) & 2.3(7) & 8.570(32)  & 8.778(32) & 2.2(7) & 8.511(26)  & 8.761(27) & 2.2(7)\\ 
 24 & 4.10 & 500 & 8.070(53)  & 8.236(54) & 4(1) & 7.988(43)  & 8.182(44) & 4(1) & 7.911(34)  & 8.144(35) & 4(1)\\ 
 24 & 4.15 & 203 & 7.263(44)  & 7.412(45) & 2.1(9) & 7.201(36)  & 7.375(37) & 1.9(7) & 7.140(29)  & 7.350(30) & 1.8(7)\\ 
 24 & 4.20 & 191 & 6.712(48)  & 6.849(49) & 2.3(9) & 6.643(37)  & 6.805(38) & 2.0(8) & 6.576(28)  & 6.769(29) & 1.7(6)\\ 
 24 & 4.25 & 315 & 6.242(50)  & 6.370(51) & 3(1) & 6.184(38)  & 6.334(39) & 3(1) & 6.125(28)  & 6.305(29) & 2.3(8)\\ 
 24 & 4.30 & 307 & 5.838(33)  & 5.958(33) & 2.1(7) & 5.791(27)  & 5.932(27) & 1.9(6) & 5.741(20)  & 5.909(21) & 1.7(5)\\ 
 24 & 4.40 & 324 & 5.299(33)  & 5.407(34) & 3(1) & 5.237(26)  & 5.364(27) & 3(1) & 5.176(21)  & 5.329(21) & 2.5(9)\\ 
 24 & 4.50 & 257 & 4.783(39)  & 4.881(40) & 3(1) & 4.742(31)  & 4.857(31) & 3(1) & 4.698(24)  & 4.836(25) & 3(1)\\ 
 24 & 4.70 & 282 & 4.053(23)  & 4.136(24) & 2.4(9) & 4.019(19)  & 4.116(19) & 2.3(8) & 3.983(14)  & 4.100(15) & 2.0(7)\\ 
 24 & 5.00 & 334 & 3.355(24)  & 3.424(25) & 3(1) & 3.323(19)  & 3.403(19) & 2.6(10) & 3.290(14)  & 3.387(15) & 2.4(8)\\ 
 24 & 5.50 & 241 & 2.568(16)  & 2.621(17) & 2.1(8) & 2.553(12)  & 2.615(13) & 1.8(6) & 2.5373(93)  & 2.6118(96) & 1.6(5)\\ 
 24 & 6.00 & 430 & 2.1055(78)  & 2.1487(79) & 1.4(4) & 2.0933(62)  & 2.1441(64) & 1.2(3) & 2.0802(49)  & 2.1413(51) & 1.2(3)\\ 
 24 & 6.50 & 227 & 1.7862(90)  & 1.8228(92) & 1.6(6) & 1.7762(69)  & 1.8193(71) & 1.3(4) & 1.7656(53)  & 1.8175(54) & 1.1(3)\\ 
 24 & 7.00 & 301 & 1.5286(44)  & 1.5599(44) & 0.8(2) & 1.5262(35)  & 1.5633(36) & 0.7(2) & 1.5225(28)  & 1.5672(29) & 0.6(1)\\ 
 \hline 
28 & 4.02 & 397 & 10.893(67)  & 11.056(68) & 3(1) & 10.774(53)  & 10.964(54) & 2.6(9) & 10.682(41)  & 10.910(42) & 2.3(8)\\ 
 28 & 4.03 & 383 & 10.223(55)  & 10.375(56) & 2.7(10) & 10.117(42)  & 10.295(43) & 2.3(8) & 10.031(32)  & 10.245(33) & 2.0(7)\\ 
 28 & 4.05 & 376 & 9.497(65)  & 9.639(66) & 4(2) & 9.367(45)  & 9.532(45) & 2.8(10) & 9.254(32)  & 9.452(33) & 2.2(7)\\ 
 28 & 4.07 & 374 & 8.870(64)  & 9.002(65) & 4(2) & 8.765(50)  & 8.919(51) & 4(2) & 8.666(36)  & 8.851(37) & 3(1)\\ 
 28 & 4.10 & 377 & 8.265(48)  & 8.388(49) & 3(1) & 8.145(40)  & 8.288(40) & 3(1) & 8.034(31)  & 8.205(32) & 2.7(10)\\ 
 28 & 4.15 & 353 & 7.420(51)  & 7.531(52) & 4(1) & 7.331(43)  & 7.460(43) & 3(1) & 7.246(35)  & 7.400(35) & 3(1)\\ 
 28 & 4.20 & 354 & 6.858(57)  & 6.961(57) & 4(2) & 6.785(42)  & 6.904(43) & 4(1) & 6.708(29)  & 6.852(30) & 2.6(9)\\ 
 28 & 4.25 & 357 & 6.435(47)  & 6.531(48) & 4(2) & 6.354(36)  & 6.466(37) & 4(2) & 6.271(27)  & 6.405(28) & 3(1)\\ 
 28 & 4.30 & 352 & 6.040(47)  & 6.130(48) & 4(2) & 5.964(37)  & 6.069(38) & 3(1) & 5.886(28)  & 6.011(29) & 3(1)\\ 
 28 & 4.40 & 364 & 5.375(35)  & 5.456(35) & 3(1) & 5.313(26)  & 5.407(27) & 3(1) & 5.251(19)  & 5.363(20) & 2.3(8)\\ 
 28 & 4.50 & 362 & 4.936(22)  & 5.010(23) & 2.0(7) & 4.869(17)  & 4.955(18) & 1.8(6) & 4.802(13)  & 4.905(14) & 1.5(5)\\ 
 28 & 4.70 & 365 & 4.167(28)  & 4.230(29) & 3(1) & 4.117(23)  & 4.190(23) & 3(1) & 4.067(18)  & 4.154(18) & 3(1)\\ 
 28 & 5.00 & 284 & 3.363(16)  & 3.414(16) & 1.7(6) & 3.338(13)  & 3.397(13) & 1.5(5) & 3.311(10)  & 3.382(10) & 1.3(4)\\ 
 28 & 5.50 & 272 & 2.600(14)  & 2.639(14) & 2.0(7) & 2.584(10)  & 2.629(10) & 1.6(5) & 2.5654(76)  & 2.6202(77) & 1.3(4)\\ 
 28 & 6.00 & 211 & 2.096(14)  & 2.127(14) & 2.3(9) & 2.090(11)  & 2.126(12) & 2.0(8) & 2.0816(87)  & 2.1260(89) & 1.7(7)\\ 
 28 & 6.50 & 201 & 1.790(14)  & 1.817(14) & 3(1) & 1.783(11)  & 1.814(11) & 2(1) & 1.7744(83)  & 1.8122(84) & 2.1(9)\\ 
 28 & 7.00 & 211 & 1.541(12)  & 1.564(12) & 3(1) & 1.5374(87)  & 1.5645(89) & 2.3(9) & 1.5326(66)  & 1.5653(68) & 2.0(7)\\ 
 \hline 
32 & 4.02 & 372 & 10.905(97)  & 11.029(98) & 5(2) & 10.790(75)  & 10.935(76) & 5(2) & 10.691(55)  & 10.864(56) & 4(2)\\ 
 32 & 4.03 & 372 & 10.387(64)  & 10.505(65) & 3(1) & 10.262(54)  & 10.400(54) & 3(1) & 10.150(44)  & 10.314(45) & 3(1)\\ 
 32 & 4.05 & 373 & 9.672(69)  & 9.782(70) & 4(2) & 9.526(55)  & 9.654(56) & 4(2) & 9.391(43)  & 9.543(44) & 4(1)\\ 
 32 & 4.07 & 372 & 9.018(67)  & 9.121(68) & 5(2) & 8.903(54)  & 9.023(55) & 5(2) & 8.791(42)  & 8.934(43) & 4(2)\\ 
 32 & 4.10 & 387 & 8.443(85)  & 8.539(85) & 8(4) & 8.302(61)  & 8.414(62) & 7(3) & 8.173(44)  & 8.306(45) & 5(2)\\ 
 32 & 4.15 & 361 & 7.702(42)  & 7.790(42) & 3(1) & 7.575(34)  & 7.677(35) & 3(1) & 7.450(28)  & 7.571(28) & 3(1)\\ 
 32 & 4.20 & 403 & 6.983(51)  & 7.062(52) & 5(2) & 6.888(40)  & 6.980(40) & 4(2) & 6.796(30)  & 6.906(30) & 4(1)\\ 
 32 & 4.25 & 340 & 6.525(37)  & 6.599(37) & 2.6(9) & 6.446(30)  & 6.532(30) & 2.4(8) & 6.362(24)  & 6.465(24) & 2.2(8)\\ 
 32 & 4.30 & 342 & 6.103(49)  & 6.173(49) & 5(2) & 6.024(38)  & 6.105(39) & 4(2) & 5.947(29)  & 6.043(30) & 3(1)\\ 
 32 & 4.40 & 318 & 5.525(38)  & 5.588(38) & 3(1) & 5.441(29)  & 5.514(30) & 2.5(9) & 5.359(23)  & 5.446(23) & 2.3(8)\\ 
 32 & 4.50 & 341 & 5.001(35)  & 5.058(35) & 4(2) & 4.934(26)  & 5.000(27) & 3(1) & 4.867(20)  & 4.946(20) & 3(1)\\ 
 32 & 4.70 & 361 & 4.192(35)  & 4.239(36) & 6(3) & 4.148(29)  & 4.203(29) & 6(3) & 4.102(23)  & 4.168(23) & 6(2)\\ 
 32 & 5.00 & 361 & 3.449(29)  & 3.488(29) & 5(2) & 3.413(22)  & 3.459(23) & 4(2) & 3.376(17)  & 3.431(17) & 4(1)\\ 
 32 & 5.50 & 301 & 2.621(25)  & 2.651(26) & 6(3) & 2.607(20)  & 2.642(20) & 5(3) & 2.589(15)  & 2.631(16) & 5(2)\\ 
 32 & 6.00 & 281 & 2.166(19)  & 2.191(19) & 5(2) & 2.147(14)  & 2.176(14) & 4(2) & 2.128(10)  & 2.162(11) & 3(1)\\ 
 32 & 6.50 & 201 & 1.8208(92)  & 1.8415(93) & 1.7(6) & 1.8093(75)  & 1.8336(76) & 1.5(6) & 1.7969(60)  & 1.8260(61) & 1.4(5)\\ 
 32 & 7.00 & 201 & 1.571(14)  & 1.589(14) & 4(2) & 1.567(10)  & 1.588(10) & 4(2) & 1.5604(76)  & 1.5857(77) & 3(1)\\ 
\end{longtable*}

\section{Alternative Zeuthen flow analysis}
\label{Sec.AltFlow}
\begin{figure*}[t]
  \includegraphics[width=0.99\columnwidth]{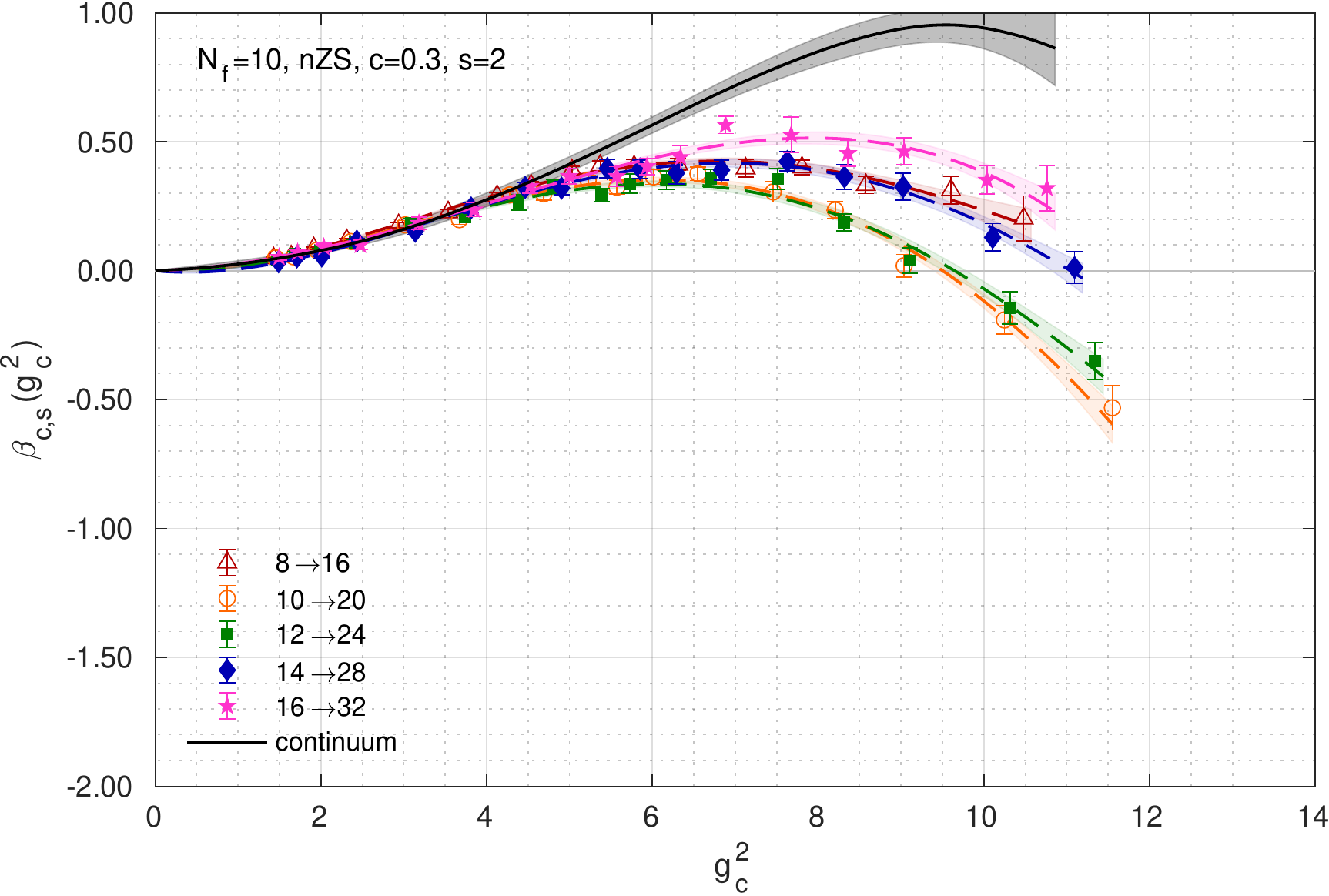}\hfill
  \includegraphics[width=0.99\columnwidth]{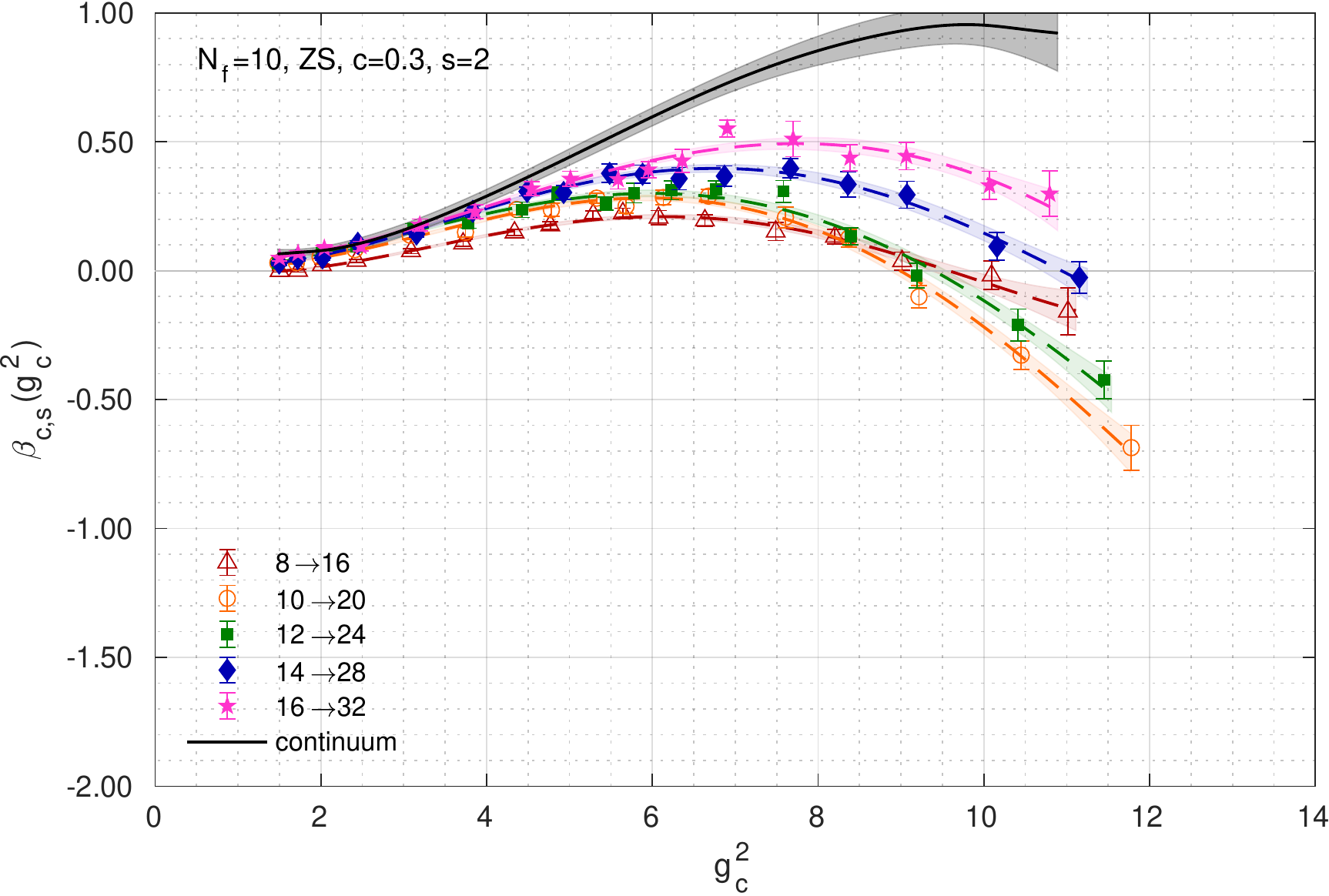}
  \includegraphics[width=0.99\columnwidth]{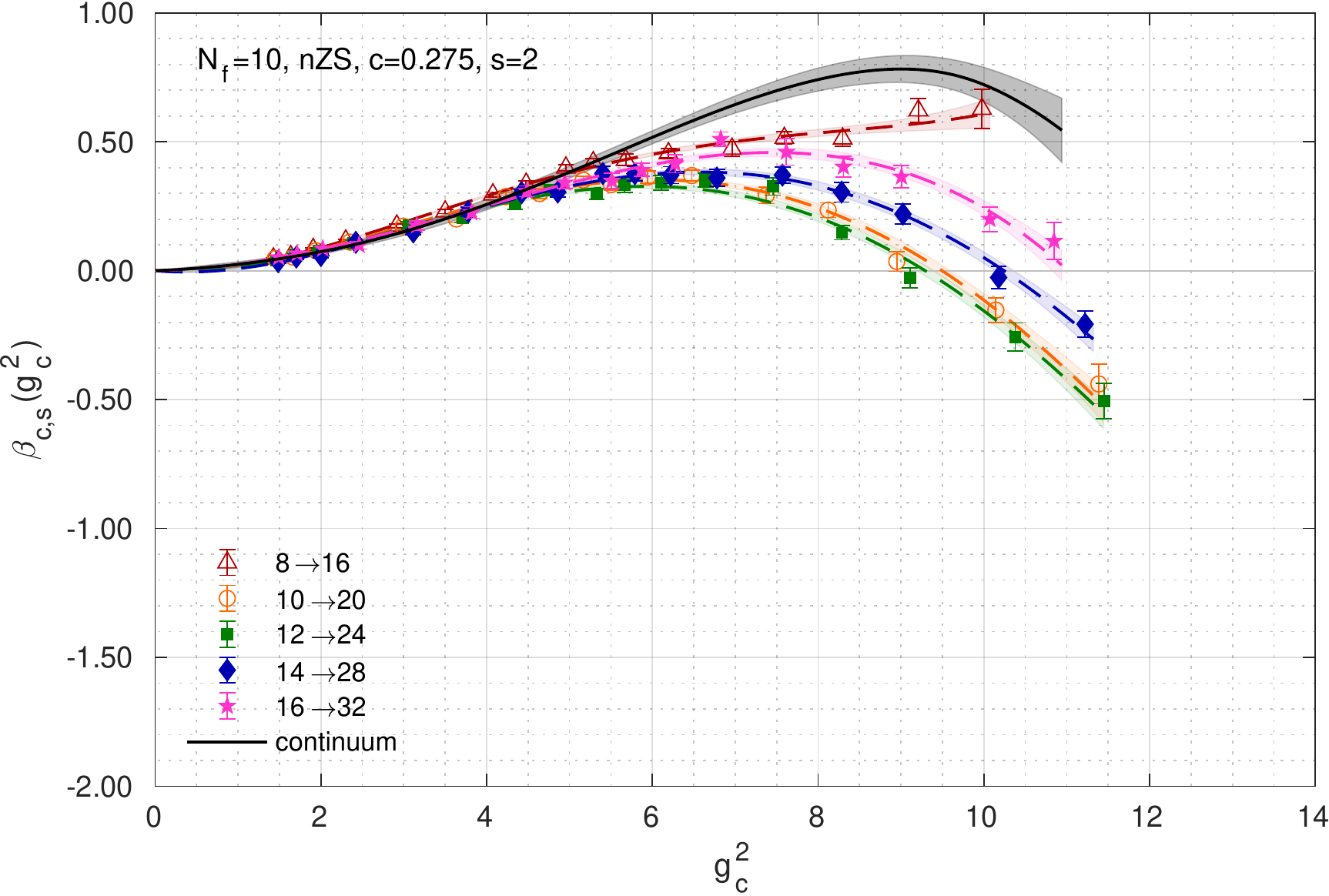}\hfill
  \includegraphics[width=0.99\columnwidth]{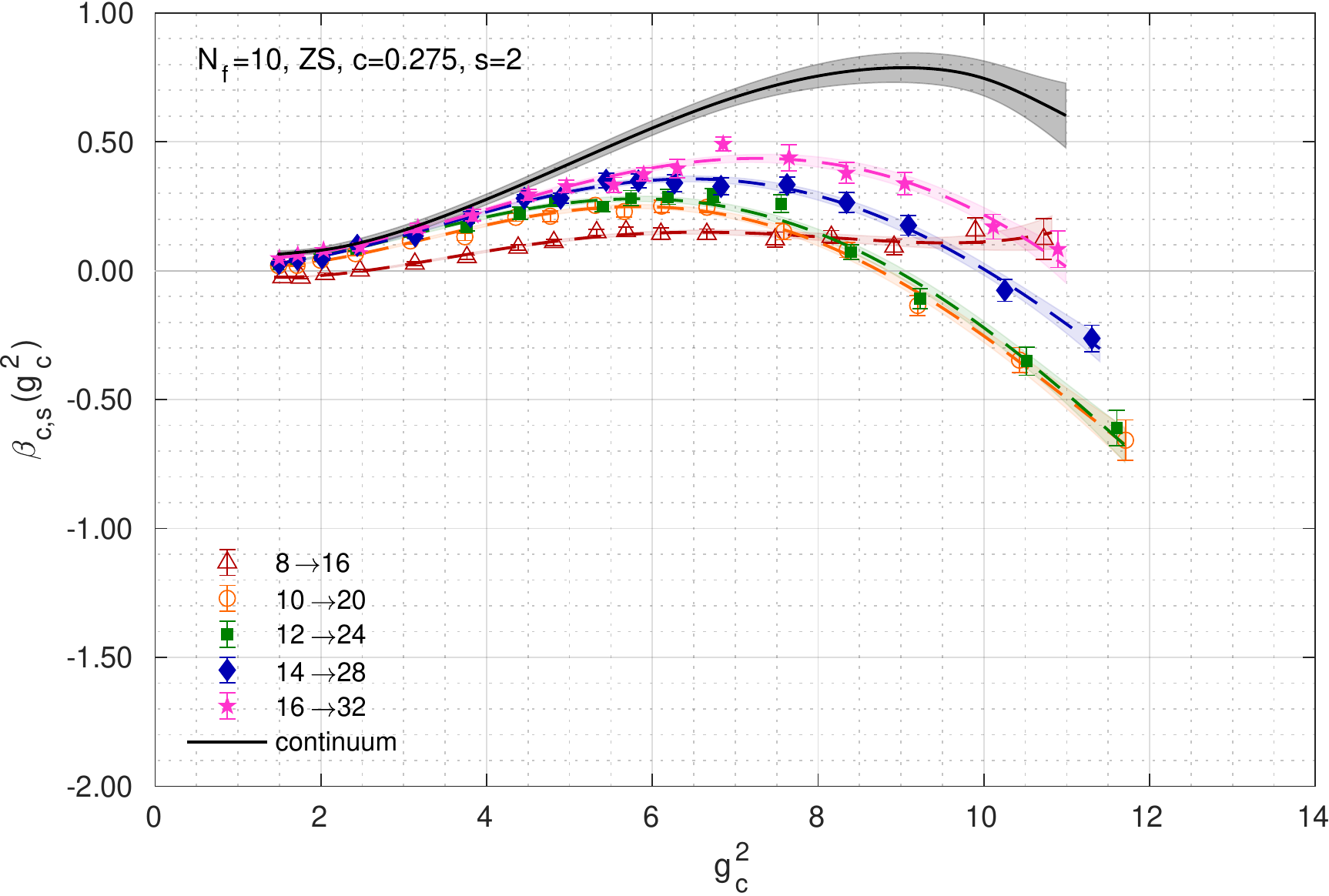}\\
  \includegraphics[width=0.99\columnwidth]{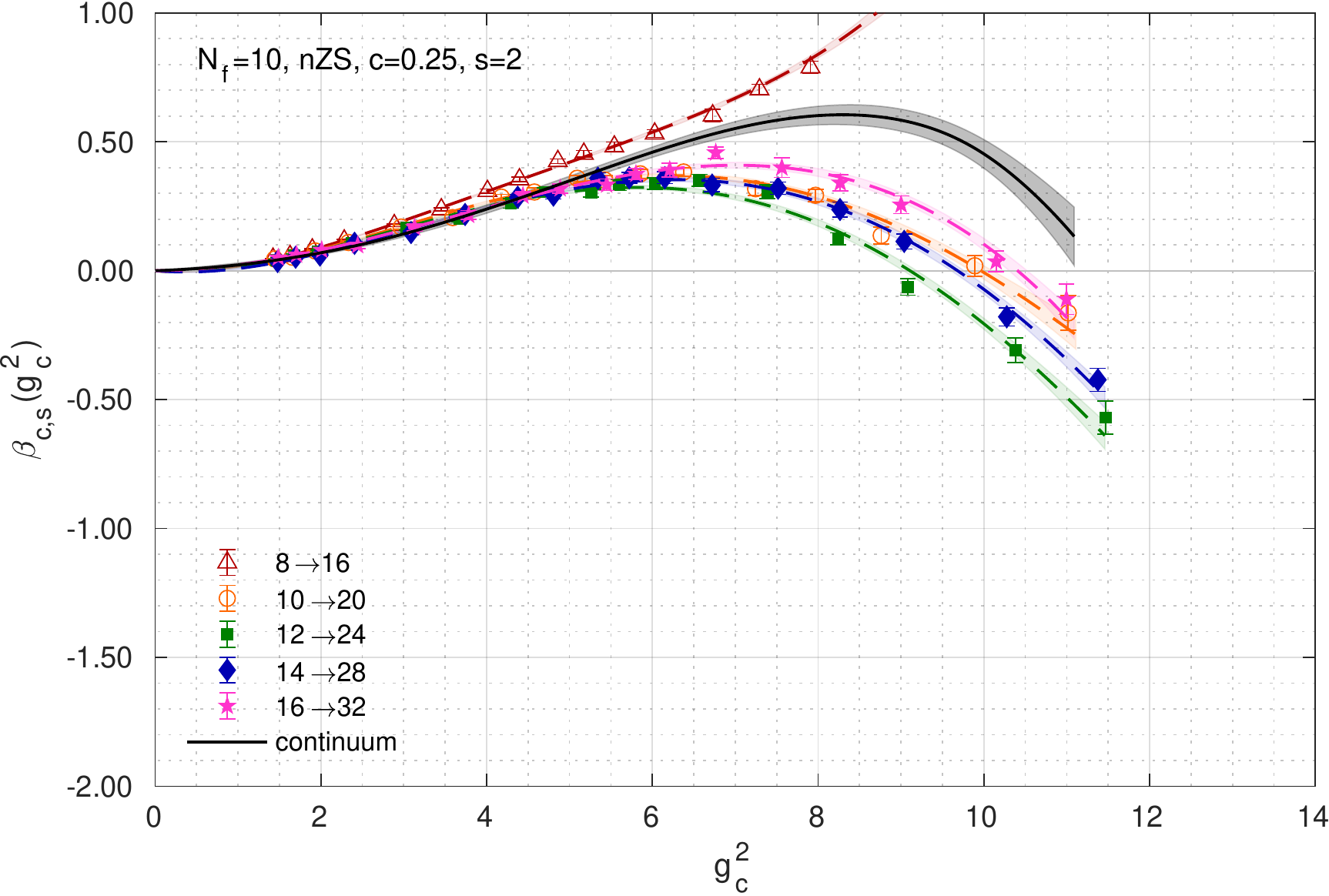}\hfill
  \includegraphics[width=0.99\columnwidth]{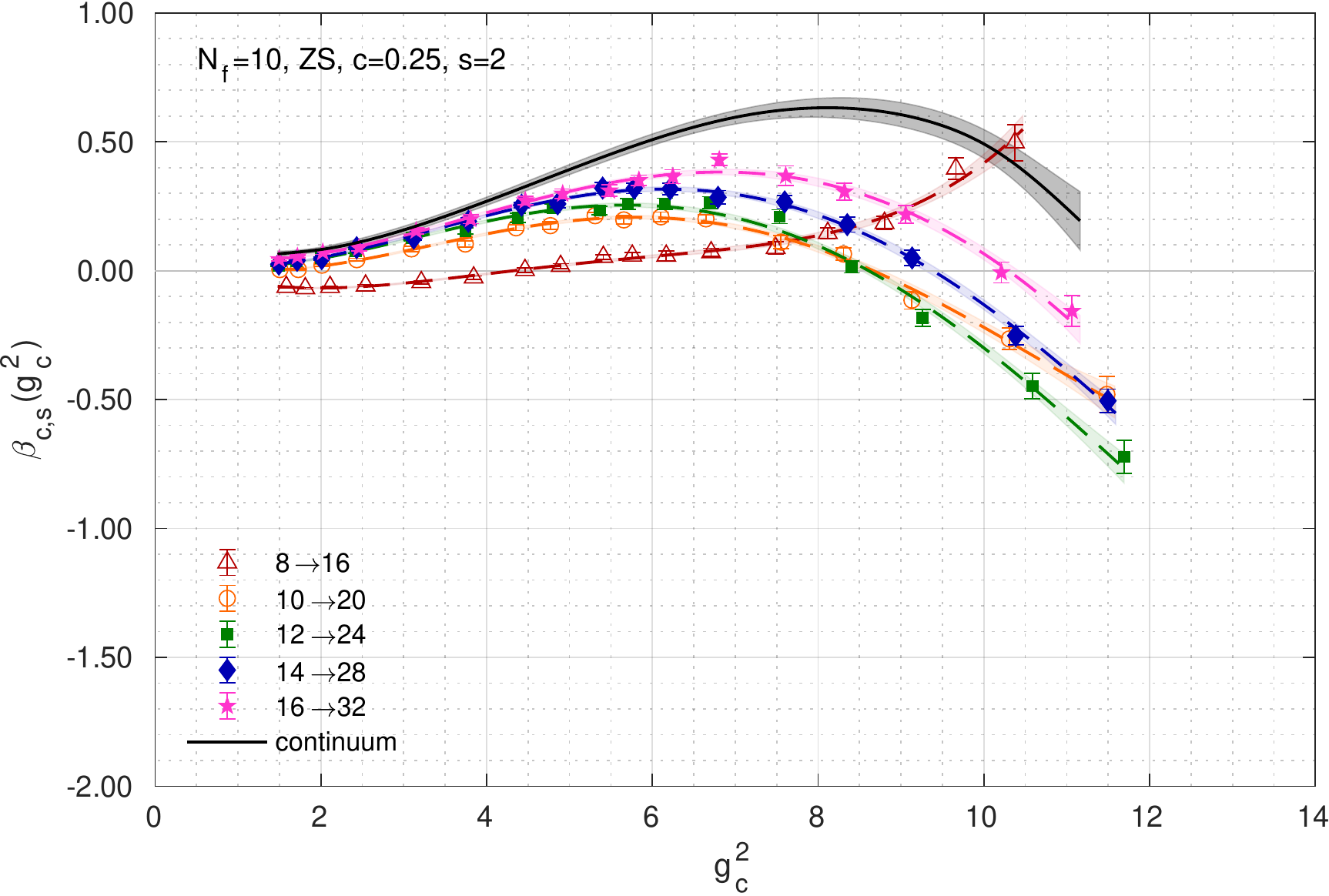}\\
  \caption{Alternative determination of the discrete $\beta$ function using Zeuthen flow with Symanzik operator (without topological charge filtering). Plots on the left show the analysis including the tree-level improvement (nZS), plots on the right without (ZS). From top to bottom we present results for the renormalization scheme $c=0.300$, 0.275, and 0.250. Only statistical errors are shown. }
  \label{Fig.beta_alt_ZS}
\end{figure*}

\begin{figure*}[t]
  \includegraphics[width=0.99\columnwidth]{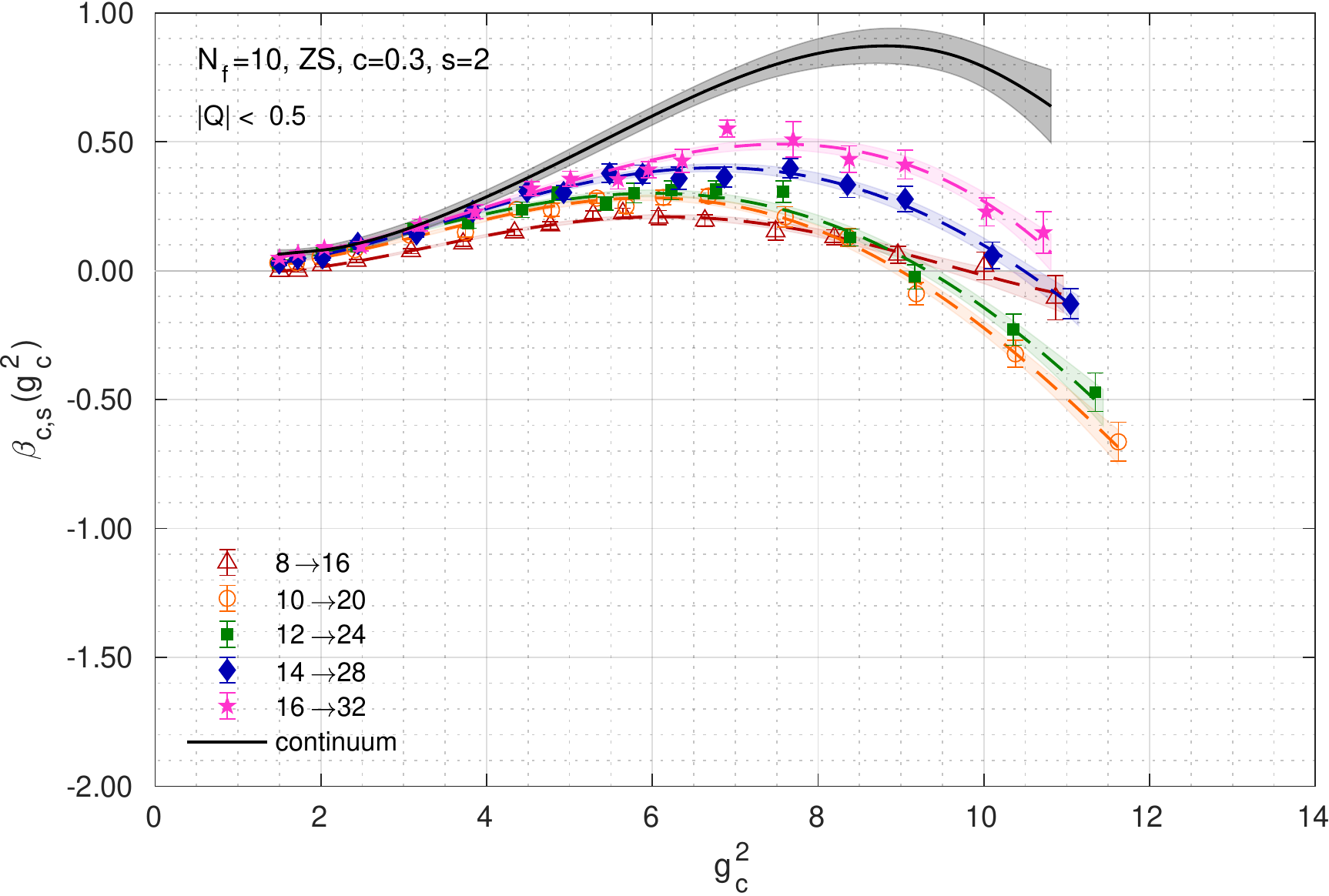}\hfill
  \includegraphics[width=0.99\columnwidth]{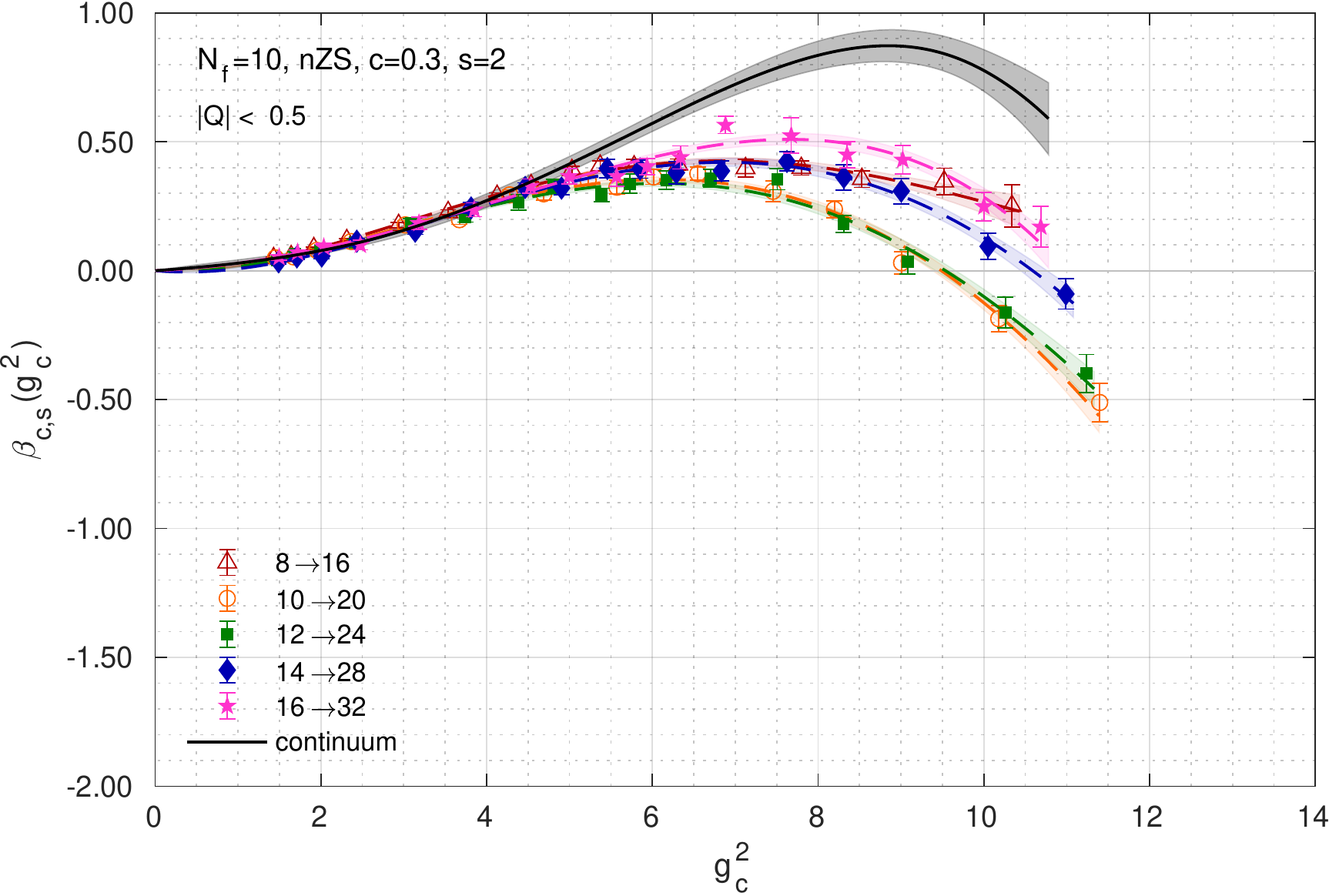}  
  \includegraphics[width=0.99\columnwidth]{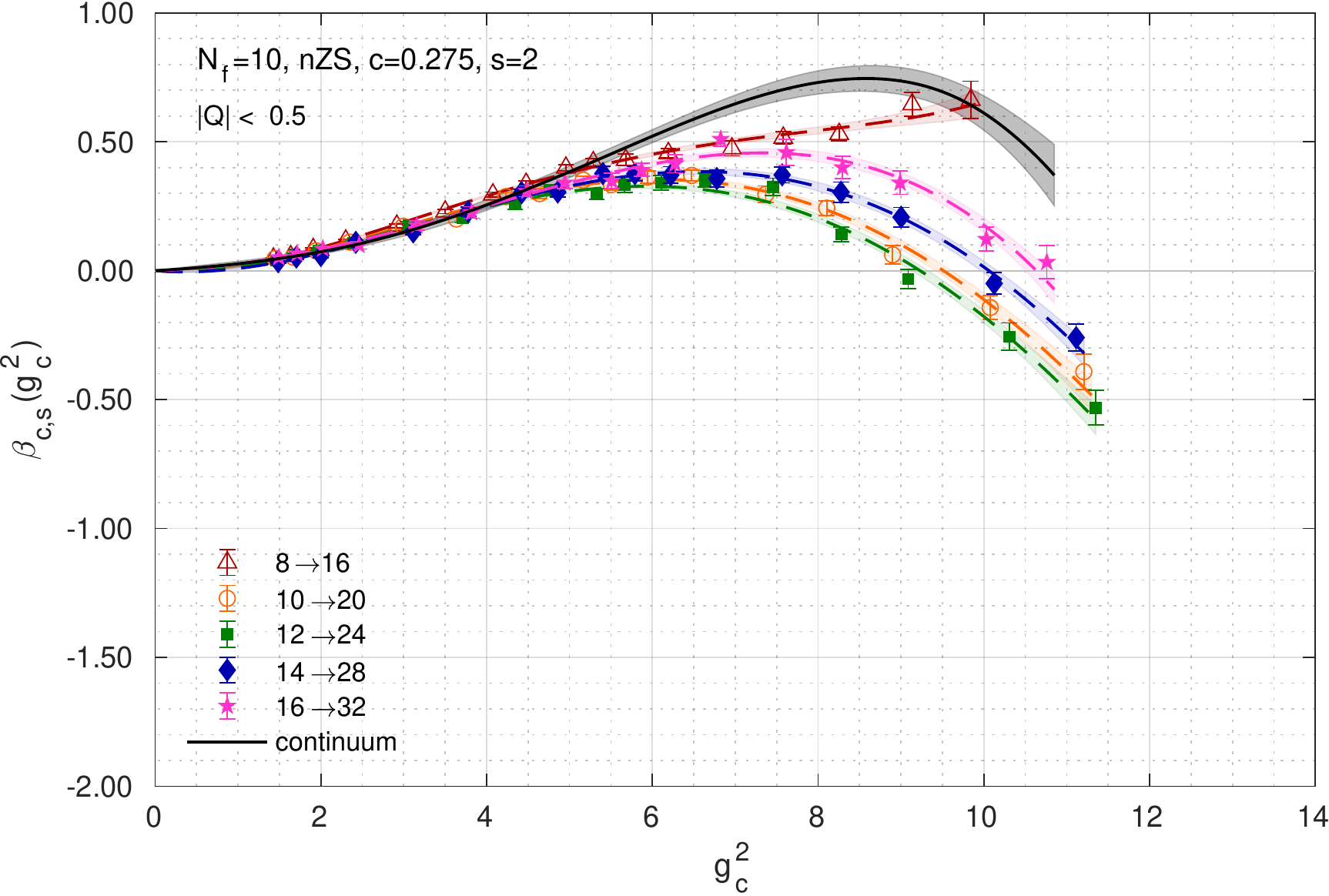}\hfill  
  \includegraphics[width=0.99\columnwidth]{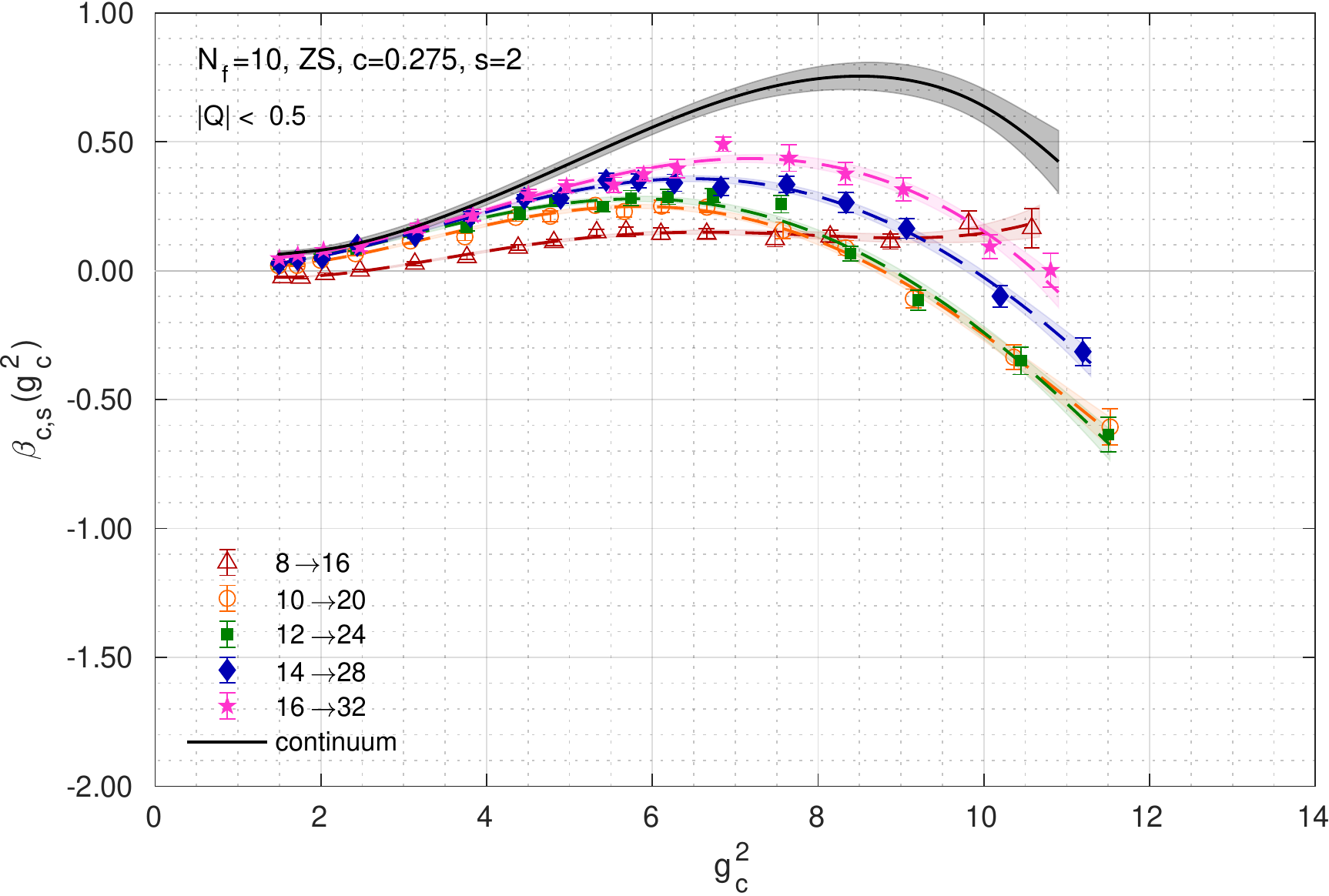}\\
  \includegraphics[width=0.99\columnwidth]{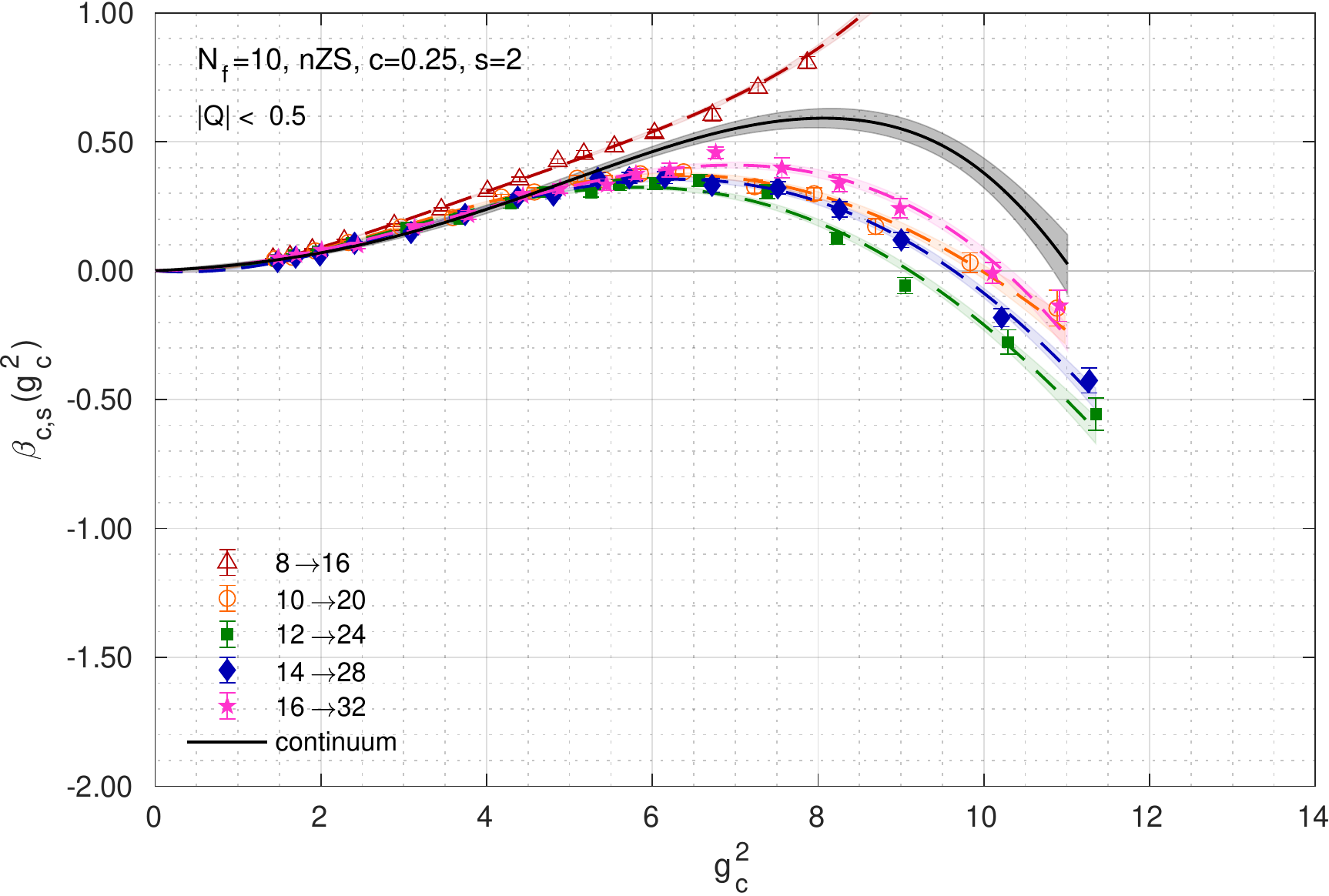}\hfill
  \includegraphics[width=0.99\columnwidth]{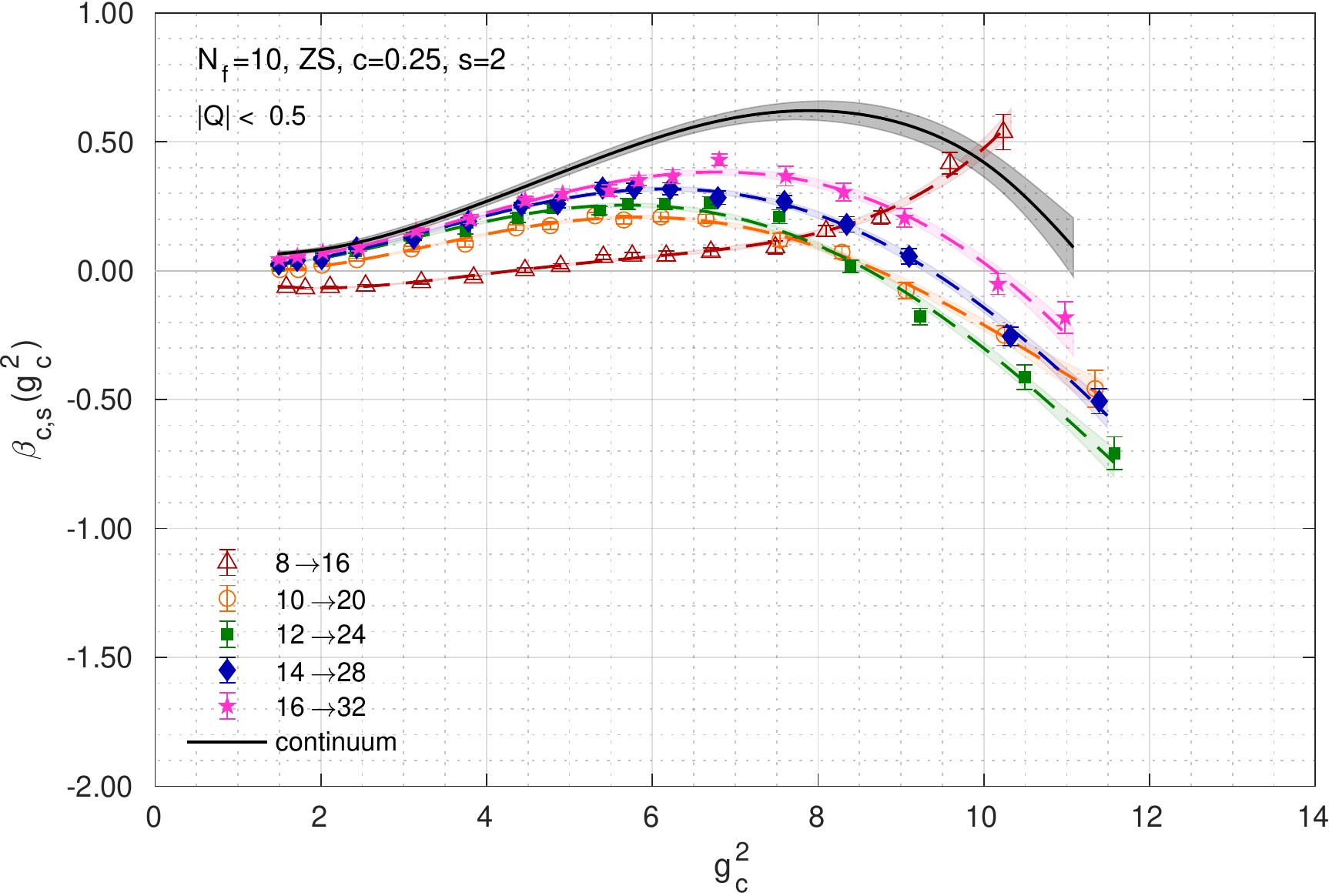}\\
  \caption{Alternative determination of the discrete $\beta$ function using Zeuthen flow with Symanzik operator with topological charge filtering. Plots on the left show the analysis including the tree-level improvement (nZS), plots on the right without (ZS). From top to bottom we present results for the renormalization scheme $c=0.300$, 0.275, and 0.250. Only statistical errors are shown.}
  \label{Fig.beta_alt_ZS_topo}
\end{figure*}

Figure \ref{Fig.beta_alt_ZS}  shows the determination of the discrete beta function using Zeuthen flow and Symanzik operator both with and without tree-level improvement in the  $c=0.300$,  0.275, and 0.250 renormalization schemes. Fig~\ref{Fig.beta_alt_ZS_topo} shows the result of the same analysis on topologically filtered $|Q|<0.5$ data set. When filtering we discard all configurations with topological charge $|Q_\text{geom}|>0.5$ at flow time $t=(cL)^2/8$. We compare the filtered and direct results in Fig.~\ref{Fig.gettingfinal}.

\bibliography{../General/BSM}
\bibliographystyle{apsrev4-1} 
\end{document}